\documentclass[aps,prb,reprint,twocolumn,superscriptaddress]{revtex4-2}

\usepackage{graphicx}
\usepackage{bm}
\usepackage{hyperref}
\usepackage{amsmath}
\usepackage{amssymb}

\usepackage{comment}
\usepackage[table]{xcolor}
\usepackage{array}
\usepackage{multirow}
\usepackage{physics}
\usepackage{siunitx}

\begin{document}
\title{Designing topology and fractionalization in narrow gap semiconductor films \\ via electrostatic engineering}
\date{\today}
\author{Tixuan Tan}
\thanks{These authors contributed equally to this work.}
\affiliation{Department of Physics, Stanford University, Stanford, CA 94305, USA}
\author{Aidan P. Reddy}
\thanks{These authors contributed equally to this work.}
\affiliation{Department of Physics, Massachusetts Institute of Technology, Cambridge, Massachusetts 02139, USA}
\author{Liang Fu}
\affiliation{Department of Physics, Massachusetts Institute of Technology, Cambridge, Massachusetts 02139, USA}
\author{Trithep Devakul}
\affiliation{Department of Physics, Stanford University, Stanford, CA 94305, USA}
\begin{abstract}
We show that topological flat minibands can be engineered in a class of narrow gap semiconductor films using only an external electrostatic superlattice potential.
We demonstrate that, for realistic material parameters, these bands are capable of hosting correlated topological phases such as integer and fractional quantum anomalous Hall states and composite Fermi liquid phases at zero magnetic field. Our results provide a path towards the realization of fractionalized topological states in a broad range of materials.

\end{abstract}
\maketitle

\begin{figure}[t]
\begin{center}
\includegraphics[width=0.48\textwidth]{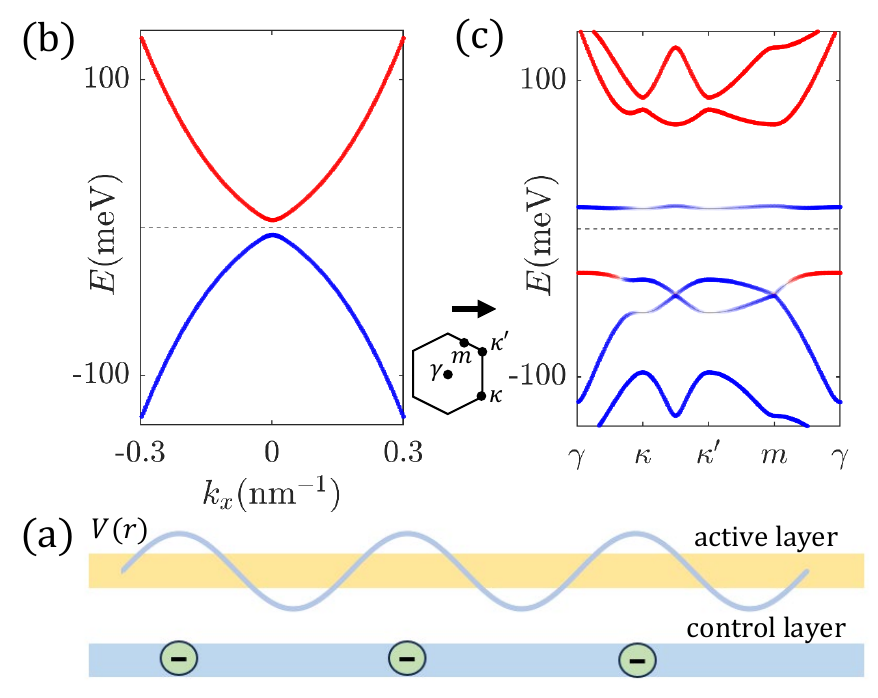}
\end{center}
\caption{ Illustration of the proposed setup. 
\textbf{(a)} A control layer, hosting a periodic charge distribution, imparts an electrostatic superlattice potential on a nearby active layer composed of a narrow gap semiconductor film.
\textbf{(b,c)} The electronic bands in the active layer is illustrated without and with the effect of the superlattice potential.
For an appropriately chosen superlattice potential, topological flat bands can be achieved.
Color of the bands (red and blue) indicates the orbital content of the band.  
}
\label{fig:schematic}
\end{figure}

\emph{Introduction ---} 
Recent years have seen a great effort in the design of quantum materials with tailored electronic properties.  
Topological bands with tunable Chern numbers are particularly sought after  
because 
electron-electron interaction in partially filled flat 
Chern bands can give rise to 
fractional Chern insulators (FCI) \cite{sheng2011fractional,neupert2011fractional,regnault2011fractional,sun2011nearly,tang2011high,parameswaran2013fractional,liu2022recent} that host fractionalized quasiparticle excitations. 
To date, much study has been focused on two-dimensional van der Waals (vdW) heterostructures, which provide a highly versatile platform for realizing topological moir\'e bands~\cite{andrei2021marvels,mak2022semiconductor,bistritzer2011moire,morell2010flat,wu2019topological}.
Various graphene and semiconductor moir\'e superlattices host a time-reversed pair of Chern bands of opposite spin and/or valley polarization.  
In these systems, FCI states     
can form 
at zero magnetic field when Coulomb interaction drives spontaneous spin/valley ferromagnetism \cite{li2021spontaneous, crepel2023anomalous}. 
In twisted bilayer MoTe$_2$ \cite{cai2023signatures,zeng2023thermodynamic,park2023observation,xu2023observation} and hBN-aligned pentalayer rhombohedral graphene \cite{lu2023fractional}, these zero-field FCI states showing fractionally quantized anomalous Hall (FQAH) effects have been experimentally observed.  

Despite their remarkable tunability, moir\'e devices made from exfoliated 2D material flakes face challenges in terms of robustness and scalability in production \cite{lau2022reproducibility}.
In contrast, epitaxially grown semiconductor films and heterostructures have 
historically provided high-mobility electron gases for quantum Hall research~\cite{stormer1999fractional}. 
A natural and pressing question is whether it is possible to create fractionalized electron phases in semiconductor films at zero magnetic field.          


In this work, we present a general mechanism for realizing topological and fractionalized phases in thin films of narrow gap semiconductors through electrostatic engineering.
We show, using a universal low-energy model, that an externally applied electrostatic superlattice potential can lead to strong hybridization of the parent conduction and valence bands, resulting in the formation of topological minibands with highly tunable bandwith and topology. Through many-body calculations, we demonstrate that robust integer and fractionalized topological states can be realized with realistic model parameters.




The proposed setup is illustrated in Fig~\ref{fig:schematic}. 
The ``active layer'' is a narrow gap semiconductor film, whose low-energy electronic structure is illustrated in Fig~\ref{fig:schematic} (b).
It is placed in close proximity to a ``control layer'', featuring a periodic modulation of charge density with a period on the order of $10$ nanometers and experimentally feasible. 
The periodic charge density may be realized, for instance, via a doped moir\'e superlattice~\cite{gu2023remote,zhang2024engineering,he2024dynamically}.
The modulated charge density in the control layer produces an electrostatic superlattice potential acting on electrons/holes in the active layer, resulting in the formation of topological minibands, Fig~\ref{fig:schematic}(b,c).
This method 
for realizing an electrostatic superlattice potential has been demonstrated experimentally in several available platforms~\cite{gu2023remote,zhang2024engineering,he2024dynamically,forsythe2018band,yasuda2021stacking,wang2022interfacial,zhao2021universal,kim2023electrostatic,woods2021charge,shi2019gate,xu2021creation,barcons2022engineering,sun2023signature,li2021anisotropic,wang2024bandstructureengineeringusing}; and
such setups have been explored with graphene~\cite{song2015topological,huber2020gate,ghorashi2023topological,ghorashi2023multilayer,zeng2024gatetunable} { and 2D electron gases~\cite{albrecht1999fermiology,wang2024tuning} as active layers.}

\emph{Designing topological bands ---}
We take as our starting point the two-band $k\cdot p$  Hamiltonian ~\cite{fu2007topological,bernevig2006quantum}, 
\begin{equation}\label{eq:TopoinversionHam}
H_{0}^{\tau}(\bm{k})=\begin{pmatrix}
\alpha_1 |\bm{k}|^2+\frac{\delta}{2}& v( \tau k_x-i k_y)\\
v( \tau k_x+i k_y)& -\alpha_2 |\bm{k}|^2-\frac{\delta}{2}
\end{pmatrix}
\end{equation}
 $H_0^\tau$ describes a large class of narrow gap semiconductor films in which the conduction and valence bands have opposite parity, containing all terms up to $k^2$ allowed by inversion, time reversal and rotational symmetry~\cite{fu2007topological,supp}.
Eq~\ref{eq:TopoinversionHam} is written in the eigenbasis 
of $\sigma_z$, labeled $\{\ket{1},\ket{2}\}$, which 
represents the orbitals making up the conduction and valence bands. 
$\tau=\pm 1$ denotes Kramers degeneracy associated with electron's spin.  
The effective parameters $\delta,v,\alpha_1,\alpha_2$ are material and thickness dependent.
{ Thus far, $H_0^{\tau}(\bm{k})$ describes the low-energy electronic properties of a pristine 2D semiconductor film.} 


We consider the effect of an electrostatic superlattice potential. 
We focus on the $C_3$ symmetric potential, 
\begin{equation}
V(\bm{r})=2V_0\sum_{n=1}^{3} \cos(\bm{g}_n\cdot\bm{r}+\phi)\label{eq:VrC3}
\end{equation}
where $\bm{g}_n=g(-\cos\frac{2\pi n}{3},\sin\frac{2\pi n}{3})$. The potential is characterized by its strength $V_0$, reciprocal lattice vector $g=\frac{4\pi}{\sqrt{3}a}$ where $a$ is the superlattice period, and the parameter $\phi$ which controls the shape of the potential.  
We fix $\phi=\pi$, which gives 
a superlattice potential with minima forming a triangular lattice and maxima forming a honeycomb lattice. 
We take $V(r)$ to couple only to the in-plane density, i.e. the coupling is proportional to identity in $\sigma$.   
{ effect of the finite thickness of the film, as well as other potential shapes $\phi$, are discussed in the Supplemental material~\cite{supp}.
The potential}
$V(r)$ 
produces mini-bands in the mini-Brillouin zone defined by $\bm{g}$, as illustrated in Fig~\ref{fig:schematic}b.

At very low energy or very long wavelengths, the $k^2$ term in Eq~\ref{eq:TopoinversionHam} can be neglected ($\alpha_1,\alpha_2\rightarrow 0$), in which case $H_0^\tau$ describes a massive Dirac fermion.
In this limit, an appropriate choice of superlattice potential $V(r)$ can be used to realize topological single-particle bands~\cite{Su_2022,suri2023superlattice}.
However, as shown in the Supplementary Material~\cite{supp}, we find that Coulomb interactions are unable to stabilize a spin-polarized Chern insulator at integer filling in this limit. 
We therefore turn to the full $k \cdot p$ Hamiltonian $H_0^\tau$ including the quadratic terms.

For simplicity, we set $\alpha_1=\alpha_2\equiv\alpha$. 
We define an energy scale $E_D=v^2/\alpha$ and momentum scale $k_D=v/\alpha$, which characterizes the energy and momentum above which the massive Dirac fermion description breaks down and the quadratic terms in $H_0$ becomes important.
Up to these overall energy and momentum scales, the single particle physics of our model is controlled entirely by three dimensionless parameters $\delta/E_D$, $V_0/E_D$, and $g/k_D$.

\begin{figure}
    \centering
    \includegraphics[width=1\linewidth]{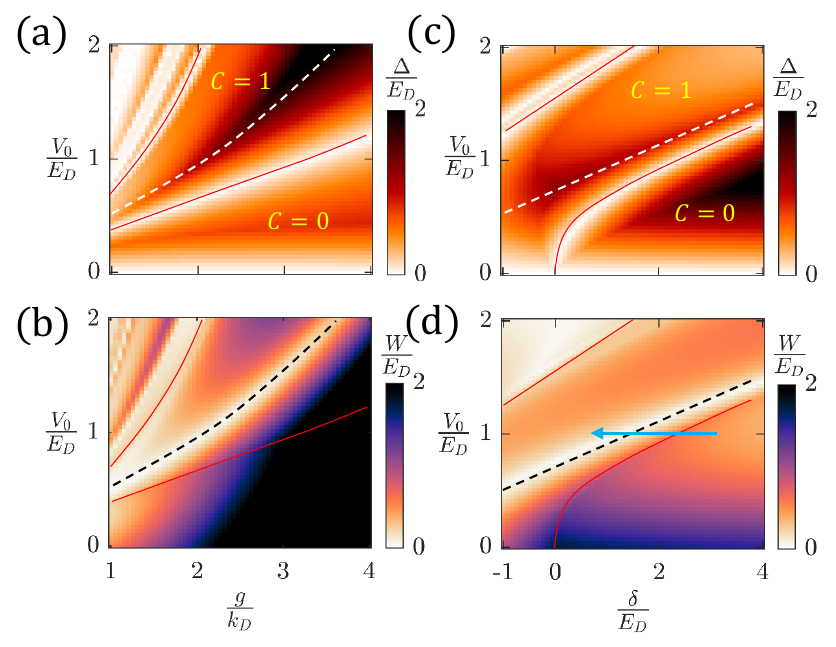}
    \caption{ 
    Phase diagram of the model.
    The \textbf{(a,c)} direct gap $\Delta$, \textbf{(b,d)} bandwidth $W$ of the first conduction miniband as a function of superlattice potential strength $V_0$, reciprocal wave vector $g$, and pristine gap $\delta$. 
    \textbf{(a,b)} is plotted at $\frac{\delta}{E_D}=1$, and \textbf{(c,d)} is plotted at $\frac{g}{k_D}=2$. 
    The solid lines delineates the boundary of topological $C=1$ phase, and 
    the dashed line indicates where the bandwidth is minimized. 
    The blue arrow in \textbf{(d)} is discussed in the main text.
    }
    \label{fig:TBmagic}
\end{figure}

We focus on the properties of the first conduction miniband of $\tau=+$ (bands of 
$\tau=-$ are related by time reversal symmetry).
In Fig~\ref{fig:TBmagic}(a,b), we show the bandwidth ($W$) and minimum direct gap ($\Delta$) as a function of the dimensionless superlattice potential parameters $V_0/E_D$ and $g/k_D$, for a fixed $\delta/E_D=1$.
We highlight a robust $C=1$ region spanning a wide parameter range. Importantly, topological minibands are formed even when the pristine semiconductor films have topologically trivial band structure ($\delta>0$).   
In a large portion of this region, $W\ll\Delta$, implying that Coulomb interactions are likely to result in a spin polarized quantum anomalous Hall (QAH) insulator at filling $n=1$ electron per superlattice unit cell -- a scenario we will soon confirm. 

Within this region, we further find an optimal line (indicated by the dashed line) along which the band becomes almost perfectly flat. 
This flat band is nearly ``ideal'' for the realization of fractionalized topological phases at partial fillings~\cite{parameswaran2012fractional,roy2014band,claassen2015position,jackson2015geometric,wang2021exact,ledwith2020fractional,ledwith2022vortexability}, and is similar to those previously found in certain moir\'e materials~\cite{bistritzer2011moire,khalaf2019magic,devakul2021magic,devakul2023magic,morales2023magic,reddy2023toward,crepel2023chiral} (we discuss various quantum geometric indicators in the Supplemental material~\cite{supp}).  
In Fig~\ref{fig:TBmagic}(c,d) we show the same quantities, but as a function of $V_0/E_D$ and $\delta/E_D$, for a fixed value of $g/k_D=2$.
We again find a robust $C=1$ phase accompanied by a line of almost vanishing bandwidth.
Collectively, our findings imply that the topological $C=1$ phase, with $W\ll \Delta$, is robust over a wide region of the three-dimensional parameter space, and that the bandwidth nearly vanishes along an optimal two-dimensional surface in this space.





\begin{figure}
    \centering
    \includegraphics[width=\linewidth]{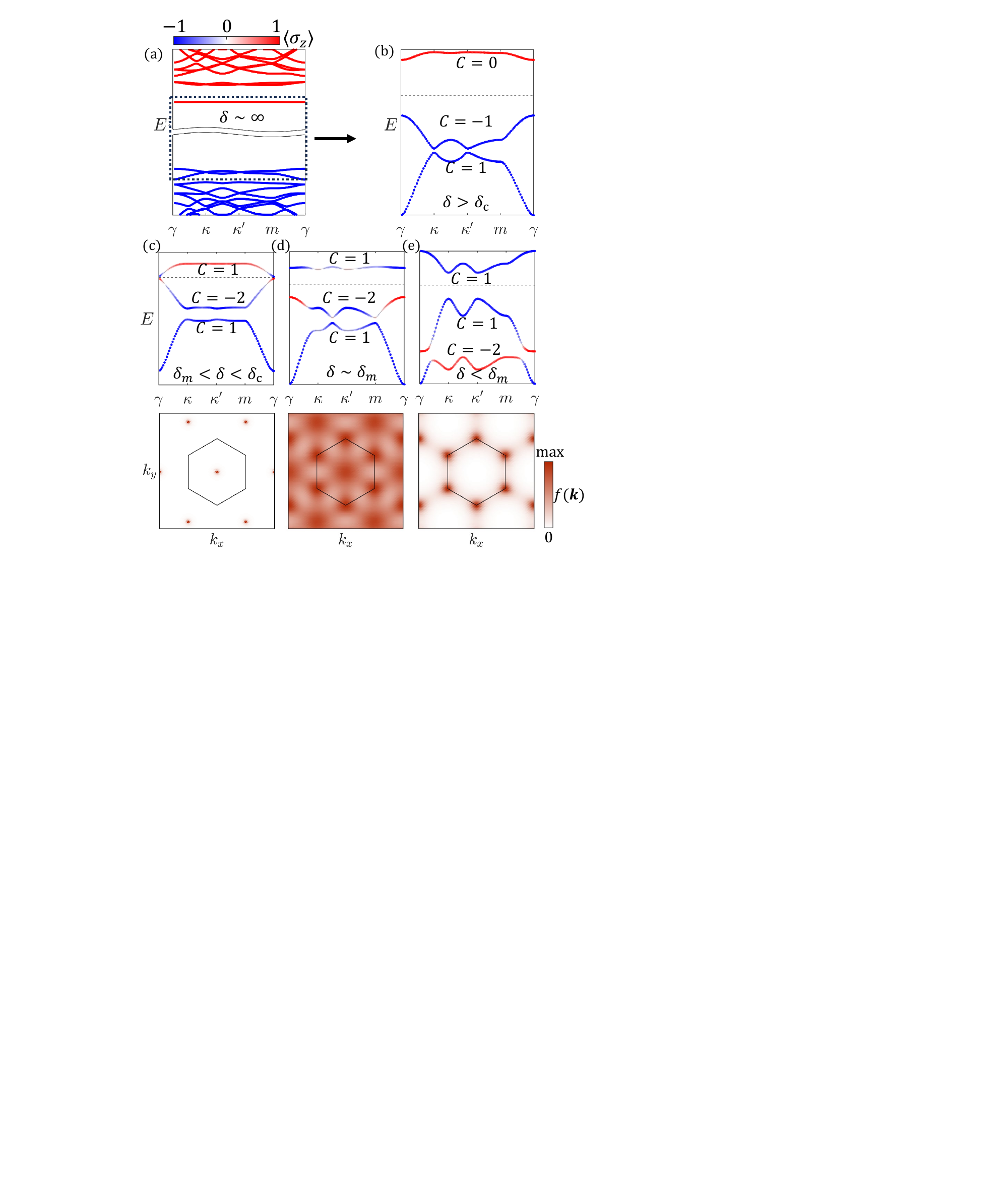}
    \caption{Illustration of the first minibands as $\delta$ is tuned.
    The color indicates the $\langle\sigma_z\rangle$ expectation value.
    The Chern numbers of each miniband are labeled. 
    We show
    \textbf{(a)} the large $\delta$ limit,
    \textbf{(b)} $\delta>\delta_c$ before inversion,
    \textbf{(c)} $\delta_m<\delta<\delta_c$ immediately after inversion,
    \textbf{(d)} $\delta=\delta_m$ at the magic value, and
    \textbf{(e)} $\delta<\delta_m$ below the magic value.
    Bottom panels of \textbf{(c,d,e)}
     show the corresponding Berry curvature $f(\bm{k})$ of the first conduction miniband, with boundary of mBZ indicated by the black solid line. 
     The bandwidth is minimized and the Berry curvature distribution is most uniform near $\delta_m$.
     }
    \label{fig:TBinvert}
\end{figure}

To understand the emergence of the topological flat band, we consider tuning $\delta$ while keeping all other parameters fixed.
First, consider the limit of large $\delta\rightarrow\infty$, in which the conduction band (from orbital $\ket{1}$) and valence band (from orbital $\ket{2}$) of $H_0^{\tau}$ are separated in energy and essentially unhybridized.
As illustrated in Fig~\ref{fig:TBinvert}a, in the presence of the superlattice potential $V(r)$, the first (lowest-energy) $\ket{1}$ miniband resembles that of a triangular lattice tight binding model centered at the potential minima; while the first two (highest-energy) valence minibands from $\ket{2}$ resembles those 
of a honeycomb lattice centered at the potential maxima.
At large but finite $\delta$, Fig~\ref{fig:TBinvert}b, the mini-Dirac points at $\kappa$ and $\kappa^\prime$ in the $\ket{2}$ bands are gapped out by an effective Haldane type term, generated by virtual tunneling processes with the $\ket{1}$ band, resulting in topological bands.

We now consider reducing $\delta$, along the trajectory marked by the blue arrow in Fig~\ref{fig:TBmagic}d.
There are two notable points: a critical $\delta_c$ at which the gap closes, and a ``magic'' $\delta_m$ at which the bandwidth is minimized, corresponding to the solid and dashed lines.

The emergence of the nearly flat band can be understood by considering $\delta$ slightly 
above and below $\delta_m$.
At $\delta=\delta_c > \delta_m$, 
the first conduction and valence minibands undergo a topological band inversion at $\gamma$, resulting in a topological first conduction band with an energy 
minimum at $\gamma$, shown in Fig~\ref{fig:TBinvert}c.
When 
$\delta<\delta_m$, Fig~\ref{fig:TBinvert}e, this band evolves smoothly into an upper honeycomb band of majority $\ket{2}$ character, with energy 
minima at $\kappa$ and $\kappa^\prime$.
The magic $\delta_m$, Fig~\ref{fig:TBinvert}d, occurs between these two limits: the energy dispersion is suppressed as the energy extrema swap positions in the Brillouin zone.

Beyond just bandwidth, the Berry curvature $f(k)$ also becomes extremely uniform near $\delta_m$. 
Immediately following the mini-band inversion, $\delta\approx\delta_c>\delta_m$ (Fig~\ref{fig:TBinvert}c), Berry curvature concentrates near the inversion point $\gamma$. In contrast, when $\delta<\delta_m$ (Fig~\ref{fig:TBinvert}e), the band's honeycomb character leads to Berry curvature hotspots at $\kappa$, $\kappa$'. Near the crossover $\delta \sim {\delta}_m$ between these extremes (Fig~\ref{fig:TBinvert}d), Berry curvature fluctuations are suppressed.
Additionally, due to the crossover from triangular to honeycomb-like bands, the charge density in real space also becomes evenly distributed near $\delta_m$~\cite{supp}.

We remark that this crossover is highly reminiscent of a similar mechanism leading to nearly ideal flat bands in twisted TMD homobilayers \cite{devakul2021magic,reddy2023fractional,morales2023magic}, a platform in which FQAH has been observed experimentally \cite{cai2023signatures,zeng2023thermodynamic,park2023observation,xu2023observation}.
Similar flat bands have also been observed in the context of quasiperiodicity~\cite{fu2020magic,fu2021flat}.
{Finally, although we have considered tuning $\delta$ to make the physical origin of the flat bands clear, any other parameter can also be used to tune to the flat band.}

\begin{figure}
    \centering
    \includegraphics[width=1.0\columnwidth]{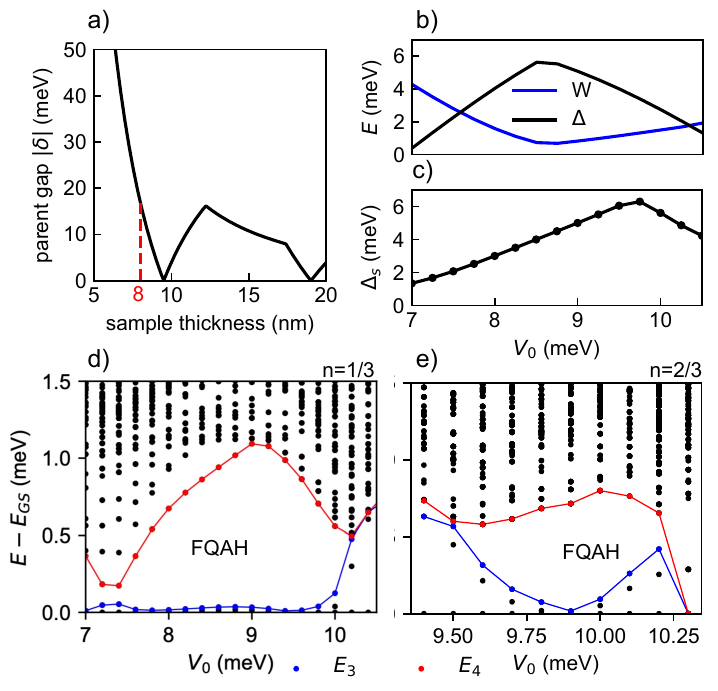}
    \caption{{
    Results for Cd$_3$As$_2$ thin film.
    (a) Parent bandgap $|\delta|$ as a function of its thickness, $L=\SI{8}{nm}$ is chosen based on the criterion discussed in the main text. (b) Lowest valence mini-bandwidth $W$ and minimum direct gap $\Delta$ as a function of superlattice potential strength $V_0$. (c) Magnon gap $\Delta_s = E(S_{z}^{\mathrm{max}}-1) -E(S_{z}^{\mathrm{max}})$ at $n=1$ as a function of $V_0$, demonstrating a robust $|C|=1$ 
    QAH state. 
    Evolution of low-energy spectrum within the maximum $S_z$ sector at (d) $n=\frac{1}{3}$ and (e) $\frac{2}{3}$ as a function of $V_0$, showing FQAH states. 
    The effective parameters are: $\delta=\SI{16.9}{meV},v=\SI{88.9}{meV nm},\alpha_1=\SI{250}{nm^2},\alpha_2=\SI{20}{meV nm^2}$~\cite{supp};
    $a=\SI{20}{nm}$, $\phi=0$, and $\epsilon=10$.
    The finite system used contains 27 superlattice unit cells. 
    }}
    \label{fig:TIED}
\end{figure}

{
\emph{Example ---} As an example of our theory, we consider Cd$_3$As$_2$.
In bulk form, Cd$_3$As$_2$ is a 3D Dirac semimetal~\cite{wang2013three,armitage2018weyl,borisenko2014experimental,neupane2014observation,liu2014stable,ali2014crystal,crassee20183d}.
In films, it
oscillates between a 2D trivial or topological insulator~\cite{wang2013three,lygo2023two} depending on thickness, in good agreement with low-energy $k\cdot p$ modeling~\cite{cano2017chiral,baidya2020first,wang2013three,miao2023engineering}.
Fig.~\ref{fig:TIED}(a) shows the parent bandgap $|\delta|$ (that is, the bandgap in the absence of a superlattice potential) of Cd$_3$As$_2$ thin film as a function of its thickness $L$ obtained by solving the 3D $k\cdot p$ model~\cite{cano2017chiral} in the thin film geometry~\cite{supp}.
The film is a non-trivial 2D $Z_2$ topological insulator for $\SI{10}{nm}\lessapprox L \lessapprox \SI{19}{nm}$.

To identify optimal film thicknesses for realizing topological flat minibands by our mechanism, we observe from Fig~\ref{fig:TBmagic}(c,d) that the bandwidth-to-bandgap ratio is most optimal when $\delta$ is small and positive.
This suggests that one should choose $L$ to be close to a topological band inversion, on the trivial side.
Based on this criterion and Fig.~\ref{fig:TIED}(a), we take $L=\SI{8}{nm}$.
We remark that this criterion may be useful as an experimental ``shortcut'' to identifying optimal film thicknesses even in the absence of knowledge of full microscopic band structure.

We derive an effective two-band $k\cdot p$ Hamiltonian for $\SI{8}{nm}$ Cd$_3$As$_2$ thin film of the form of Eq~\ref{eq:TopoinversionHam} with parameters listed in Fig~\ref{fig:TIED}~\cite{supp}.
Note that because $\alpha_1\neq \alpha_2$, this model is particle-hole asymmetric and does not map exactly to Fig~\ref{fig:TBmagic}.
Nevertheless, we still expect the topological flat band based on our physical picture for their origin. 
As the valence band is significantly more massive ($\alpha_2 \ll \alpha_1$), we consider the first valence miniband in the presence of an electrostatic potential of the form Eq.~\ref{eq:VrC3} with $\phi=0$
and a realistic superlattice period $a=\SI{20}{nm}$.
For a range of $V_0$, 
the first valence miniband becomes topological ($C=\pm 1$), flat, and well isolated as shown in Fig.~\ref{fig:TIED}(b).
}

\emph{Many-body analysis ---} Having established the existence of flat Chern bands, we now demonstrate their potential for realizing 
topological and fractionalized phases via a fully interacting many-body analysis.
{Here, we focus the first valence miniband of Cd$_3$As$_2$ thin films as introduced above.} We define the filling factor $n$ as the number of holes per superlattice unit cell relative to charge neutrality. 
Focusing on the filling range $0<n\leq 1$, we introduce a two-body Coulomb interaction $V_C(r)=\frac{e^2}{\epsilon|r|}$ {between holes} and 
perform exact diagonalization within the Hilbert subspace of the time-reversed pair of {first} valence minibands. 
Finally, we define the total spin $S_z = \frac{1}{2}(N_{+} - N_{-})$, where $N_{\tau}$ is the number of electrons in the $\tau$ Chern band, which 
is a conserved quantum number.




In Fig.~\ref{fig:TIED}(c), we show the magnon gap $\Delta_s$ at $n=1$, defined as the energy difference between the lowest-energy states in the fully polarized $S_z=S_{z}^{\mathrm{max}}$ sector and the $S_z=S_{z}^{\mathrm{max}}-1$ sector.
We find that, 
{throughout a range of $V_0$}, $\Delta_s>0$ indicating a fully spin-polarized Chern insulator ground state~\cite{supp}.
This robust integer QAH effect can be attributed to flat band ferromagnetism, arising due to the topological band with $W\ll \Delta$.

Next, we examine 
fractional fillings $n=\frac{1}{3}$ and $\frac{2}{3}$.  
 Fig.~\ref{fig:TIED} (d) and (e) demonstrate the evolution of the low energy spectra within the fully spin-polarized sector at $n=\frac{1}{3},\frac{2}{3}$ with varying $V_0$. 
The large gap between the fourth and third lowest energy states marked 
by the red and blue lines, 
 indicates a gapped, three-fold quasi-degenerate ground state. 
By examining their center-of-mass momenta, we confirm that these ground states 
are fractional Chern insulators, as opposed to a charge density waves with tripled unit cells~\cite{supp,reddy2023fractional,reddy2023toward,morales2023pressure,wilhelm2021interplay}. We have verified on smaller systems that the ground state at these filling factors is indeed fully spin-polarized~\cite{supp}. {As shown in the Supplemental Material, we also find evidence for an anomalous composite Fermi liquid at $n=\frac{1}{2}$ \cite{supp,goldman2023zero,dong2023composite}.}





\emph{Discussion ---}
We have therefore demonstrated a mechanism by which topological flat bands and fractionalized phases can be realized in narrow gap semiconductor films using only an electrostatic superlattice potential.
Due to the generality of our mechanism, and high degree of flexibility in terms of material choice, film thickness, and superlattice potential, our theory provides a potential pathway to realizing fractionalized topological phases in new platforms and potentially higher temperatures.
In moir\'e materials, the chemistry at the van der Waals interface plays a key role: properties such as the interlayer tunneling are difficult to tune (requiring, for instance, high pressure~\cite{yankowitz2018dynamic,yankowitz2019tuning,carr2018pressure,morales2023pressure}).
As a result, models are typically material-specific and it is not clear how to extend the observed fractionalized states to higher temperatures.
In contrast, our proposed mechanism is not tied to a specific material realization,
allowing for a great degree of freedom in device design and optimization.

{
The critical temperature of fractionalized states is mostly set by the Coulomb interaction scale $E_C \sim \frac{e^2}{\epsilon a}$. 
While we have focused on $a=\SI{20}{nm}$ superlattice period for Cd$_3$As$_2$, this is not a fine-tuned choice since flat bands can be realized for a broad range of $g$ (see Fig~\ref{fig:TBmagic}(a,b)), and we find similar physics at smaller periods as well~\cite{supp}.
Thus, depending on the material properties and experimentally accessible superlattice potential strengths and periods, higher temperature fractionalized states can, in principle, be engineered through this mechanism.
}


{The idea of using of superlattice potential for realizing FCIs has been explored in the context of multilayer graphene \cite{ghorashi2023topological,ghorashi2023multilayer}. 
The main distinction of our work is that we go beyond simply showing the existence of FCIs: 
we provide a robust physical picture for why a topological flat band with optimized quantum geometry must emerge in this model.
This approach allows us to generalize our findings beyond any one specific material.
}

The essential ingredient for the active layer in our theory is a 2D narrow gap semiconductor described at low energy by the universal Hamiltonian,  Eq~\ref{eq:TopoinversionHam}.  
Potential candidates include thin films of: 3D topological insulators~\cite{hasan2010colloquium} (e.g. Bi${_2}$Te${_3}$, Bi$_2$Se$_3$, Sb$_2$Te$_3$,  Sb${_2}$Se${_3}$),
IV-VI semiconductors (SnTe, PbTe)~\cite{hsieh2012topological,liu2015electrically}, 3D Dirac semimetals (Cd$_3$As$_2$)~\cite{liu2014stable,lygo2023two}, or Bi~\cite{murakami2006quantum,fu2007topological,chen2023exceptional}.  
{ In this work, we have used our theory to identify and predict fractionalized states in Cd$_3$As$_2$.
We expect that such physics is possible in many other materials as well (similar results for Sb$_2$Te$_3$ are shown in the supplemental material~\cite{supp}).}
A full systematic analysis of candidate materials is an important subject for future study.
 
There are many potential avenues to realize the control layer.
One possibility is a doped moir\'e superlattice, such as a TMD heterobilayer~\cite{wu2018hubbard}, 
in which the doped electrons or holes are localized at the most energetically favorable stacking configurations, typically forming a triangular lattice.
The system exhibits a modulated charge density, leading to a periodic electrostatic potential felt by the nearby active layer~\cite{gu2023remote,zhang2024engineering,he2024dynamically}.
The period can be tuned by choosing materials with different atomic mismatch or applying a twist; and the potential strength can be adjusted by tuning the density of doped electrons or holes.
Another possibility for realizing a superlattice potential is via the twist interface of hBN or TMD~\cite{yasuda2021stacking,wang2022interfacial,zhao2021universal,kim2023electrostatic,woods2021charge,wang2024bandstructureengineeringusing}.
Another interesting approach is that the periodic potential may be generated spontaneously via a charge-ordering induced at the interface~\cite{lu2023synergistic,tseng2022gate,yang2023unconventional,wang2022quantum}.
An effective potential can also be generated via a moir\'e at the interface with a monolayer material~\cite{yang2023mathbb}.
Other ways to engineer superlattice potentials include patterned dielectrics~\cite{forsythe2018band,shi2019gate,xu2021creation,li2021anisotropic,barcons2022engineering,sun2023signature}, or LaAlO$_3$/SrTiO$_3$ interface patterned by conductive atomic force miscroscope lithography~\cite{pai2018physics}.

In conclusion, we have presented a route by which topological and fractionalized phases may be realized in a large class of narrow gap semiconductor films. 
Our results opens up new possibilities for engineered quantum materials that may offer even greater flexibility and richness than existing moir\'e materials.

\acknowledgments
We thank Aviram Uri, Aaron Sharpe, Ahmed Abouelkomsan, Jia Li, and Long Ju for helpful comments on the manuscript and for valuable discussions. TD acknowledges support from a startup fund at Stanford University. TT is supported by the Stanford Graduate fellowship.  The work at Massachusetts Institute of Technology was supported by the Air Force Office of Scientific Research (AFOSR) under Award No. FA9550-22-1-0432. The authors acknowledge the MIT SuperCloud and Lincoln Laboratory Supercomputing Center for providing HPC resources that have contributed to the research results reported within this paper.

\bibliography{arxiv_final_ref}

\begin{thebibliography}{107}%
\makeatletter
\providecommand \@ifxundefined [1]{%
 \@ifx{#1\undefined}
}%
\providecommand \@ifnum [1]{%
 \ifnum #1\expandafter \@firstoftwo
 \else \expandafter \@secondoftwo
 \fi
}%
\providecommand \@ifx [1]{%
 \ifx #1\expandafter \@firstoftwo
 \else \expandafter \@secondoftwo
 \fi
}%
\providecommand \natexlab [1]{#1}%
\providecommand \enquote  [1]{``#1''}%
\providecommand \bibnamefont  [1]{#1}%
\providecommand \bibfnamefont [1]{#1}%
\providecommand \citenamefont [1]{#1}%
\providecommand \href@noop [0]{\@secondoftwo}%
\providecommand \href [0]{\begingroup \@sanitize@url \@href}%
\providecommand \@href[1]{\@@startlink{#1}\@@href}%
\providecommand \@@href[1]{\endgroup#1\@@endlink}%
\providecommand \@sanitize@url [0]{\catcode `\\12\catcode `\$12\catcode `\&12\catcode `\#12\catcode `\^12\catcode `\_12\catcode `\%12\relax}%
\providecommand \@@startlink[1]{}%
\providecommand \@@endlink[0]{}%
\providecommand \url  [0]{\begingroup\@sanitize@url \@url }%
\providecommand \@url [1]{\endgroup\@href {#1}{\urlprefix }}%
\providecommand \urlprefix  [0]{URL }%
\providecommand \Eprint [0]{\href }%
\providecommand \doibase [0]{https://doi.org/}%
\providecommand \selectlanguage [0]{\@gobble}%
\providecommand \bibinfo  [0]{\@secondoftwo}%
\providecommand \bibfield  [0]{\@secondoftwo}%
\providecommand \translation [1]{[#1]}%
\providecommand \BibitemOpen [0]{}%
\providecommand \bibitemStop [0]{}%
\providecommand \bibitemNoStop [0]{.\EOS\space}%
\providecommand \EOS [0]{\spacefactor3000\relax}%
\providecommand \BibitemShut  [1]{\csname bibitem#1\endcsname}%
\let\auto@bib@innerbib\@empty
\bibitem [{\citenamefont {Sheng}\ \emph {et~al.}(2011)\citenamefont {Sheng}, \citenamefont {Gu}, \citenamefont {Sun},\ and\ \citenamefont {Sheng}}]{sheng2011fractional}%
  \BibitemOpen
  \bibfield  {author} {\bibinfo {author} {\bibfnamefont {D.}~\bibnamefont {Sheng}}, \bibinfo {author} {\bibfnamefont {Z.-C.}\ \bibnamefont {Gu}}, \bibinfo {author} {\bibfnamefont {K.}~\bibnamefont {Sun}},\ and\ \bibinfo {author} {\bibfnamefont {L.}~\bibnamefont {Sheng}},\ }\bibfield  {title} {\bibinfo {title} {Fractional quantum hall effect in the absence of landau levels},\ }\href@noop {} {\bibfield  {journal} {\bibinfo  {journal} {Nature communications}\ }\textbf {\bibinfo {volume} {2}},\ \bibinfo {pages} {389} (\bibinfo {year} {2011})}\BibitemShut {NoStop}%
\bibitem [{\citenamefont {Neupert}\ \emph {et~al.}(2011)\citenamefont {Neupert}, \citenamefont {Santos}, \citenamefont {Chamon},\ and\ \citenamefont {Mudry}}]{neupert2011fractional}%
  \BibitemOpen
  \bibfield  {author} {\bibinfo {author} {\bibfnamefont {T.}~\bibnamefont {Neupert}}, \bibinfo {author} {\bibfnamefont {L.}~\bibnamefont {Santos}}, \bibinfo {author} {\bibfnamefont {C.}~\bibnamefont {Chamon}},\ and\ \bibinfo {author} {\bibfnamefont {C.}~\bibnamefont {Mudry}},\ }\bibfield  {title} {\bibinfo {title} {Fractional quantum hall states at zero magnetic field},\ }\href@noop {} {\bibfield  {journal} {\bibinfo  {journal} {Physical review letters}\ }\textbf {\bibinfo {volume} {106}},\ \bibinfo {pages} {236804} (\bibinfo {year} {2011})}\BibitemShut {NoStop}%
\bibitem [{\citenamefont {Regnault}\ and\ \citenamefont {Bernevig}(2011)}]{regnault2011fractional}%
  \BibitemOpen
  \bibfield  {author} {\bibinfo {author} {\bibfnamefont {N.}~\bibnamefont {Regnault}}\ and\ \bibinfo {author} {\bibfnamefont {B.~A.}\ \bibnamefont {Bernevig}},\ }\bibfield  {title} {\bibinfo {title} {Fractional chern insulator},\ }\href@noop {} {\bibfield  {journal} {\bibinfo  {journal} {Physical Review X}\ }\textbf {\bibinfo {volume} {1}},\ \bibinfo {pages} {021014} (\bibinfo {year} {2011})}\BibitemShut {NoStop}%
\bibitem [{\citenamefont {Sun}\ \emph {et~al.}(2011)\citenamefont {Sun}, \citenamefont {Gu}, \citenamefont {Katsura},\ and\ \citenamefont {Sarma}}]{sun2011nearly}%
  \BibitemOpen
  \bibfield  {author} {\bibinfo {author} {\bibfnamefont {K.}~\bibnamefont {Sun}}, \bibinfo {author} {\bibfnamefont {Z.}~\bibnamefont {Gu}}, \bibinfo {author} {\bibfnamefont {H.}~\bibnamefont {Katsura}},\ and\ \bibinfo {author} {\bibfnamefont {S.~D.}\ \bibnamefont {Sarma}},\ }\bibfield  {title} {\bibinfo {title} {Nearly flatbands with nontrivial topology},\ }\href@noop {} {\bibfield  {journal} {\bibinfo  {journal} {Physical review letters}\ }\textbf {\bibinfo {volume} {106}},\ \bibinfo {pages} {236803} (\bibinfo {year} {2011})}\BibitemShut {NoStop}%
\bibitem [{\citenamefont {Tang}\ \emph {et~al.}(2011)\citenamefont {Tang}, \citenamefont {Mei},\ and\ \citenamefont {Wen}}]{tang2011high}%
  \BibitemOpen
  \bibfield  {author} {\bibinfo {author} {\bibfnamefont {E.}~\bibnamefont {Tang}}, \bibinfo {author} {\bibfnamefont {J.-W.}\ \bibnamefont {Mei}},\ and\ \bibinfo {author} {\bibfnamefont {X.-G.}\ \bibnamefont {Wen}},\ }\bibfield  {title} {\bibinfo {title} {High-temperature fractional quantum hall states},\ }\href@noop {} {\bibfield  {journal} {\bibinfo  {journal} {Physical review letters}\ }\textbf {\bibinfo {volume} {106}},\ \bibinfo {pages} {236802} (\bibinfo {year} {2011})}\BibitemShut {NoStop}%
\bibitem [{\citenamefont {Parameswaran}\ \emph {et~al.}(2013)\citenamefont {Parameswaran}, \citenamefont {Roy},\ and\ \citenamefont {Sondhi}}]{parameswaran2013fractional}%
  \BibitemOpen
  \bibfield  {author} {\bibinfo {author} {\bibfnamefont {S.~A.}\ \bibnamefont {Parameswaran}}, \bibinfo {author} {\bibfnamefont {R.}~\bibnamefont {Roy}},\ and\ \bibinfo {author} {\bibfnamefont {S.~L.}\ \bibnamefont {Sondhi}},\ }\bibfield  {title} {\bibinfo {title} {Fractional quantum hall physics in topological flat bands},\ }\href@noop {} {\bibfield  {journal} {\bibinfo  {journal} {Comptes Rendus Physique}\ }\textbf {\bibinfo {volume} {14}},\ \bibinfo {pages} {816} (\bibinfo {year} {2013})}\BibitemShut {NoStop}%
\bibitem [{\citenamefont {Liu}\ and\ \citenamefont {Bergholtz}(2022)}]{liu2022recent}%
  \BibitemOpen
  \bibfield  {author} {\bibinfo {author} {\bibfnamefont {Z.}~\bibnamefont {Liu}}\ and\ \bibinfo {author} {\bibfnamefont {E.~J.}\ \bibnamefont {Bergholtz}},\ }\bibfield  {title} {\bibinfo {title} {Recent developments in fractional chern insulators},\ }\href@noop {} {\bibfield  {journal} {\bibinfo  {journal} {arXiv preprint arXiv:2208.08449}\ } (\bibinfo {year} {2022})}\BibitemShut {NoStop}%
\bibitem [{\citenamefont {Andrei}\ \emph {et~al.}(2021)\citenamefont {Andrei}, \citenamefont {Efetov}, \citenamefont {Jarillo-Herrero}, \citenamefont {MacDonald}, \citenamefont {Mak}, \citenamefont {Senthil}, \citenamefont {Tutuc}, \citenamefont {Yazdani},\ and\ \citenamefont {Young}}]{andrei2021marvels}%
  \BibitemOpen
  \bibfield  {author} {\bibinfo {author} {\bibfnamefont {E.~Y.}\ \bibnamefont {Andrei}}, \bibinfo {author} {\bibfnamefont {D.~K.}\ \bibnamefont {Efetov}}, \bibinfo {author} {\bibfnamefont {P.}~\bibnamefont {Jarillo-Herrero}}, \bibinfo {author} {\bibfnamefont {A.~H.}\ \bibnamefont {MacDonald}}, \bibinfo {author} {\bibfnamefont {K.~F.}\ \bibnamefont {Mak}}, \bibinfo {author} {\bibfnamefont {T.}~\bibnamefont {Senthil}}, \bibinfo {author} {\bibfnamefont {E.}~\bibnamefont {Tutuc}}, \bibinfo {author} {\bibfnamefont {A.}~\bibnamefont {Yazdani}},\ and\ \bibinfo {author} {\bibfnamefont {A.~F.}\ \bibnamefont {Young}},\ }\bibfield  {title} {\bibinfo {title} {The marvels of moir{\'e} materials},\ }\href@noop {} {\bibfield  {journal} {\bibinfo  {journal} {Nature Reviews Materials}\ }\textbf {\bibinfo {volume} {6}},\ \bibinfo {pages} {201} (\bibinfo {year} {2021})}\BibitemShut {NoStop}%
\bibitem [{\citenamefont {Mak}\ and\ \citenamefont {Shan}(2022)}]{mak2022semiconductor}%
  \BibitemOpen
  \bibfield  {author} {\bibinfo {author} {\bibfnamefont {K.~F.}\ \bibnamefont {Mak}}\ and\ \bibinfo {author} {\bibfnamefont {J.}~\bibnamefont {Shan}},\ }\bibfield  {title} {\bibinfo {title} {Semiconductor moir{\'e} materials},\ }\href@noop {} {\bibfield  {journal} {\bibinfo  {journal} {Nature Nanotechnology}\ }\textbf {\bibinfo {volume} {17}},\ \bibinfo {pages} {686} (\bibinfo {year} {2022})}\BibitemShut {NoStop}%
\bibitem [{\citenamefont {Bistritzer}\ and\ \citenamefont {MacDonald}(2011)}]{bistritzer2011moire}%
  \BibitemOpen
  \bibfield  {author} {\bibinfo {author} {\bibfnamefont {R.}~\bibnamefont {Bistritzer}}\ and\ \bibinfo {author} {\bibfnamefont {A.~H.}\ \bibnamefont {MacDonald}},\ }\bibfield  {title} {\bibinfo {title} {Moir{\'e} bands in twisted double-layer graphene},\ }\href@noop {} {\bibfield  {journal} {\bibinfo  {journal} {Proceedings of the National Academy of Sciences}\ }\textbf {\bibinfo {volume} {108}},\ \bibinfo {pages} {12233} (\bibinfo {year} {2011})}\BibitemShut {NoStop}%
\bibitem [{\citenamefont {Morell}\ \emph {et~al.}(2010)\citenamefont {Morell}, \citenamefont {Correa}, \citenamefont {Vargas}, \citenamefont {Pacheco},\ and\ \citenamefont {Barticevic}}]{morell2010flat}%
  \BibitemOpen
  \bibfield  {author} {\bibinfo {author} {\bibfnamefont {E.~S.}\ \bibnamefont {Morell}}, \bibinfo {author} {\bibfnamefont {J.}~\bibnamefont {Correa}}, \bibinfo {author} {\bibfnamefont {P.}~\bibnamefont {Vargas}}, \bibinfo {author} {\bibfnamefont {M.}~\bibnamefont {Pacheco}},\ and\ \bibinfo {author} {\bibfnamefont {Z.}~\bibnamefont {Barticevic}},\ }\bibfield  {title} {\bibinfo {title} {Flat bands in slightly twisted bilayer graphene: Tight-binding calculations},\ }\href@noop {} {\bibfield  {journal} {\bibinfo  {journal} {Physical Review B}\ }\textbf {\bibinfo {volume} {82}},\ \bibinfo {pages} {121407} (\bibinfo {year} {2010})}\BibitemShut {NoStop}%
\bibitem [{\citenamefont {Wu}\ \emph {et~al.}(2019)\citenamefont {Wu}, \citenamefont {Lovorn}, \citenamefont {Tutuc}, \citenamefont {Martin},\ and\ \citenamefont {MacDonald}}]{wu2019topological}%
  \BibitemOpen
  \bibfield  {author} {\bibinfo {author} {\bibfnamefont {F.}~\bibnamefont {Wu}}, \bibinfo {author} {\bibfnamefont {T.}~\bibnamefont {Lovorn}}, \bibinfo {author} {\bibfnamefont {E.}~\bibnamefont {Tutuc}}, \bibinfo {author} {\bibfnamefont {I.}~\bibnamefont {Martin}},\ and\ \bibinfo {author} {\bibfnamefont {A.}~\bibnamefont {MacDonald}},\ }\bibfield  {title} {\bibinfo {title} {Topological insulators in twisted transition metal dichalcogenide homobilayers},\ }\href@noop {} {\bibfield  {journal} {\bibinfo  {journal} {Physical review letters}\ }\textbf {\bibinfo {volume} {122}},\ \bibinfo {pages} {086402} (\bibinfo {year} {2019})}\BibitemShut {NoStop}%
\bibitem [{\citenamefont {Li}\ \emph {et~al.}(2021{\natexlab{a}})\citenamefont {Li}, \citenamefont {Kumar}, \citenamefont {Sun},\ and\ \citenamefont {Lin}}]{li2021spontaneous}%
  \BibitemOpen
  \bibfield  {author} {\bibinfo {author} {\bibfnamefont {H.}~\bibnamefont {Li}}, \bibinfo {author} {\bibfnamefont {U.}~\bibnamefont {Kumar}}, \bibinfo {author} {\bibfnamefont {K.}~\bibnamefont {Sun}},\ and\ \bibinfo {author} {\bibfnamefont {S.-Z.}\ \bibnamefont {Lin}},\ }\bibfield  {title} {\bibinfo {title} {Spontaneous fractional chern insulators in transition metal dichalcogenide moir{\'e} superlattices},\ }\href@noop {} {\bibfield  {journal} {\bibinfo  {journal} {Physical Review Research}\ }\textbf {\bibinfo {volume} {3}},\ \bibinfo {pages} {L032070} (\bibinfo {year} {2021}{\natexlab{a}})}\BibitemShut {NoStop}%
\bibitem [{\citenamefont {Cr{\'e}pel}\ and\ \citenamefont {Fu}(2023)}]{crepel2023anomalous}%
  \BibitemOpen
  \bibfield  {author} {\bibinfo {author} {\bibfnamefont {V.}~\bibnamefont {Cr{\'e}pel}}\ and\ \bibinfo {author} {\bibfnamefont {L.}~\bibnamefont {Fu}},\ }\bibfield  {title} {\bibinfo {title} {Anomalous hall metal and fractional chern insulator in twisted transition metal dichalcogenides},\ }\href@noop {} {\bibfield  {journal} {\bibinfo  {journal} {Physical Review B}\ }\textbf {\bibinfo {volume} {107}},\ \bibinfo {pages} {L201109} (\bibinfo {year} {2023})}\BibitemShut {NoStop}%
\bibitem [{\citenamefont {Cai}\ \emph {et~al.}(2023)\citenamefont {Cai}, \citenamefont {Anderson}, \citenamefont {Wang}, \citenamefont {Zhang}, \citenamefont {Liu}, \citenamefont {Holtzmann}, \citenamefont {Zhang}, \citenamefont {Fan}, \citenamefont {Taniguchi}, \citenamefont {Watanabe} \emph {et~al.}}]{cai2023signatures}%
  \BibitemOpen
  \bibfield  {author} {\bibinfo {author} {\bibfnamefont {J.}~\bibnamefont {Cai}}, \bibinfo {author} {\bibfnamefont {E.}~\bibnamefont {Anderson}}, \bibinfo {author} {\bibfnamefont {C.}~\bibnamefont {Wang}}, \bibinfo {author} {\bibfnamefont {X.}~\bibnamefont {Zhang}}, \bibinfo {author} {\bibfnamefont {X.}~\bibnamefont {Liu}}, \bibinfo {author} {\bibfnamefont {W.}~\bibnamefont {Holtzmann}}, \bibinfo {author} {\bibfnamefont {Y.}~\bibnamefont {Zhang}}, \bibinfo {author} {\bibfnamefont {F.}~\bibnamefont {Fan}}, \bibinfo {author} {\bibfnamefont {T.}~\bibnamefont {Taniguchi}}, \bibinfo {author} {\bibfnamefont {K.}~\bibnamefont {Watanabe}}, \emph {et~al.},\ }\bibfield  {title} {\bibinfo {title} {Signatures of fractional quantum anomalous hall states in twisted mote2},\ }\href@noop {} {\bibfield  {journal} {\bibinfo  {journal} {Nature}\ }\textbf {\bibinfo {volume} {622}},\ \bibinfo {pages} {63} (\bibinfo {year} {2023})}\BibitemShut {NoStop}%
\bibitem [{\citenamefont {Zeng}\ \emph {et~al.}(2023)\citenamefont {Zeng}, \citenamefont {Xia}, \citenamefont {Kang}, \citenamefont {Zhu}, \citenamefont {Kn{\"u}ppel}, \citenamefont {Vaswani}, \citenamefont {Watanabe}, \citenamefont {Taniguchi}, \citenamefont {Mak},\ and\ \citenamefont {Shan}}]{zeng2023thermodynamic}%
  \BibitemOpen
  \bibfield  {author} {\bibinfo {author} {\bibfnamefont {Y.}~\bibnamefont {Zeng}}, \bibinfo {author} {\bibfnamefont {Z.}~\bibnamefont {Xia}}, \bibinfo {author} {\bibfnamefont {K.}~\bibnamefont {Kang}}, \bibinfo {author} {\bibfnamefont {J.}~\bibnamefont {Zhu}}, \bibinfo {author} {\bibfnamefont {P.}~\bibnamefont {Kn{\"u}ppel}}, \bibinfo {author} {\bibfnamefont {C.}~\bibnamefont {Vaswani}}, \bibinfo {author} {\bibfnamefont {K.}~\bibnamefont {Watanabe}}, \bibinfo {author} {\bibfnamefont {T.}~\bibnamefont {Taniguchi}}, \bibinfo {author} {\bibfnamefont {K.~F.}\ \bibnamefont {Mak}},\ and\ \bibinfo {author} {\bibfnamefont {J.}~\bibnamefont {Shan}},\ }\bibfield  {title} {\bibinfo {title} {Thermodynamic evidence of fractional chern insulator in moir{\'e} mote2},\ }\href@noop {} {\bibfield  {journal} {\bibinfo  {journal} {Nature}\ ,\ \bibinfo {pages} {1}} (\bibinfo {year} {2023})}\BibitemShut {NoStop}%
\bibitem [{\citenamefont {Park}\ \emph {et~al.}(2023)\citenamefont {Park}, \citenamefont {Cai}, \citenamefont {Anderson}, \citenamefont {Zhang}, \citenamefont {Zhu}, \citenamefont {Liu}, \citenamefont {Wang}, \citenamefont {Holtzmann}, \citenamefont {Hu}, \citenamefont {Liu} \emph {et~al.}}]{park2023observation}%
  \BibitemOpen
  \bibfield  {author} {\bibinfo {author} {\bibfnamefont {H.}~\bibnamefont {Park}}, \bibinfo {author} {\bibfnamefont {J.}~\bibnamefont {Cai}}, \bibinfo {author} {\bibfnamefont {E.}~\bibnamefont {Anderson}}, \bibinfo {author} {\bibfnamefont {Y.}~\bibnamefont {Zhang}}, \bibinfo {author} {\bibfnamefont {J.}~\bibnamefont {Zhu}}, \bibinfo {author} {\bibfnamefont {X.}~\bibnamefont {Liu}}, \bibinfo {author} {\bibfnamefont {C.}~\bibnamefont {Wang}}, \bibinfo {author} {\bibfnamefont {W.}~\bibnamefont {Holtzmann}}, \bibinfo {author} {\bibfnamefont {C.}~\bibnamefont {Hu}}, \bibinfo {author} {\bibfnamefont {Z.}~\bibnamefont {Liu}}, \emph {et~al.},\ }\bibfield  {title} {\bibinfo {title} {Observation of fractionally quantized anomalous hall effect},\ }\href@noop {} {\bibfield  {journal} {\bibinfo  {journal} {Nature}\ }\textbf {\bibinfo {volume} {622}},\ \bibinfo {pages} {74} (\bibinfo {year} {2023})}\BibitemShut {NoStop}%
\bibitem [{\citenamefont {Xu}\ \emph {et~al.}(2023)\citenamefont {Xu}, \citenamefont {Sun}, \citenamefont {Jia}, \citenamefont {Liu}, \citenamefont {Xu}, \citenamefont {Li}, \citenamefont {Gu}, \citenamefont {Watanabe}, \citenamefont {Taniguchi}, \citenamefont {Tong} \emph {et~al.}}]{xu2023observation}%
  \BibitemOpen
  \bibfield  {author} {\bibinfo {author} {\bibfnamefont {F.}~\bibnamefont {Xu}}, \bibinfo {author} {\bibfnamefont {Z.}~\bibnamefont {Sun}}, \bibinfo {author} {\bibfnamefont {T.}~\bibnamefont {Jia}}, \bibinfo {author} {\bibfnamefont {C.}~\bibnamefont {Liu}}, \bibinfo {author} {\bibfnamefont {C.}~\bibnamefont {Xu}}, \bibinfo {author} {\bibfnamefont {C.}~\bibnamefont {Li}}, \bibinfo {author} {\bibfnamefont {Y.}~\bibnamefont {Gu}}, \bibinfo {author} {\bibfnamefont {K.}~\bibnamefont {Watanabe}}, \bibinfo {author} {\bibfnamefont {T.}~\bibnamefont {Taniguchi}}, \bibinfo {author} {\bibfnamefont {B.}~\bibnamefont {Tong}}, \emph {et~al.},\ }\bibfield  {title} {\bibinfo {title} {Observation of integer and fractional quantum anomalous hall states in twisted bilayer mote2},\ }\href@noop {} {\bibfield  {journal} {\bibinfo  {journal} {arXiv preprint arXiv:2308.06177}\ } (\bibinfo {year} {2023})}\BibitemShut {NoStop}%
\bibitem [{\citenamefont {Lu}\ \emph {et~al.}(2023{\natexlab{a}})\citenamefont {Lu}, \citenamefont {Han}, \citenamefont {Yao}, \citenamefont {Reddy}, \citenamefont {Yang}, \citenamefont {Seo}, \citenamefont {Watanabe}, \citenamefont {Taniguchi}, \citenamefont {Fu},\ and\ \citenamefont {Ju}}]{lu2023fractional}%
  \BibitemOpen
  \bibfield  {author} {\bibinfo {author} {\bibfnamefont {Z.}~\bibnamefont {Lu}}, \bibinfo {author} {\bibfnamefont {T.}~\bibnamefont {Han}}, \bibinfo {author} {\bibfnamefont {Y.}~\bibnamefont {Yao}}, \bibinfo {author} {\bibfnamefont {A.~P.}\ \bibnamefont {Reddy}}, \bibinfo {author} {\bibfnamefont {J.}~\bibnamefont {Yang}}, \bibinfo {author} {\bibfnamefont {J.}~\bibnamefont {Seo}}, \bibinfo {author} {\bibfnamefont {K.}~\bibnamefont {Watanabe}}, \bibinfo {author} {\bibfnamefont {T.}~\bibnamefont {Taniguchi}}, \bibinfo {author} {\bibfnamefont {L.}~\bibnamefont {Fu}},\ and\ \bibinfo {author} {\bibfnamefont {L.}~\bibnamefont {Ju}},\ }\bibfield  {title} {\bibinfo {title} {Fractional quantum anomalous hall effect in a graphene moire superlattice},\ }\href@noop {} {\bibfield  {journal} {\bibinfo  {journal} {arXiv preprint arXiv:2309.17436}\ } (\bibinfo {year} {2023}{\natexlab{a}})}\BibitemShut {NoStop}%
\bibitem [{\citenamefont {Lau}\ \emph {et~al.}(2022)\citenamefont {Lau}, \citenamefont {Bockrath}, \citenamefont {Mak},\ and\ \citenamefont {Zhang}}]{lau2022reproducibility}%
  \BibitemOpen
  \bibfield  {author} {\bibinfo {author} {\bibfnamefont {C.~N.}\ \bibnamefont {Lau}}, \bibinfo {author} {\bibfnamefont {M.~W.}\ \bibnamefont {Bockrath}}, \bibinfo {author} {\bibfnamefont {K.~F.}\ \bibnamefont {Mak}},\ and\ \bibinfo {author} {\bibfnamefont {F.}~\bibnamefont {Zhang}},\ }\bibfield  {title} {\bibinfo {title} {Reproducibility in the fabrication and physics of moir{\'e} materials},\ }\href@noop {} {\bibfield  {journal} {\bibinfo  {journal} {Nature}\ }\textbf {\bibinfo {volume} {602}},\ \bibinfo {pages} {41} (\bibinfo {year} {2022})}\BibitemShut {NoStop}%
\bibitem [{\citenamefont {Stormer}\ \emph {et~al.}(1999)\citenamefont {Stormer}, \citenamefont {Tsui},\ and\ \citenamefont {Gossard}}]{stormer1999fractional}%
  \BibitemOpen
  \bibfield  {author} {\bibinfo {author} {\bibfnamefont {H.~L.}\ \bibnamefont {Stormer}}, \bibinfo {author} {\bibfnamefont {D.~C.}\ \bibnamefont {Tsui}},\ and\ \bibinfo {author} {\bibfnamefont {A.~C.}\ \bibnamefont {Gossard}},\ }\bibfield  {title} {\bibinfo {title} {The fractional quantum hall effect},\ }\href@noop {} {\bibfield  {journal} {\bibinfo  {journal} {Reviews of Modern Physics}\ }\textbf {\bibinfo {volume} {71}},\ \bibinfo {pages} {S298} (\bibinfo {year} {1999})}\BibitemShut {NoStop}%
\bibitem [{\citenamefont {Gu}\ \emph {et~al.}(2023)\citenamefont {Gu}, \citenamefont {Watanabe}, \citenamefont {Taniguchi}, \citenamefont {Shan},\ and\ \citenamefont {Mak}}]{gu2023remote}%
  \BibitemOpen
  \bibfield  {author} {\bibinfo {author} {\bibfnamefont {J.}~\bibnamefont {Gu}}, \bibinfo {author} {\bibfnamefont {K.}~\bibnamefont {Watanabe}}, \bibinfo {author} {\bibfnamefont {T.}~\bibnamefont {Taniguchi}}, \bibinfo {author} {\bibfnamefont {J.}~\bibnamefont {Shan}},\ and\ \bibinfo {author} {\bibfnamefont {K.~F.}\ \bibnamefont {Mak}},\ }\bibfield  {title} {\bibinfo {title} {Remote imprinting of moir{\'e} lattices},\ }\href@noop {} {\bibfield  {journal} {\bibinfo  {journal} {Research Square}\ } (\bibinfo {year} {2023})}\BibitemShut {NoStop}%
\bibitem [{\citenamefont {Zhang}\ \emph {et~al.}(2024)\citenamefont {Zhang}, \citenamefont {Xie}, \citenamefont {Zhao}, \citenamefont {Qi}, \citenamefont {Sanborn}, \citenamefont {Wang}, \citenamefont {Kahn}, \citenamefont {Watanabe}, \citenamefont {Taniguchi}, \citenamefont {Zettl} \emph {et~al.}}]{zhang2024engineering}%
  \BibitemOpen
  \bibfield  {author} {\bibinfo {author} {\bibfnamefont {Z.}~\bibnamefont {Zhang}}, \bibinfo {author} {\bibfnamefont {J.}~\bibnamefont {Xie}}, \bibinfo {author} {\bibfnamefont {W.}~\bibnamefont {Zhao}}, \bibinfo {author} {\bibfnamefont {R.}~\bibnamefont {Qi}}, \bibinfo {author} {\bibfnamefont {C.}~\bibnamefont {Sanborn}}, \bibinfo {author} {\bibfnamefont {S.}~\bibnamefont {Wang}}, \bibinfo {author} {\bibfnamefont {S.}~\bibnamefont {Kahn}}, \bibinfo {author} {\bibfnamefont {K.}~\bibnamefont {Watanabe}}, \bibinfo {author} {\bibfnamefont {T.}~\bibnamefont {Taniguchi}}, \bibinfo {author} {\bibfnamefont {A.}~\bibnamefont {Zettl}}, \emph {et~al.},\ }\bibfield  {title} {\bibinfo {title} {Engineering correlated insulators in bilayer graphene with a remote coulomb superlattice},\ }\href@noop {} {\bibfield  {journal} {\bibinfo  {journal} {Nature Materials}\ ,\ \bibinfo {pages} {1}} (\bibinfo {year} {2024})}\BibitemShut {NoStop}%
\bibitem [{\citenamefont {He}\ \emph {et~al.}(2024)\citenamefont {He}, \citenamefont {Cai}, \citenamefont {Zheng}, \citenamefont {Seewald}, \citenamefont {Taniguchi}, \citenamefont {Watanabe}, \citenamefont {Yan}, \citenamefont {Yankowitz}, \citenamefont {Pasupathy}, \citenamefont {Yao} \emph {et~al.}}]{he2024dynamically}%
  \BibitemOpen
  \bibfield  {author} {\bibinfo {author} {\bibfnamefont {M.}~\bibnamefont {He}}, \bibinfo {author} {\bibfnamefont {J.}~\bibnamefont {Cai}}, \bibinfo {author} {\bibfnamefont {H.}~\bibnamefont {Zheng}}, \bibinfo {author} {\bibfnamefont {E.}~\bibnamefont {Seewald}}, \bibinfo {author} {\bibfnamefont {T.}~\bibnamefont {Taniguchi}}, \bibinfo {author} {\bibfnamefont {K.}~\bibnamefont {Watanabe}}, \bibinfo {author} {\bibfnamefont {J.}~\bibnamefont {Yan}}, \bibinfo {author} {\bibfnamefont {M.}~\bibnamefont {Yankowitz}}, \bibinfo {author} {\bibfnamefont {A.}~\bibnamefont {Pasupathy}}, \bibinfo {author} {\bibfnamefont {W.}~\bibnamefont {Yao}}, \emph {et~al.},\ }\bibfield  {title} {\bibinfo {title} {Dynamically tunable moir{\'e} exciton rydberg states in a monolayer semiconductor on twisted bilayer graphene},\ }\href@noop {} {\bibfield  {journal} {\bibinfo  {journal} {Nature Materials}\ ,\ \bibinfo {pages} {1}} (\bibinfo {year} {2024})}\BibitemShut {NoStop}%
\bibitem [{\citenamefont {Forsythe}\ \emph {et~al.}(2018)\citenamefont {Forsythe}, \citenamefont {Zhou}, \citenamefont {Watanabe}, \citenamefont {Taniguchi}, \citenamefont {Pasupathy}, \citenamefont {Moon}, \citenamefont {Koshino}, \citenamefont {Kim},\ and\ \citenamefont {Dean}}]{forsythe2018band}%
  \BibitemOpen
  \bibfield  {author} {\bibinfo {author} {\bibfnamefont {C.}~\bibnamefont {Forsythe}}, \bibinfo {author} {\bibfnamefont {X.}~\bibnamefont {Zhou}}, \bibinfo {author} {\bibfnamefont {K.}~\bibnamefont {Watanabe}}, \bibinfo {author} {\bibfnamefont {T.}~\bibnamefont {Taniguchi}}, \bibinfo {author} {\bibfnamefont {A.}~\bibnamefont {Pasupathy}}, \bibinfo {author} {\bibfnamefont {P.}~\bibnamefont {Moon}}, \bibinfo {author} {\bibfnamefont {M.}~\bibnamefont {Koshino}}, \bibinfo {author} {\bibfnamefont {P.}~\bibnamefont {Kim}},\ and\ \bibinfo {author} {\bibfnamefont {C.~R.}\ \bibnamefont {Dean}},\ }\bibfield  {title} {\bibinfo {title} {Band structure engineering of 2d materials using patterned dielectric superlattices},\ }\href@noop {} {\bibfield  {journal} {\bibinfo  {journal} {Nature nanotechnology}\ }\textbf {\bibinfo {volume} {13}},\ \bibinfo {pages} {566} (\bibinfo {year} {2018})}\BibitemShut {NoStop}%
\bibitem [{\citenamefont {Yasuda}\ \emph {et~al.}(2021)\citenamefont {Yasuda}, \citenamefont {Wang}, \citenamefont {Watanabe}, \citenamefont {Taniguchi},\ and\ \citenamefont {Jarillo-Herrero}}]{yasuda2021stacking}%
  \BibitemOpen
  \bibfield  {author} {\bibinfo {author} {\bibfnamefont {K.}~\bibnamefont {Yasuda}}, \bibinfo {author} {\bibfnamefont {X.}~\bibnamefont {Wang}}, \bibinfo {author} {\bibfnamefont {K.}~\bibnamefont {Watanabe}}, \bibinfo {author} {\bibfnamefont {T.}~\bibnamefont {Taniguchi}},\ and\ \bibinfo {author} {\bibfnamefont {P.}~\bibnamefont {Jarillo-Herrero}},\ }\bibfield  {title} {\bibinfo {title} {Stacking-engineered ferroelectricity in bilayer boron nitride},\ }\href@noop {} {\bibfield  {journal} {\bibinfo  {journal} {Science}\ }\textbf {\bibinfo {volume} {372}},\ \bibinfo {pages} {1458} (\bibinfo {year} {2021})}\BibitemShut {NoStop}%
\bibitem [{\citenamefont {Wang}\ \emph {et~al.}(2022{\natexlab{a}})\citenamefont {Wang}, \citenamefont {Yasuda}, \citenamefont {Zhang}, \citenamefont {Liu}, \citenamefont {Watanabe}, \citenamefont {Taniguchi}, \citenamefont {Hone}, \citenamefont {Fu},\ and\ \citenamefont {Jarillo-Herrero}}]{wang2022interfacial}%
  \BibitemOpen
  \bibfield  {author} {\bibinfo {author} {\bibfnamefont {X.}~\bibnamefont {Wang}}, \bibinfo {author} {\bibfnamefont {K.}~\bibnamefont {Yasuda}}, \bibinfo {author} {\bibfnamefont {Y.}~\bibnamefont {Zhang}}, \bibinfo {author} {\bibfnamefont {S.}~\bibnamefont {Liu}}, \bibinfo {author} {\bibfnamefont {K.}~\bibnamefont {Watanabe}}, \bibinfo {author} {\bibfnamefont {T.}~\bibnamefont {Taniguchi}}, \bibinfo {author} {\bibfnamefont {J.}~\bibnamefont {Hone}}, \bibinfo {author} {\bibfnamefont {L.}~\bibnamefont {Fu}},\ and\ \bibinfo {author} {\bibfnamefont {P.}~\bibnamefont {Jarillo-Herrero}},\ }\bibfield  {title} {\bibinfo {title} {Interfacial ferroelectricity in rhombohedral-stacked bilayer transition metal dichalcogenides},\ }\href@noop {} {\bibfield  {journal} {\bibinfo  {journal} {Nature nanotechnology}\ }\textbf {\bibinfo {volume} {17}},\ \bibinfo {pages} {367} (\bibinfo {year} {2022}{\natexlab{a}})}\BibitemShut {NoStop}%
\bibitem [{\citenamefont {Zhao}\ \emph {et~al.}(2021)\citenamefont {Zhao}, \citenamefont {Xiao},\ and\ \citenamefont {Yao}}]{zhao2021universal}%
  \BibitemOpen
  \bibfield  {author} {\bibinfo {author} {\bibfnamefont {P.}~\bibnamefont {Zhao}}, \bibinfo {author} {\bibfnamefont {C.}~\bibnamefont {Xiao}},\ and\ \bibinfo {author} {\bibfnamefont {W.}~\bibnamefont {Yao}},\ }\bibfield  {title} {\bibinfo {title} {Universal superlattice potential for 2d materials from twisted interface inside h-bn substrate},\ }\href@noop {} {\bibfield  {journal} {\bibinfo  {journal} {npj 2D Materials and Applications}\ }\textbf {\bibinfo {volume} {5}},\ \bibinfo {pages} {38} (\bibinfo {year} {2021})}\BibitemShut {NoStop}%
\bibitem [{\citenamefont {Kim}\ \emph {et~al.}(2023)\citenamefont {Kim}, \citenamefont {Dominguez}, \citenamefont {Mayorga-Luna}, \citenamefont {Ye}, \citenamefont {Embley}, \citenamefont {Tan}, \citenamefont {Ni}, \citenamefont {Liu}, \citenamefont {Ford}, \citenamefont {Gao} \emph {et~al.}}]{kim2023electrostatic}%
  \BibitemOpen
  \bibfield  {author} {\bibinfo {author} {\bibfnamefont {D.~S.}\ \bibnamefont {Kim}}, \bibinfo {author} {\bibfnamefont {R.~C.}\ \bibnamefont {Dominguez}}, \bibinfo {author} {\bibfnamefont {R.}~\bibnamefont {Mayorga-Luna}}, \bibinfo {author} {\bibfnamefont {D.}~\bibnamefont {Ye}}, \bibinfo {author} {\bibfnamefont {J.}~\bibnamefont {Embley}}, \bibinfo {author} {\bibfnamefont {T.}~\bibnamefont {Tan}}, \bibinfo {author} {\bibfnamefont {Y.}~\bibnamefont {Ni}}, \bibinfo {author} {\bibfnamefont {Z.}~\bibnamefont {Liu}}, \bibinfo {author} {\bibfnamefont {M.}~\bibnamefont {Ford}}, \bibinfo {author} {\bibfnamefont {F.~Y.}\ \bibnamefont {Gao}}, \emph {et~al.},\ }\bibfield  {title} {\bibinfo {title} {Electrostatic moir{\'e} potential from twisted hexagonal boron nitride layers},\ }\href@noop {} {\bibfield  {journal} {\bibinfo  {journal} {Nature materials}\ ,\ \bibinfo {pages} {1}} (\bibinfo {year} {2023})}\BibitemShut {NoStop}%
\bibitem [{\citenamefont {Woods}\ \emph {et~al.}(2021)\citenamefont {Woods}, \citenamefont {Ares}, \citenamefont {Nevison-Andrews}, \citenamefont {Holwill}, \citenamefont {Fabregas}, \citenamefont {Guinea}, \citenamefont {Geim}, \citenamefont {Novoselov}, \citenamefont {Walet},\ and\ \citenamefont {Fumagalli}}]{woods2021charge}%
  \BibitemOpen
  \bibfield  {author} {\bibinfo {author} {\bibfnamefont {C.}~\bibnamefont {Woods}}, \bibinfo {author} {\bibfnamefont {P.}~\bibnamefont {Ares}}, \bibinfo {author} {\bibfnamefont {H.}~\bibnamefont {Nevison-Andrews}}, \bibinfo {author} {\bibfnamefont {M.}~\bibnamefont {Holwill}}, \bibinfo {author} {\bibfnamefont {R.}~\bibnamefont {Fabregas}}, \bibinfo {author} {\bibfnamefont {F.}~\bibnamefont {Guinea}}, \bibinfo {author} {\bibfnamefont {A.}~\bibnamefont {Geim}}, \bibinfo {author} {\bibfnamefont {K.}~\bibnamefont {Novoselov}}, \bibinfo {author} {\bibfnamefont {N.}~\bibnamefont {Walet}},\ and\ \bibinfo {author} {\bibfnamefont {L.}~\bibnamefont {Fumagalli}},\ }\bibfield  {title} {\bibinfo {title} {Charge-polarized interfacial superlattices in marginally twisted hexagonal boron nitride},\ }\href@noop {} {\bibfield  {journal} {\bibinfo  {journal} {Nature communications}\ }\textbf {\bibinfo {volume} {12}},\ \bibinfo {pages} {347} (\bibinfo {year} {2021})}\BibitemShut {NoStop}%
\bibitem [{\citenamefont {Shi}\ \emph {et~al.}(2019)\citenamefont {Shi}, \citenamefont {Ma},\ and\ \citenamefont {Song}}]{shi2019gate}%
  \BibitemOpen
  \bibfield  {author} {\bibinfo {author} {\bibfnamefont {L.-k.}\ \bibnamefont {Shi}}, \bibinfo {author} {\bibfnamefont {J.}~\bibnamefont {Ma}},\ and\ \bibinfo {author} {\bibfnamefont {J.~C.}\ \bibnamefont {Song}},\ }\bibfield  {title} {\bibinfo {title} {Gate-tunable flat bands in van der waals patterned dielectric superlattices},\ }\href@noop {} {\bibfield  {journal} {\bibinfo  {journal} {2D Materials}\ }\textbf {\bibinfo {volume} {7}},\ \bibinfo {pages} {015028} (\bibinfo {year} {2019})}\BibitemShut {NoStop}%
\bibitem [{\citenamefont {Xu}\ \emph {et~al.}(2021)\citenamefont {Xu}, \citenamefont {Horn}, \citenamefont {Zhu}, \citenamefont {Tang}, \citenamefont {Ma}, \citenamefont {Li}, \citenamefont {Liu}, \citenamefont {Watanabe}, \citenamefont {Taniguchi}, \citenamefont {Hone} \emph {et~al.}}]{xu2021creation}%
  \BibitemOpen
  \bibfield  {author} {\bibinfo {author} {\bibfnamefont {Y.}~\bibnamefont {Xu}}, \bibinfo {author} {\bibfnamefont {C.}~\bibnamefont {Horn}}, \bibinfo {author} {\bibfnamefont {J.}~\bibnamefont {Zhu}}, \bibinfo {author} {\bibfnamefont {Y.}~\bibnamefont {Tang}}, \bibinfo {author} {\bibfnamefont {L.}~\bibnamefont {Ma}}, \bibinfo {author} {\bibfnamefont {L.}~\bibnamefont {Li}}, \bibinfo {author} {\bibfnamefont {S.}~\bibnamefont {Liu}}, \bibinfo {author} {\bibfnamefont {K.}~\bibnamefont {Watanabe}}, \bibinfo {author} {\bibfnamefont {T.}~\bibnamefont {Taniguchi}}, \bibinfo {author} {\bibfnamefont {J.~C.}\ \bibnamefont {Hone}}, \emph {et~al.},\ }\bibfield  {title} {\bibinfo {title} {Creation of moir{\'e} bands in a monolayer semiconductor by spatially periodic dielectric screening},\ }\href@noop {} {\bibfield  {journal} {\bibinfo  {journal} {Nature Materials}\ }\textbf {\bibinfo {volume} {20}},\ \bibinfo {pages} {645} (\bibinfo {year} {2021})}\BibitemShut {NoStop}%
\bibitem [{\citenamefont {Barcons~Ruiz}\ \emph {et~al.}(2022)\citenamefont {Barcons~Ruiz}, \citenamefont {Herzig~Sheinfux}, \citenamefont {Hoffmann}, \citenamefont {Torre}, \citenamefont {Agarwal}, \citenamefont {Kumar}, \citenamefont {Vistoli}, \citenamefont {Taniguchi}, \citenamefont {Watanabe}, \citenamefont {Bachtold} \emph {et~al.}}]{barcons2022engineering}%
  \BibitemOpen
  \bibfield  {author} {\bibinfo {author} {\bibfnamefont {D.}~\bibnamefont {Barcons~Ruiz}}, \bibinfo {author} {\bibfnamefont {H.}~\bibnamefont {Herzig~Sheinfux}}, \bibinfo {author} {\bibfnamefont {R.}~\bibnamefont {Hoffmann}}, \bibinfo {author} {\bibfnamefont {I.}~\bibnamefont {Torre}}, \bibinfo {author} {\bibfnamefont {H.}~\bibnamefont {Agarwal}}, \bibinfo {author} {\bibfnamefont {R.~K.}\ \bibnamefont {Kumar}}, \bibinfo {author} {\bibfnamefont {L.}~\bibnamefont {Vistoli}}, \bibinfo {author} {\bibfnamefont {T.}~\bibnamefont {Taniguchi}}, \bibinfo {author} {\bibfnamefont {K.}~\bibnamefont {Watanabe}}, \bibinfo {author} {\bibfnamefont {A.}~\bibnamefont {Bachtold}}, \emph {et~al.},\ }\bibfield  {title} {\bibinfo {title} {Engineering high quality graphene superlattices via ion milled ultra-thin etching masks},\ }\href@noop {} {\bibfield  {journal} {\bibinfo  {journal} {Nature Communications}\ }\textbf {\bibinfo {volume} {13}},\ \bibinfo {pages} {6926} (\bibinfo {year} {2022})}\BibitemShut {NoStop}%
\bibitem [{\citenamefont {Sun}\ \emph {et~al.}(2023)\citenamefont {Sun}, \citenamefont {Ghorashi}, \citenamefont {Watanabe}, \citenamefont {Taniguchi}, \citenamefont {Camino}, \citenamefont {Cano},\ and\ \citenamefont {Du}}]{sun2023signature}%
  \BibitemOpen
  \bibfield  {author} {\bibinfo {author} {\bibfnamefont {J.}~\bibnamefont {Sun}}, \bibinfo {author} {\bibfnamefont {S.~A.~A.}\ \bibnamefont {Ghorashi}}, \bibinfo {author} {\bibfnamefont {K.}~\bibnamefont {Watanabe}}, \bibinfo {author} {\bibfnamefont {T.}~\bibnamefont {Taniguchi}}, \bibinfo {author} {\bibfnamefont {F.}~\bibnamefont {Camino}}, \bibinfo {author} {\bibfnamefont {J.}~\bibnamefont {Cano}},\ and\ \bibinfo {author} {\bibfnamefont {X.}~\bibnamefont {Du}},\ }\bibfield  {title} {\bibinfo {title} {Signature of correlated insulator in electric field controlled superlattice},\ }\href@noop {} {\bibfield  {journal} {\bibinfo  {journal} {arXiv preprint arXiv:2306.06848}\ } (\bibinfo {year} {2023})}\BibitemShut {NoStop}%
\bibitem [{\citenamefont {Li}\ \emph {et~al.}(2021{\natexlab{b}})\citenamefont {Li}, \citenamefont {Dietrich}, \citenamefont {Forsythe}, \citenamefont {Taniguchi}, \citenamefont {Watanabe}, \citenamefont {Moon},\ and\ \citenamefont {Dean}}]{li2021anisotropic}%
  \BibitemOpen
  \bibfield  {author} {\bibinfo {author} {\bibfnamefont {Y.}~\bibnamefont {Li}}, \bibinfo {author} {\bibfnamefont {S.}~\bibnamefont {Dietrich}}, \bibinfo {author} {\bibfnamefont {C.}~\bibnamefont {Forsythe}}, \bibinfo {author} {\bibfnamefont {T.}~\bibnamefont {Taniguchi}}, \bibinfo {author} {\bibfnamefont {K.}~\bibnamefont {Watanabe}}, \bibinfo {author} {\bibfnamefont {P.}~\bibnamefont {Moon}},\ and\ \bibinfo {author} {\bibfnamefont {C.~R.}\ \bibnamefont {Dean}},\ }\bibfield  {title} {\bibinfo {title} {Anisotropic band flattening in graphene with one-dimensional superlattices},\ }\href@noop {} {\bibfield  {journal} {\bibinfo  {journal} {Nature Nanotechnology}\ }\textbf {\bibinfo {volume} {16}},\ \bibinfo {pages} {525} (\bibinfo {year} {2021}{\natexlab{b}})}\BibitemShut {NoStop}%
\bibitem [{\citenamefont {Wang}\ \emph {et~al.}(2024{\natexlab{a}})\citenamefont {Wang}, \citenamefont {Xu}, \citenamefont {Aronson}, \citenamefont {Bennett}, \citenamefont {Paul}, \citenamefont {Crowley}, \citenamefont {Collignon}, \citenamefont {Watanabe}, \citenamefont {Taniguchi}, \citenamefont {Ashoori}, \citenamefont {Kaxiras}, \citenamefont {Zhang}, \citenamefont {Jarillo-Herrero},\ and\ \citenamefont {Yasuda}}]{wang2024bandstructureengineeringusing}%
  \BibitemOpen
  \bibfield  {author} {\bibinfo {author} {\bibfnamefont {X.}~\bibnamefont {Wang}}, \bibinfo {author} {\bibfnamefont {C.}~\bibnamefont {Xu}}, \bibinfo {author} {\bibfnamefont {S.}~\bibnamefont {Aronson}}, \bibinfo {author} {\bibfnamefont {D.}~\bibnamefont {Bennett}}, \bibinfo {author} {\bibfnamefont {N.}~\bibnamefont {Paul}}, \bibinfo {author} {\bibfnamefont {P.~J.~D.}\ \bibnamefont {Crowley}}, \bibinfo {author} {\bibfnamefont {C.}~\bibnamefont {Collignon}}, \bibinfo {author} {\bibfnamefont {K.}~\bibnamefont {Watanabe}}, \bibinfo {author} {\bibfnamefont {T.}~\bibnamefont {Taniguchi}}, \bibinfo {author} {\bibfnamefont {R.}~\bibnamefont {Ashoori}}, \bibinfo {author} {\bibfnamefont {E.}~\bibnamefont {Kaxiras}}, \bibinfo {author} {\bibfnamefont {Y.}~\bibnamefont {Zhang}}, \bibinfo {author} {\bibfnamefont {P.}~\bibnamefont {Jarillo-Herrero}},\ and\ \bibinfo {author} {\bibfnamefont {K.}~\bibnamefont {Yasuda}},\ }\href {https://arxiv.org/abs/2405.03761} {\bibinfo {title} {Band structure engineering using a moir\'e
  polar substrate}} (\bibinfo {year} {2024}{\natexlab{a}}),\ \Eprint {https://arxiv.org/abs/2405.03761} {arXiv:2405.03761 [cond-mat.mes-hall]} \BibitemShut {NoStop}%
\bibitem [{\citenamefont {Song}\ \emph {et~al.}(2015)\citenamefont {Song}, \citenamefont {Samutpraphoot},\ and\ \citenamefont {Levitov}}]{song2015topological}%
  \BibitemOpen
  \bibfield  {author} {\bibinfo {author} {\bibfnamefont {J.~C.}\ \bibnamefont {Song}}, \bibinfo {author} {\bibfnamefont {P.}~\bibnamefont {Samutpraphoot}},\ and\ \bibinfo {author} {\bibfnamefont {L.~S.}\ \bibnamefont {Levitov}},\ }\bibfield  {title} {\bibinfo {title} {Topological bloch bands in graphene superlattices},\ }\href@noop {} {\bibfield  {journal} {\bibinfo  {journal} {Proceedings of the National Academy of Sciences}\ }\textbf {\bibinfo {volume} {112}},\ \bibinfo {pages} {10879} (\bibinfo {year} {2015})}\BibitemShut {NoStop}%
\bibitem [{\citenamefont {Huber}\ \emph {et~al.}(2020)\citenamefont {Huber}, \citenamefont {Liu}, \citenamefont {Chen}, \citenamefont {Drienovsky}, \citenamefont {Sandner}, \citenamefont {Watanabe}, \citenamefont {Taniguchi}, \citenamefont {Richter}, \citenamefont {Weiss},\ and\ \citenamefont {Eroms}}]{huber2020gate}%
  \BibitemOpen
  \bibfield  {author} {\bibinfo {author} {\bibfnamefont {R.}~\bibnamefont {Huber}}, \bibinfo {author} {\bibfnamefont {M.-H.}\ \bibnamefont {Liu}}, \bibinfo {author} {\bibfnamefont {S.-C.}\ \bibnamefont {Chen}}, \bibinfo {author} {\bibfnamefont {M.}~\bibnamefont {Drienovsky}}, \bibinfo {author} {\bibfnamefont {A.}~\bibnamefont {Sandner}}, \bibinfo {author} {\bibfnamefont {K.}~\bibnamefont {Watanabe}}, \bibinfo {author} {\bibfnamefont {T.}~\bibnamefont {Taniguchi}}, \bibinfo {author} {\bibfnamefont {K.}~\bibnamefont {Richter}}, \bibinfo {author} {\bibfnamefont {D.}~\bibnamefont {Weiss}},\ and\ \bibinfo {author} {\bibfnamefont {J.}~\bibnamefont {Eroms}},\ }\bibfield  {title} {\bibinfo {title} {Gate-tunable two-dimensional superlattices in graphene},\ }\href@noop {} {\bibfield  {journal} {\bibinfo  {journal} {Nano letters}\ }\textbf {\bibinfo {volume} {20}},\ \bibinfo {pages} {8046} (\bibinfo {year} {2020})}\BibitemShut {NoStop}%
\bibitem [{\citenamefont {Ghorashi}\ \emph {et~al.}(2023)\citenamefont {Ghorashi}, \citenamefont {Dunbrack}, \citenamefont {Abouelkomsan}, \citenamefont {Sun}, \citenamefont {Du},\ and\ \citenamefont {Cano}}]{ghorashi2023topological}%
  \BibitemOpen
  \bibfield  {author} {\bibinfo {author} {\bibfnamefont {S.~A.~A.}\ \bibnamefont {Ghorashi}}, \bibinfo {author} {\bibfnamefont {A.}~\bibnamefont {Dunbrack}}, \bibinfo {author} {\bibfnamefont {A.}~\bibnamefont {Abouelkomsan}}, \bibinfo {author} {\bibfnamefont {J.}~\bibnamefont {Sun}}, \bibinfo {author} {\bibfnamefont {X.}~\bibnamefont {Du}},\ and\ \bibinfo {author} {\bibfnamefont {J.}~\bibnamefont {Cano}},\ }\bibfield  {title} {\bibinfo {title} {Topological and stacked flat bands in bilayer graphene with a superlattice potential},\ }\href@noop {} {\bibfield  {journal} {\bibinfo  {journal} {Physical Review Letters}\ }\textbf {\bibinfo {volume} {130}},\ \bibinfo {pages} {196201} (\bibinfo {year} {2023})}\BibitemShut {NoStop}%
\bibitem [{\citenamefont {Ghorashi}\ and\ \citenamefont {Cano}(2023)}]{ghorashi2023multilayer}%
  \BibitemOpen
  \bibfield  {author} {\bibinfo {author} {\bibfnamefont {S.~A.~A.}\ \bibnamefont {Ghorashi}}\ and\ \bibinfo {author} {\bibfnamefont {J.}~\bibnamefont {Cano}},\ }\bibfield  {title} {\bibinfo {title} {Multilayer graphene with a superlattice potential},\ }\href@noop {} {\bibfield  {journal} {\bibinfo  {journal} {Physical Review B}\ }\textbf {\bibinfo {volume} {107}},\ \bibinfo {pages} {195423} (\bibinfo {year} {2023})}\BibitemShut {NoStop}%
\bibitem [{\citenamefont {Zeng}\ \emph {et~al.}(2024)\citenamefont {Zeng}, \citenamefont {Wolf}, \citenamefont {Huang}, \citenamefont {Wei}, \citenamefont {Ghorashi}, \citenamefont {MacDonald},\ and\ \citenamefont {Cano}}]{zeng2024gatetunable}%
  \BibitemOpen
  \bibfield  {author} {\bibinfo {author} {\bibfnamefont {Y.}~\bibnamefont {Zeng}}, \bibinfo {author} {\bibfnamefont {T.~M.~R.}\ \bibnamefont {Wolf}}, \bibinfo {author} {\bibfnamefont {C.}~\bibnamefont {Huang}}, \bibinfo {author} {\bibfnamefont {N.}~\bibnamefont {Wei}}, \bibinfo {author} {\bibfnamefont {S.~A.~A.}\ \bibnamefont {Ghorashi}}, \bibinfo {author} {\bibfnamefont {A.~H.}\ \bibnamefont {MacDonald}},\ and\ \bibinfo {author} {\bibfnamefont {J.}~\bibnamefont {Cano}},\ }\href@noop {} {\bibinfo {title} {Gate-tunable topological phases in superlattice modulated bilayer graphene}} (\bibinfo {year} {2024}),\ \Eprint {https://arxiv.org/abs/2401.04321} {arXiv:2401.04321 [cond-mat.mes-hall]} \BibitemShut {NoStop}%
\bibitem [{\citenamefont {Albrecht}\ \emph {et~al.}(1999)\citenamefont {Albrecht}, \citenamefont {Smet}, \citenamefont {Weiss}, \citenamefont {Von~Klitzing}, \citenamefont {Hennig}, \citenamefont {Langenbuch}, \citenamefont {Suhrke}, \citenamefont {R{\"o}ssler}, \citenamefont {Umansky},\ and\ \citenamefont {Schweizer}}]{albrecht1999fermiology}%
  \BibitemOpen
  \bibfield  {author} {\bibinfo {author} {\bibfnamefont {C.}~\bibnamefont {Albrecht}}, \bibinfo {author} {\bibfnamefont {J.}~\bibnamefont {Smet}}, \bibinfo {author} {\bibfnamefont {D.}~\bibnamefont {Weiss}}, \bibinfo {author} {\bibfnamefont {K.}~\bibnamefont {Von~Klitzing}}, \bibinfo {author} {\bibfnamefont {R.}~\bibnamefont {Hennig}}, \bibinfo {author} {\bibfnamefont {M.}~\bibnamefont {Langenbuch}}, \bibinfo {author} {\bibfnamefont {M.}~\bibnamefont {Suhrke}}, \bibinfo {author} {\bibfnamefont {U.}~\bibnamefont {R{\"o}ssler}}, \bibinfo {author} {\bibfnamefont {V.}~\bibnamefont {Umansky}},\ and\ \bibinfo {author} {\bibfnamefont {H.}~\bibnamefont {Schweizer}},\ }\bibfield  {title} {\bibinfo {title} {Fermiology of two-dimensional lateral superlattices},\ }\href@noop {} {\bibfield  {journal} {\bibinfo  {journal} {Physical review letters}\ }\textbf {\bibinfo {volume} {83}},\ \bibinfo {pages} {2234} (\bibinfo {year} {1999})}\BibitemShut {NoStop}%
\bibitem [{\citenamefont {Wang}\ \emph {et~al.}(2024{\natexlab{b}})\citenamefont {Wang}, \citenamefont {Krix}, \citenamefont {Tkachenko}, \citenamefont {Tkachenko}, \citenamefont {Chen}, \citenamefont {Farrer}, \citenamefont {Ritchie}, \citenamefont {Sushkov}, \citenamefont {Hamilton},\ and\ \citenamefont {Klochan}}]{wang2024tuning}%
  \BibitemOpen
  \bibfield  {author} {\bibinfo {author} {\bibfnamefont {D.~Q.}\ \bibnamefont {Wang}}, \bibinfo {author} {\bibfnamefont {Z.}~\bibnamefont {Krix}}, \bibinfo {author} {\bibfnamefont {O.~A.}\ \bibnamefont {Tkachenko}}, \bibinfo {author} {\bibfnamefont {V.~A.}\ \bibnamefont {Tkachenko}}, \bibinfo {author} {\bibfnamefont {C.}~\bibnamefont {Chen}}, \bibinfo {author} {\bibfnamefont {I.}~\bibnamefont {Farrer}}, \bibinfo {author} {\bibfnamefont {D.~A.}\ \bibnamefont {Ritchie}}, \bibinfo {author} {\bibfnamefont {O.~P.}\ \bibnamefont {Sushkov}}, \bibinfo {author} {\bibfnamefont {A.~R.}\ \bibnamefont {Hamilton}},\ and\ \bibinfo {author} {\bibfnamefont {O.}~\bibnamefont {Klochan}},\ }\bibfield  {title} {\bibinfo {title} {Tuning the bandstructure of electrons in a two-dimensional artificial electrostatic crystal in gaas quantum wells},\ }\href@noop {} {\bibfield  {journal} {\bibinfo  {journal} {arXiv preprint arXiv:2402.12769}\ } (\bibinfo {year} {2024}{\natexlab{b}})}\BibitemShut {NoStop}%
\bibitem [{\citenamefont {Fu}\ and\ \citenamefont {Kane}(2007)}]{fu2007topological}%
  \BibitemOpen
  \bibfield  {author} {\bibinfo {author} {\bibfnamefont {L.}~\bibnamefont {Fu}}\ and\ \bibinfo {author} {\bibfnamefont {C.~L.}\ \bibnamefont {Kane}},\ }\bibfield  {title} {\bibinfo {title} {Topological insulators with inversion symmetry},\ }\href@noop {} {\bibfield  {journal} {\bibinfo  {journal} {Physical Review B}\ }\textbf {\bibinfo {volume} {76}},\ \bibinfo {pages} {045302} (\bibinfo {year} {2007})}\BibitemShut {NoStop}%
\bibitem [{\citenamefont {Bernevig}\ \emph {et~al.}(2006)\citenamefont {Bernevig}, \citenamefont {Hughes},\ and\ \citenamefont {Zhang}}]{bernevig2006quantum}%
  \BibitemOpen
  \bibfield  {author} {\bibinfo {author} {\bibfnamefont {B.~A.}\ \bibnamefont {Bernevig}}, \bibinfo {author} {\bibfnamefont {T.~L.}\ \bibnamefont {Hughes}},\ and\ \bibinfo {author} {\bibfnamefont {S.-C.}\ \bibnamefont {Zhang}},\ }\bibfield  {title} {\bibinfo {title} {Quantum spin hall effect and topological phase transition in hgte quantum wells},\ }\href@noop {} {\bibfield  {journal} {\bibinfo  {journal} {science}\ }\textbf {\bibinfo {volume} {314}},\ \bibinfo {pages} {1757} (\bibinfo {year} {2006})}\BibitemShut {NoStop}%
\bibitem [{sup()}]{supp}%
  \BibitemOpen
  \href@noop {} {}\bibinfo {note} {See the supplementary information for (1) more detailed properties of the bands near the magic point, (2) symmetry analysis of k.p model, (3) additional details on the many-body calculation, (4) detailed discussion on the massive Dirac fermion model, (5) detailed discussion of modeling real materials, which includes Refs.\cite{rezayi1994fermi, geraedts2018berry,fremling2018trial,wang2019lattice,wang2019dirac,stern2023transport,fang2012bulk,wang2021moire, cano2021moire,lu2010massive,liu2010model,Cano_2017}}\BibitemShut {NoStop}%
\bibitem [{\citenamefont {Su}\ \emph {et~al.}(2022)\citenamefont {Su}, \citenamefont {Li}, \citenamefont {Zhang}, \citenamefont {Sun},\ and\ \citenamefont {Lin}}]{Su_2022}%
  \BibitemOpen
  \bibfield  {author} {\bibinfo {author} {\bibfnamefont {Y.}~\bibnamefont {Su}}, \bibinfo {author} {\bibfnamefont {H.}~\bibnamefont {Li}}, \bibinfo {author} {\bibfnamefont {C.}~\bibnamefont {Zhang}}, \bibinfo {author} {\bibfnamefont {K.}~\bibnamefont {Sun}},\ and\ \bibinfo {author} {\bibfnamefont {S.-Z.}\ \bibnamefont {Lin}},\ }\bibfield  {title} {\bibinfo {title} {Massive dirac fermions in moiré superlattices: A route towards topological flat minibands and correlated topological insulators},\ }\bibfield  {journal} {\bibinfo  {journal} {Physical Review Research}\ }\textbf {\bibinfo {volume} {4}},\ \href {https://doi.org/10.1103/physrevresearch.4.l032024} {10.1103/physrevresearch.4.l032024} (\bibinfo {year} {2022})\BibitemShut {NoStop}%
\bibitem [{\citenamefont {Suri}\ \emph {et~al.}(2023)\citenamefont {Suri}, \citenamefont {Wang}, \citenamefont {Hunt},\ and\ \citenamefont {Xiao}}]{suri2023superlattice}%
  \BibitemOpen
  \bibfield  {author} {\bibinfo {author} {\bibfnamefont {N.}~\bibnamefont {Suri}}, \bibinfo {author} {\bibfnamefont {C.}~\bibnamefont {Wang}}, \bibinfo {author} {\bibfnamefont {B.~M.}\ \bibnamefont {Hunt}},\ and\ \bibinfo {author} {\bibfnamefont {D.}~\bibnamefont {Xiao}},\ }\bibfield  {title} {\bibinfo {title} {Superlattice engineering of topology in massive dirac fermions},\ }\href@noop {} {\bibfield  {journal} {\bibinfo  {journal} {arXiv preprint arXiv:2305.13522}\ } (\bibinfo {year} {2023})}\BibitemShut {NoStop}%
\bibitem [{\citenamefont {Parameswaran}\ \emph {et~al.}(2012)\citenamefont {Parameswaran}, \citenamefont {Roy},\ and\ \citenamefont {Sondhi}}]{parameswaran2012fractional}%
  \BibitemOpen
  \bibfield  {author} {\bibinfo {author} {\bibfnamefont {S.}~\bibnamefont {Parameswaran}}, \bibinfo {author} {\bibfnamefont {R.}~\bibnamefont {Roy}},\ and\ \bibinfo {author} {\bibfnamefont {S.~L.}\ \bibnamefont {Sondhi}},\ }\bibfield  {title} {\bibinfo {title} {Fractional chern insulators and the w$\infty$ algebra},\ }\href@noop {} {\bibfield  {journal} {\bibinfo  {journal} {Physical Review B}\ }\textbf {\bibinfo {volume} {85}},\ \bibinfo {pages} {241308} (\bibinfo {year} {2012})}\BibitemShut {NoStop}%
\bibitem [{\citenamefont {Roy}(2014)}]{roy2014band}%
  \BibitemOpen
  \bibfield  {author} {\bibinfo {author} {\bibfnamefont {R.}~\bibnamefont {Roy}},\ }\bibfield  {title} {\bibinfo {title} {Band geometry of fractional topological insulators},\ }\href@noop {} {\bibfield  {journal} {\bibinfo  {journal} {Physical Review B}\ }\textbf {\bibinfo {volume} {90}},\ \bibinfo {pages} {165139} (\bibinfo {year} {2014})}\BibitemShut {NoStop}%
\bibitem [{\citenamefont {Claassen}\ \emph {et~al.}(2015)\citenamefont {Claassen}, \citenamefont {Lee}, \citenamefont {Thomale}, \citenamefont {Qi},\ and\ \citenamefont {Devereaux}}]{claassen2015position}%
  \BibitemOpen
  \bibfield  {author} {\bibinfo {author} {\bibfnamefont {M.}~\bibnamefont {Claassen}}, \bibinfo {author} {\bibfnamefont {C.~H.}\ \bibnamefont {Lee}}, \bibinfo {author} {\bibfnamefont {R.}~\bibnamefont {Thomale}}, \bibinfo {author} {\bibfnamefont {X.-L.}\ \bibnamefont {Qi}},\ and\ \bibinfo {author} {\bibfnamefont {T.~P.}\ \bibnamefont {Devereaux}},\ }\bibfield  {title} {\bibinfo {title} {Position-momentum duality and fractional quantum hall effect in chern insulators},\ }\href@noop {} {\bibfield  {journal} {\bibinfo  {journal} {Physical review letters}\ }\textbf {\bibinfo {volume} {114}},\ \bibinfo {pages} {236802} (\bibinfo {year} {2015})}\BibitemShut {NoStop}%
\bibitem [{\citenamefont {Jackson}\ \emph {et~al.}(2015)\citenamefont {Jackson}, \citenamefont {M{\"o}ller},\ and\ \citenamefont {Roy}}]{jackson2015geometric}%
  \BibitemOpen
  \bibfield  {author} {\bibinfo {author} {\bibfnamefont {T.~S.}\ \bibnamefont {Jackson}}, \bibinfo {author} {\bibfnamefont {G.}~\bibnamefont {M{\"o}ller}},\ and\ \bibinfo {author} {\bibfnamefont {R.}~\bibnamefont {Roy}},\ }\bibfield  {title} {\bibinfo {title} {Geometric stability of topological lattice phases},\ }\href@noop {} {\bibfield  {journal} {\bibinfo  {journal} {Nature communications}\ }\textbf {\bibinfo {volume} {6}},\ \bibinfo {pages} {8629} (\bibinfo {year} {2015})}\BibitemShut {NoStop}%
\bibitem [{\citenamefont {Wang}\ \emph {et~al.}(2021{\natexlab{a}})\citenamefont {Wang}, \citenamefont {Cano}, \citenamefont {Millis}, \citenamefont {Liu},\ and\ \citenamefont {Yang}}]{wang2021exact}%
  \BibitemOpen
  \bibfield  {author} {\bibinfo {author} {\bibfnamefont {J.}~\bibnamefont {Wang}}, \bibinfo {author} {\bibfnamefont {J.}~\bibnamefont {Cano}}, \bibinfo {author} {\bibfnamefont {A.~J.}\ \bibnamefont {Millis}}, \bibinfo {author} {\bibfnamefont {Z.}~\bibnamefont {Liu}},\ and\ \bibinfo {author} {\bibfnamefont {B.}~\bibnamefont {Yang}},\ }\bibfield  {title} {\bibinfo {title} {Exact landau level description of geometry and interaction in a flatband},\ }\href@noop {} {\bibfield  {journal} {\bibinfo  {journal} {Physical review letters}\ }\textbf {\bibinfo {volume} {127}},\ \bibinfo {pages} {246403} (\bibinfo {year} {2021}{\natexlab{a}})}\BibitemShut {NoStop}%
\bibitem [{\citenamefont {Ledwith}\ \emph {et~al.}(2020)\citenamefont {Ledwith}, \citenamefont {Tarnopolsky}, \citenamefont {Khalaf},\ and\ \citenamefont {Vishwanath}}]{ledwith2020fractional}%
  \BibitemOpen
  \bibfield  {author} {\bibinfo {author} {\bibfnamefont {P.~J.}\ \bibnamefont {Ledwith}}, \bibinfo {author} {\bibfnamefont {G.}~\bibnamefont {Tarnopolsky}}, \bibinfo {author} {\bibfnamefont {E.}~\bibnamefont {Khalaf}},\ and\ \bibinfo {author} {\bibfnamefont {A.}~\bibnamefont {Vishwanath}},\ }\bibfield  {title} {\bibinfo {title} {Fractional chern insulator states in twisted bilayer graphene: An analytical approach},\ }\href@noop {} {\bibfield  {journal} {\bibinfo  {journal} {Physical Review Research}\ }\textbf {\bibinfo {volume} {2}},\ \bibinfo {pages} {023237} (\bibinfo {year} {2020})}\BibitemShut {NoStop}%
\bibitem [{\citenamefont {Ledwith}\ \emph {et~al.}(2022)\citenamefont {Ledwith}, \citenamefont {Vishwanath},\ and\ \citenamefont {Parker}}]{ledwith2022vortexability}%
  \BibitemOpen
  \bibfield  {author} {\bibinfo {author} {\bibfnamefont {P.~J.}\ \bibnamefont {Ledwith}}, \bibinfo {author} {\bibfnamefont {A.}~\bibnamefont {Vishwanath}},\ and\ \bibinfo {author} {\bibfnamefont {D.~E.}\ \bibnamefont {Parker}},\ }\bibfield  {title} {\bibinfo {title} {Vortexability: A unifying criterion for ideal fractional chern insulators},\ }\href@noop {} {\bibfield  {journal} {\bibinfo  {journal} {arXiv preprint arXiv:2209.15023}\ } (\bibinfo {year} {2022})}\BibitemShut {NoStop}%
\bibitem [{\citenamefont {Khalaf}\ \emph {et~al.}(2019)\citenamefont {Khalaf}, \citenamefont {Kruchkov}, \citenamefont {Tarnopolsky},\ and\ \citenamefont {Vishwanath}}]{khalaf2019magic}%
  \BibitemOpen
  \bibfield  {author} {\bibinfo {author} {\bibfnamefont {E.}~\bibnamefont {Khalaf}}, \bibinfo {author} {\bibfnamefont {A.~J.}\ \bibnamefont {Kruchkov}}, \bibinfo {author} {\bibfnamefont {G.}~\bibnamefont {Tarnopolsky}},\ and\ \bibinfo {author} {\bibfnamefont {A.}~\bibnamefont {Vishwanath}},\ }\bibfield  {title} {\bibinfo {title} {Magic angle hierarchy in twisted graphene multilayers},\ }\href@noop {} {\bibfield  {journal} {\bibinfo  {journal} {Physical Review B}\ }\textbf {\bibinfo {volume} {100}},\ \bibinfo {pages} {085109} (\bibinfo {year} {2019})}\BibitemShut {NoStop}%
\bibitem [{\citenamefont {Devakul}\ \emph {et~al.}(2021)\citenamefont {Devakul}, \citenamefont {Cr{\'e}pel}, \citenamefont {Zhang},\ and\ \citenamefont {Fu}}]{devakul2021magic}%
  \BibitemOpen
  \bibfield  {author} {\bibinfo {author} {\bibfnamefont {T.}~\bibnamefont {Devakul}}, \bibinfo {author} {\bibfnamefont {V.}~\bibnamefont {Cr{\'e}pel}}, \bibinfo {author} {\bibfnamefont {Y.}~\bibnamefont {Zhang}},\ and\ \bibinfo {author} {\bibfnamefont {L.}~\bibnamefont {Fu}},\ }\bibfield  {title} {\bibinfo {title} {Magic in twisted transition metal dichalcogenide bilayers},\ }\href@noop {} {\bibfield  {journal} {\bibinfo  {journal} {Nature communications}\ }\textbf {\bibinfo {volume} {12}},\ \bibinfo {pages} {6730} (\bibinfo {year} {2021})}\BibitemShut {NoStop}%
\bibitem [{\citenamefont {Devakul}\ \emph {et~al.}(2023)\citenamefont {Devakul}, \citenamefont {Ledwith}, \citenamefont {Xia}, \citenamefont {Uri}, \citenamefont {de~la Barrera}, \citenamefont {Jarillo-Herrero},\ and\ \citenamefont {Fu}}]{devakul2023magic}%
  \BibitemOpen
  \bibfield  {author} {\bibinfo {author} {\bibfnamefont {T.}~\bibnamefont {Devakul}}, \bibinfo {author} {\bibfnamefont {P.~J.}\ \bibnamefont {Ledwith}}, \bibinfo {author} {\bibfnamefont {L.-Q.}\ \bibnamefont {Xia}}, \bibinfo {author} {\bibfnamefont {A.}~\bibnamefont {Uri}}, \bibinfo {author} {\bibfnamefont {S.~C.}\ \bibnamefont {de~la Barrera}}, \bibinfo {author} {\bibfnamefont {P.}~\bibnamefont {Jarillo-Herrero}},\ and\ \bibinfo {author} {\bibfnamefont {L.}~\bibnamefont {Fu}},\ }\bibfield  {title} {\bibinfo {title} {Magic-angle helical trilayer graphene},\ }\href@noop {} {\bibfield  {journal} {\bibinfo  {journal} {Science Advances}\ }\textbf {\bibinfo {volume} {9}},\ \bibinfo {pages} {eadi6063} (\bibinfo {year} {2023})}\BibitemShut {NoStop}%
\bibitem [{\citenamefont {Morales-Dur{\'a}n}\ \emph {et~al.}(2023{\natexlab{a}})\citenamefont {Morales-Dur{\'a}n}, \citenamefont {Wei},\ and\ \citenamefont {MacDonald}}]{morales2023magic}%
  \BibitemOpen
  \bibfield  {author} {\bibinfo {author} {\bibfnamefont {N.}~\bibnamefont {Morales-Dur{\'a}n}}, \bibinfo {author} {\bibfnamefont {N.}~\bibnamefont {Wei}},\ and\ \bibinfo {author} {\bibfnamefont {A.~H.}\ \bibnamefont {MacDonald}},\ }\bibfield  {title} {\bibinfo {title} {Magic angles and fractional chern insulators in twisted homobilayer tmds},\ }\href@noop {} {\bibfield  {journal} {\bibinfo  {journal} {arXiv preprint arXiv:2308.03143}\ } (\bibinfo {year} {2023}{\natexlab{a}})}\BibitemShut {NoStop}%
\bibitem [{\citenamefont {Reddy}\ and\ \citenamefont {Fu}(2023)}]{reddy2023toward}%
  \BibitemOpen
  \bibfield  {author} {\bibinfo {author} {\bibfnamefont {A.~P.}\ \bibnamefont {Reddy}}\ and\ \bibinfo {author} {\bibfnamefont {L.}~\bibnamefont {Fu}},\ }\bibfield  {title} {\bibinfo {title} {Toward a global phase diagram of the fractional quantum anomalous hall effect},\ }\href {https://doi.org/10.1103/PhysRevB.108.245159} {\bibfield  {journal} {\bibinfo  {journal} {Phys. Rev. B}\ }\textbf {\bibinfo {volume} {108}},\ \bibinfo {pages} {245159} (\bibinfo {year} {2023})}\BibitemShut {NoStop}%
\bibitem [{\citenamefont {Cr{\'e}pel}\ \emph {et~al.}(2023)\citenamefont {Cr{\'e}pel}, \citenamefont {Regnault},\ and\ \citenamefont {Queiroz}}]{crepel2023chiral}%
  \BibitemOpen
  \bibfield  {author} {\bibinfo {author} {\bibfnamefont {V.}~\bibnamefont {Cr{\'e}pel}}, \bibinfo {author} {\bibfnamefont {N.}~\bibnamefont {Regnault}},\ and\ \bibinfo {author} {\bibfnamefont {R.}~\bibnamefont {Queiroz}},\ }\bibfield  {title} {\bibinfo {title} {The chiral limits of moir$\backslash$'e semiconductors: origin of flat bands and topology in twisted transition metal dichalcogenides homobilayers},\ }\href@noop {} {\bibfield  {journal} {\bibinfo  {journal} {arXiv preprint arXiv:2305.10477}\ } (\bibinfo {year} {2023})}\BibitemShut {NoStop}%
\bibitem [{\citenamefont {Reddy}\ \emph {et~al.}(2023)\citenamefont {Reddy}, \citenamefont {Alsallom}, \citenamefont {Zhang}, \citenamefont {Devakul},\ and\ \citenamefont {Fu}}]{reddy2023fractional}%
  \BibitemOpen
  \bibfield  {author} {\bibinfo {author} {\bibfnamefont {A.~P.}\ \bibnamefont {Reddy}}, \bibinfo {author} {\bibfnamefont {F.}~\bibnamefont {Alsallom}}, \bibinfo {author} {\bibfnamefont {Y.}~\bibnamefont {Zhang}}, \bibinfo {author} {\bibfnamefont {T.}~\bibnamefont {Devakul}},\ and\ \bibinfo {author} {\bibfnamefont {L.}~\bibnamefont {Fu}},\ }\bibfield  {title} {\bibinfo {title} {Fractional quantum anomalous hall states in twisted bilayer ${\mathrm{mote}}_{2}$ and ${\mathrm{wse}}_{2}$},\ }\href {https://doi.org/10.1103/PhysRevB.108.085117} {\bibfield  {journal} {\bibinfo  {journal} {Phys. Rev. B}\ }\textbf {\bibinfo {volume} {108}},\ \bibinfo {pages} {085117} (\bibinfo {year} {2023})}\BibitemShut {NoStop}%
\bibitem [{\citenamefont {Fu}\ \emph {et~al.}(2020)\citenamefont {Fu}, \citenamefont {K{\"o}nig}, \citenamefont {Wilson}, \citenamefont {Chou},\ and\ \citenamefont {Pixley}}]{fu2020magic}%
  \BibitemOpen
  \bibfield  {author} {\bibinfo {author} {\bibfnamefont {Y.}~\bibnamefont {Fu}}, \bibinfo {author} {\bibfnamefont {E.~J.}\ \bibnamefont {K{\"o}nig}}, \bibinfo {author} {\bibfnamefont {J.~H.}\ \bibnamefont {Wilson}}, \bibinfo {author} {\bibfnamefont {Y.-Z.}\ \bibnamefont {Chou}},\ and\ \bibinfo {author} {\bibfnamefont {J.~H.}\ \bibnamefont {Pixley}},\ }\bibfield  {title} {\bibinfo {title} {Magic-angle semimetals},\ }\href@noop {} {\bibfield  {journal} {\bibinfo  {journal} {npj Quantum Materials}\ }\textbf {\bibinfo {volume} {5}},\ \bibinfo {pages} {71} (\bibinfo {year} {2020})}\BibitemShut {NoStop}%
\bibitem [{\citenamefont {Fu}\ \emph {et~al.}(2021)\citenamefont {Fu}, \citenamefont {Wilson},\ and\ \citenamefont {Pixley}}]{fu2021flat}%
  \BibitemOpen
  \bibfield  {author} {\bibinfo {author} {\bibfnamefont {Y.}~\bibnamefont {Fu}}, \bibinfo {author} {\bibfnamefont {J.~H.}\ \bibnamefont {Wilson}},\ and\ \bibinfo {author} {\bibfnamefont {J.}~\bibnamefont {Pixley}},\ }\bibfield  {title} {\bibinfo {title} {Flat topological bands and eigenstate criticality in a quasiperiodic insulator},\ }\href@noop {} {\bibfield  {journal} {\bibinfo  {journal} {Physical Review B}\ }\textbf {\bibinfo {volume} {104}},\ \bibinfo {pages} {L041106} (\bibinfo {year} {2021})}\BibitemShut {NoStop}%
\bibitem [{\citenamefont {Wang}\ \emph {et~al.}(2013)\citenamefont {Wang}, \citenamefont {Weng}, \citenamefont {Wu}, \citenamefont {Dai},\ and\ \citenamefont {Fang}}]{wang2013three}%
  \BibitemOpen
  \bibfield  {author} {\bibinfo {author} {\bibfnamefont {Z.}~\bibnamefont {Wang}}, \bibinfo {author} {\bibfnamefont {H.}~\bibnamefont {Weng}}, \bibinfo {author} {\bibfnamefont {Q.}~\bibnamefont {Wu}}, \bibinfo {author} {\bibfnamefont {X.}~\bibnamefont {Dai}},\ and\ \bibinfo {author} {\bibfnamefont {Z.}~\bibnamefont {Fang}},\ }\bibfield  {title} {\bibinfo {title} {Three-dimensional dirac semimetal and quantum transport in cd 3 as 2},\ }\href@noop {} {\bibfield  {journal} {\bibinfo  {journal} {Physical Review B}\ }\textbf {\bibinfo {volume} {88}},\ \bibinfo {pages} {125427} (\bibinfo {year} {2013})}\BibitemShut {NoStop}%
\bibitem [{\citenamefont {Armitage}\ \emph {et~al.}(2018)\citenamefont {Armitage}, \citenamefont {Mele},\ and\ \citenamefont {Vishwanath}}]{armitage2018weyl}%
  \BibitemOpen
  \bibfield  {author} {\bibinfo {author} {\bibfnamefont {N.}~\bibnamefont {Armitage}}, \bibinfo {author} {\bibfnamefont {E.}~\bibnamefont {Mele}},\ and\ \bibinfo {author} {\bibfnamefont {A.}~\bibnamefont {Vishwanath}},\ }\bibfield  {title} {\bibinfo {title} {Weyl and dirac semimetals in three-dimensional solids},\ }\href@noop {} {\bibfield  {journal} {\bibinfo  {journal} {Reviews of Modern Physics}\ }\textbf {\bibinfo {volume} {90}},\ \bibinfo {pages} {015001} (\bibinfo {year} {2018})}\BibitemShut {NoStop}%
\bibitem [{\citenamefont {Borisenko}\ \emph {et~al.}(2014)\citenamefont {Borisenko}, \citenamefont {Gibson}, \citenamefont {Evtushinsky}, \citenamefont {Zabolotnyy}, \citenamefont {B{\"u}chner},\ and\ \citenamefont {Cava}}]{borisenko2014experimental}%
  \BibitemOpen
  \bibfield  {author} {\bibinfo {author} {\bibfnamefont {S.}~\bibnamefont {Borisenko}}, \bibinfo {author} {\bibfnamefont {Q.}~\bibnamefont {Gibson}}, \bibinfo {author} {\bibfnamefont {D.}~\bibnamefont {Evtushinsky}}, \bibinfo {author} {\bibfnamefont {V.}~\bibnamefont {Zabolotnyy}}, \bibinfo {author} {\bibfnamefont {B.}~\bibnamefont {B{\"u}chner}},\ and\ \bibinfo {author} {\bibfnamefont {R.~J.}\ \bibnamefont {Cava}},\ }\bibfield  {title} {\bibinfo {title} {Experimental realization of a three-dimensional dirac semimetal},\ }\href@noop {} {\bibfield  {journal} {\bibinfo  {journal} {Physical review letters}\ }\textbf {\bibinfo {volume} {113}},\ \bibinfo {pages} {027603} (\bibinfo {year} {2014})}\BibitemShut {NoStop}%
\bibitem [{\citenamefont {Neupane}\ \emph {et~al.}(2014)\citenamefont {Neupane}, \citenamefont {Xu}, \citenamefont {Sankar}, \citenamefont {Alidoust}, \citenamefont {Bian}, \citenamefont {Liu}, \citenamefont {Belopolski}, \citenamefont {Chang}, \citenamefont {Jeng}, \citenamefont {Lin} \emph {et~al.}}]{neupane2014observation}%
  \BibitemOpen
  \bibfield  {author} {\bibinfo {author} {\bibfnamefont {M.}~\bibnamefont {Neupane}}, \bibinfo {author} {\bibfnamefont {S.-Y.}\ \bibnamefont {Xu}}, \bibinfo {author} {\bibfnamefont {R.}~\bibnamefont {Sankar}}, \bibinfo {author} {\bibfnamefont {N.}~\bibnamefont {Alidoust}}, \bibinfo {author} {\bibfnamefont {G.}~\bibnamefont {Bian}}, \bibinfo {author} {\bibfnamefont {C.}~\bibnamefont {Liu}}, \bibinfo {author} {\bibfnamefont {I.}~\bibnamefont {Belopolski}}, \bibinfo {author} {\bibfnamefont {T.-R.}\ \bibnamefont {Chang}}, \bibinfo {author} {\bibfnamefont {H.-T.}\ \bibnamefont {Jeng}}, \bibinfo {author} {\bibfnamefont {H.}~\bibnamefont {Lin}}, \emph {et~al.},\ }\bibfield  {title} {\bibinfo {title} {Observation of a three-dimensional topological dirac semimetal phase in high-mobility cd3as2},\ }\href@noop {} {\bibfield  {journal} {\bibinfo  {journal} {Nature communications}\ }\textbf {\bibinfo {volume} {5}},\ \bibinfo {pages} {3786} (\bibinfo {year} {2014})}\BibitemShut {NoStop}%
\bibitem [{\citenamefont {Liu}\ \emph {et~al.}(2014)\citenamefont {Liu}, \citenamefont {Jiang}, \citenamefont {Zhou}, \citenamefont {Wang}, \citenamefont {Zhang}, \citenamefont {Weng}, \citenamefont {Prabhakaran}, \citenamefont {Mo}, \citenamefont {Peng}, \citenamefont {Dudin} \emph {et~al.}}]{liu2014stable}%
  \BibitemOpen
  \bibfield  {author} {\bibinfo {author} {\bibfnamefont {Z.}~\bibnamefont {Liu}}, \bibinfo {author} {\bibfnamefont {J.}~\bibnamefont {Jiang}}, \bibinfo {author} {\bibfnamefont {B.}~\bibnamefont {Zhou}}, \bibinfo {author} {\bibfnamefont {Z.}~\bibnamefont {Wang}}, \bibinfo {author} {\bibfnamefont {Y.}~\bibnamefont {Zhang}}, \bibinfo {author} {\bibfnamefont {H.}~\bibnamefont {Weng}}, \bibinfo {author} {\bibfnamefont {D.}~\bibnamefont {Prabhakaran}}, \bibinfo {author} {\bibfnamefont {S.~K.}\ \bibnamefont {Mo}}, \bibinfo {author} {\bibfnamefont {H.}~\bibnamefont {Peng}}, \bibinfo {author} {\bibfnamefont {P.}~\bibnamefont {Dudin}}, \emph {et~al.},\ }\bibfield  {title} {\bibinfo {title} {A stable three-dimensional topological dirac semimetal cd 3 as 2},\ }\href@noop {} {\bibfield  {journal} {\bibinfo  {journal} {Nature materials}\ }\textbf {\bibinfo {volume} {13}},\ \bibinfo {pages} {677} (\bibinfo {year} {2014})}\BibitemShut {NoStop}%
\bibitem [{\citenamefont {Ali}\ \emph {et~al.}(2014)\citenamefont {Ali}, \citenamefont {Gibson}, \citenamefont {Jeon}, \citenamefont {Zhou}, \citenamefont {Yazdani},\ and\ \citenamefont {Cava}}]{ali2014crystal}%
  \BibitemOpen
  \bibfield  {author} {\bibinfo {author} {\bibfnamefont {M.~N.}\ \bibnamefont {Ali}}, \bibinfo {author} {\bibfnamefont {Q.}~\bibnamefont {Gibson}}, \bibinfo {author} {\bibfnamefont {S.}~\bibnamefont {Jeon}}, \bibinfo {author} {\bibfnamefont {B.~B.}\ \bibnamefont {Zhou}}, \bibinfo {author} {\bibfnamefont {A.}~\bibnamefont {Yazdani}},\ and\ \bibinfo {author} {\bibfnamefont {R.~J.}\ \bibnamefont {Cava}},\ }\bibfield  {title} {\bibinfo {title} {The crystal and electronic structures of cd3as2, the three-dimensional electronic analogue of graphene},\ }\href@noop {} {\bibfield  {journal} {\bibinfo  {journal} {Inorganic chemistry}\ }\textbf {\bibinfo {volume} {53}},\ \bibinfo {pages} {4062} (\bibinfo {year} {2014})}\BibitemShut {NoStop}%
\bibitem [{\citenamefont {Crassee}\ \emph {et~al.}(2018)\citenamefont {Crassee}, \citenamefont {Sankar}, \citenamefont {Lee}, \citenamefont {Akrap},\ and\ \citenamefont {Orlita}}]{crassee20183d}%
  \BibitemOpen
  \bibfield  {author} {\bibinfo {author} {\bibfnamefont {I.}~\bibnamefont {Crassee}}, \bibinfo {author} {\bibfnamefont {R.}~\bibnamefont {Sankar}}, \bibinfo {author} {\bibfnamefont {W.-L.}\ \bibnamefont {Lee}}, \bibinfo {author} {\bibfnamefont {A.}~\bibnamefont {Akrap}},\ and\ \bibinfo {author} {\bibfnamefont {M.}~\bibnamefont {Orlita}},\ }\bibfield  {title} {\bibinfo {title} {3d dirac semimetal cd 3 as 2: A review of material properties},\ }\href@noop {} {\bibfield  {journal} {\bibinfo  {journal} {Physical Review Materials}\ }\textbf {\bibinfo {volume} {2}},\ \bibinfo {pages} {120302} (\bibinfo {year} {2018})}\BibitemShut {NoStop}%
\bibitem [{\citenamefont {Lygo}\ \emph {et~al.}(2023)\citenamefont {Lygo}, \citenamefont {Guo}, \citenamefont {Rashidi}, \citenamefont {Huang}, \citenamefont {Cuadros-Romero},\ and\ \citenamefont {Stemmer}}]{lygo2023two}%
  \BibitemOpen
  \bibfield  {author} {\bibinfo {author} {\bibfnamefont {A.~C.}\ \bibnamefont {Lygo}}, \bibinfo {author} {\bibfnamefont {B.}~\bibnamefont {Guo}}, \bibinfo {author} {\bibfnamefont {A.}~\bibnamefont {Rashidi}}, \bibinfo {author} {\bibfnamefont {V.}~\bibnamefont {Huang}}, \bibinfo {author} {\bibfnamefont {P.}~\bibnamefont {Cuadros-Romero}},\ and\ \bibinfo {author} {\bibfnamefont {S.}~\bibnamefont {Stemmer}},\ }\bibfield  {title} {\bibinfo {title} {Two-dimensional topological insulator state in cadmium arsenide thin films},\ }\href@noop {} {\bibfield  {journal} {\bibinfo  {journal} {Physical Review Letters}\ }\textbf {\bibinfo {volume} {130}},\ \bibinfo {pages} {046201} (\bibinfo {year} {2023})}\BibitemShut {NoStop}%
\bibitem [{\citenamefont {Cano}\ \emph {et~al.}(2017{\natexlab{a}})\citenamefont {Cano}, \citenamefont {Bradlyn}, \citenamefont {Wang}, \citenamefont {Hirschberger}, \citenamefont {Ong},\ and\ \citenamefont {Bernevig}}]{cano2017chiral}%
  \BibitemOpen
  \bibfield  {author} {\bibinfo {author} {\bibfnamefont {J.}~\bibnamefont {Cano}}, \bibinfo {author} {\bibfnamefont {B.}~\bibnamefont {Bradlyn}}, \bibinfo {author} {\bibfnamefont {Z.}~\bibnamefont {Wang}}, \bibinfo {author} {\bibfnamefont {M.}~\bibnamefont {Hirschberger}}, \bibinfo {author} {\bibfnamefont {N.~P.}\ \bibnamefont {Ong}},\ and\ \bibinfo {author} {\bibfnamefont {B.~A.}\ \bibnamefont {Bernevig}},\ }\bibfield  {title} {\bibinfo {title} {Chiral anomaly factory: Creating weyl fermions with a magnetic field},\ }\href@noop {} {\bibfield  {journal} {\bibinfo  {journal} {Physical Review B}\ }\textbf {\bibinfo {volume} {95}},\ \bibinfo {pages} {161306} (\bibinfo {year} {2017}{\natexlab{a}})}\BibitemShut {NoStop}%
\bibitem [{\citenamefont {Baidya}\ and\ \citenamefont {Vanderbilt}(2020)}]{baidya2020first}%
  \BibitemOpen
  \bibfield  {author} {\bibinfo {author} {\bibfnamefont {S.}~\bibnamefont {Baidya}}\ and\ \bibinfo {author} {\bibfnamefont {D.}~\bibnamefont {Vanderbilt}},\ }\bibfield  {title} {\bibinfo {title} {First-principles theory of the dirac semimetal cd 3 as 2 under zeeman magnetic field},\ }\href@noop {} {\bibfield  {journal} {\bibinfo  {journal} {Physical Review B}\ }\textbf {\bibinfo {volume} {102}},\ \bibinfo {pages} {165115} (\bibinfo {year} {2020})}\BibitemShut {NoStop}%
\bibitem [{\citenamefont {Miao}\ \emph {et~al.}(2023)\citenamefont {Miao}, \citenamefont {Guo}, \citenamefont {Stemmer},\ and\ \citenamefont {Dai}}]{miao2023engineering}%
  \BibitemOpen
  \bibfield  {author} {\bibinfo {author} {\bibfnamefont {W.}~\bibnamefont {Miao}}, \bibinfo {author} {\bibfnamefont {B.}~\bibnamefont {Guo}}, \bibinfo {author} {\bibfnamefont {S.}~\bibnamefont {Stemmer}},\ and\ \bibinfo {author} {\bibfnamefont {X.}~\bibnamefont {Dai}},\ }\bibfield  {title} {\bibinfo {title} {Engineering the in-plane anomalous hall effect in cd $ \_3 $ as $ \_2 $ thin films},\ }\href@noop {} {\bibfield  {journal} {\bibinfo  {journal} {arXiv preprint arXiv:2309.15457}\ } (\bibinfo {year} {2023})}\BibitemShut {NoStop}%
\bibitem [{\citenamefont {Morales-Dur{\'a}n}\ \emph {et~al.}(2023{\natexlab{b}})\citenamefont {Morales-Dur{\'a}n}, \citenamefont {Wang}, \citenamefont {Schleder}, \citenamefont {Angeli}, \citenamefont {Zhu}, \citenamefont {Kaxiras}, \citenamefont {Repellin},\ and\ \citenamefont {Cano}}]{morales2023pressure}%
  \BibitemOpen
  \bibfield  {author} {\bibinfo {author} {\bibfnamefont {N.}~\bibnamefont {Morales-Dur{\'a}n}}, \bibinfo {author} {\bibfnamefont {J.}~\bibnamefont {Wang}}, \bibinfo {author} {\bibfnamefont {G.~R.}\ \bibnamefont {Schleder}}, \bibinfo {author} {\bibfnamefont {M.}~\bibnamefont {Angeli}}, \bibinfo {author} {\bibfnamefont {Z.}~\bibnamefont {Zhu}}, \bibinfo {author} {\bibfnamefont {E.}~\bibnamefont {Kaxiras}}, \bibinfo {author} {\bibfnamefont {C.}~\bibnamefont {Repellin}},\ and\ \bibinfo {author} {\bibfnamefont {J.}~\bibnamefont {Cano}},\ }\bibfield  {title} {\bibinfo {title} {Pressure-enhanced fractional chern insulators along a magic line in moir{\'e} transition metal dichalcogenides},\ }\href@noop {} {\bibfield  {journal} {\bibinfo  {journal} {Physical Review Research}\ }\textbf {\bibinfo {volume} {5}},\ \bibinfo {pages} {L032022} (\bibinfo {year} {2023}{\natexlab{b}})}\BibitemShut {NoStop}%
\bibitem [{\citenamefont {Wilhelm}\ \emph {et~al.}(2021)\citenamefont {Wilhelm}, \citenamefont {Lang},\ and\ \citenamefont {L{\"a}uchli}}]{wilhelm2021interplay}%
  \BibitemOpen
  \bibfield  {author} {\bibinfo {author} {\bibfnamefont {P.}~\bibnamefont {Wilhelm}}, \bibinfo {author} {\bibfnamefont {T.~C.}\ \bibnamefont {Lang}},\ and\ \bibinfo {author} {\bibfnamefont {A.~M.}\ \bibnamefont {L{\"a}uchli}},\ }\bibfield  {title} {\bibinfo {title} {Interplay of fractional chern insulator and charge density wave phases in twisted bilayer graphene},\ }\href@noop {} {\bibfield  {journal} {\bibinfo  {journal} {Physical Review B}\ }\textbf {\bibinfo {volume} {103}},\ \bibinfo {pages} {125406} (\bibinfo {year} {2021})}\BibitemShut {NoStop}%
\bibitem [{\citenamefont {Goldman}\ \emph {et~al.}(2023)\citenamefont {Goldman}, \citenamefont {Reddy}, \citenamefont {Paul},\ and\ \citenamefont {Fu}}]{goldman2023zero}%
  \BibitemOpen
  \bibfield  {author} {\bibinfo {author} {\bibfnamefont {H.}~\bibnamefont {Goldman}}, \bibinfo {author} {\bibfnamefont {A.~P.}\ \bibnamefont {Reddy}}, \bibinfo {author} {\bibfnamefont {N.}~\bibnamefont {Paul}},\ and\ \bibinfo {author} {\bibfnamefont {L.}~\bibnamefont {Fu}},\ }\bibfield  {title} {\bibinfo {title} {Zero-field composite fermi liquid in twisted semiconductor bilayers},\ }\href@noop {} {\bibfield  {journal} {\bibinfo  {journal} {arXiv preprint arXiv:2306.02513 (PRL in press)}\ } (\bibinfo {year} {2023})}\BibitemShut {NoStop}%
\bibitem [{\citenamefont {Dong}\ \emph {et~al.}(2023)\citenamefont {Dong}, \citenamefont {Wang}, \citenamefont {Ledwith}, \citenamefont {Vishwanath},\ and\ \citenamefont {Parker}}]{dong2023composite}%
  \BibitemOpen
  \bibfield  {author} {\bibinfo {author} {\bibfnamefont {J.}~\bibnamefont {Dong}}, \bibinfo {author} {\bibfnamefont {J.}~\bibnamefont {Wang}}, \bibinfo {author} {\bibfnamefont {P.~J.}\ \bibnamefont {Ledwith}}, \bibinfo {author} {\bibfnamefont {A.}~\bibnamefont {Vishwanath}},\ and\ \bibinfo {author} {\bibfnamefont {D.~E.}\ \bibnamefont {Parker}},\ }\bibfield  {title} {\bibinfo {title} {Composite fermi liquid at zero magnetic field in twisted mote $ \_2$},\ }\href@noop {} {\bibfield  {journal} {\bibinfo  {journal} {arXiv preprint arXiv:2306.01719}\ } (\bibinfo {year} {2023})}\BibitemShut {NoStop}%
\bibitem [{\citenamefont {Yankowitz}\ \emph {et~al.}(2018)\citenamefont {Yankowitz}, \citenamefont {Jung}, \citenamefont {Laksono}, \citenamefont {Leconte}, \citenamefont {Chittari}, \citenamefont {Watanabe}, \citenamefont {Taniguchi}, \citenamefont {Adam}, \citenamefont {Graf},\ and\ \citenamefont {Dean}}]{yankowitz2018dynamic}%
  \BibitemOpen
  \bibfield  {author} {\bibinfo {author} {\bibfnamefont {M.}~\bibnamefont {Yankowitz}}, \bibinfo {author} {\bibfnamefont {J.}~\bibnamefont {Jung}}, \bibinfo {author} {\bibfnamefont {E.}~\bibnamefont {Laksono}}, \bibinfo {author} {\bibfnamefont {N.}~\bibnamefont {Leconte}}, \bibinfo {author} {\bibfnamefont {B.~L.}\ \bibnamefont {Chittari}}, \bibinfo {author} {\bibfnamefont {K.}~\bibnamefont {Watanabe}}, \bibinfo {author} {\bibfnamefont {T.}~\bibnamefont {Taniguchi}}, \bibinfo {author} {\bibfnamefont {S.}~\bibnamefont {Adam}}, \bibinfo {author} {\bibfnamefont {D.}~\bibnamefont {Graf}},\ and\ \bibinfo {author} {\bibfnamefont {C.~R.}\ \bibnamefont {Dean}},\ }\bibfield  {title} {\bibinfo {title} {Dynamic band-structure tuning of graphene moir{\'e} superlattices with pressure},\ }\href@noop {} {\bibfield  {journal} {\bibinfo  {journal} {Nature}\ }\textbf {\bibinfo {volume} {557}},\ \bibinfo {pages} {404} (\bibinfo {year} {2018})}\BibitemShut {NoStop}%
\bibitem [{\citenamefont {Yankowitz}\ \emph {et~al.}(2019)\citenamefont {Yankowitz}, \citenamefont {Chen}, \citenamefont {Polshyn}, \citenamefont {Zhang}, \citenamefont {Watanabe}, \citenamefont {Taniguchi}, \citenamefont {Graf}, \citenamefont {Young},\ and\ \citenamefont {Dean}}]{yankowitz2019tuning}%
  \BibitemOpen
  \bibfield  {author} {\bibinfo {author} {\bibfnamefont {M.}~\bibnamefont {Yankowitz}}, \bibinfo {author} {\bibfnamefont {S.}~\bibnamefont {Chen}}, \bibinfo {author} {\bibfnamefont {H.}~\bibnamefont {Polshyn}}, \bibinfo {author} {\bibfnamefont {Y.}~\bibnamefont {Zhang}}, \bibinfo {author} {\bibfnamefont {K.}~\bibnamefont {Watanabe}}, \bibinfo {author} {\bibfnamefont {T.}~\bibnamefont {Taniguchi}}, \bibinfo {author} {\bibfnamefont {D.}~\bibnamefont {Graf}}, \bibinfo {author} {\bibfnamefont {A.~F.}\ \bibnamefont {Young}},\ and\ \bibinfo {author} {\bibfnamefont {C.~R.}\ \bibnamefont {Dean}},\ }\bibfield  {title} {\bibinfo {title} {Tuning superconductivity in twisted bilayer graphene},\ }\href@noop {} {\bibfield  {journal} {\bibinfo  {journal} {Science}\ }\textbf {\bibinfo {volume} {363}},\ \bibinfo {pages} {1059} (\bibinfo {year} {2019})}\BibitemShut {NoStop}%
\bibitem [{\citenamefont {Carr}\ \emph {et~al.}(2018)\citenamefont {Carr}, \citenamefont {Fang}, \citenamefont {Jarillo-Herrero},\ and\ \citenamefont {Kaxiras}}]{carr2018pressure}%
  \BibitemOpen
  \bibfield  {author} {\bibinfo {author} {\bibfnamefont {S.}~\bibnamefont {Carr}}, \bibinfo {author} {\bibfnamefont {S.}~\bibnamefont {Fang}}, \bibinfo {author} {\bibfnamefont {P.}~\bibnamefont {Jarillo-Herrero}},\ and\ \bibinfo {author} {\bibfnamefont {E.}~\bibnamefont {Kaxiras}},\ }\bibfield  {title} {\bibinfo {title} {Pressure dependence of the magic twist angle in graphene superlattices},\ }\href@noop {} {\bibfield  {journal} {\bibinfo  {journal} {Physical Review B}\ }\textbf {\bibinfo {volume} {98}},\ \bibinfo {pages} {085144} (\bibinfo {year} {2018})}\BibitemShut {NoStop}%
\bibitem [{\citenamefont {Hasan}\ and\ \citenamefont {Kane}(2010)}]{hasan2010colloquium}%
  \BibitemOpen
  \bibfield  {author} {\bibinfo {author} {\bibfnamefont {M.~Z.}\ \bibnamefont {Hasan}}\ and\ \bibinfo {author} {\bibfnamefont {C.~L.}\ \bibnamefont {Kane}},\ }\bibfield  {title} {\bibinfo {title} {Colloquium: topological insulators},\ }\href@noop {} {\bibfield  {journal} {\bibinfo  {journal} {Reviews of modern physics}\ }\textbf {\bibinfo {volume} {82}},\ \bibinfo {pages} {3045} (\bibinfo {year} {2010})}\BibitemShut {NoStop}%
\bibitem [{\citenamefont {Hsieh}\ \emph {et~al.}(2012)\citenamefont {Hsieh}, \citenamefont {Lin}, \citenamefont {Liu}, \citenamefont {Duan}, \citenamefont {Bansil},\ and\ \citenamefont {Fu}}]{hsieh2012topological}%
  \BibitemOpen
  \bibfield  {author} {\bibinfo {author} {\bibfnamefont {T.~H.}\ \bibnamefont {Hsieh}}, \bibinfo {author} {\bibfnamefont {H.}~\bibnamefont {Lin}}, \bibinfo {author} {\bibfnamefont {J.}~\bibnamefont {Liu}}, \bibinfo {author} {\bibfnamefont {W.}~\bibnamefont {Duan}}, \bibinfo {author} {\bibfnamefont {A.}~\bibnamefont {Bansil}},\ and\ \bibinfo {author} {\bibfnamefont {L.}~\bibnamefont {Fu}},\ }\bibfield  {title} {\bibinfo {title} {Topological crystalline insulators in the snte material class},\ }\href@noop {} {\bibfield  {journal} {\bibinfo  {journal} {Nature communications}\ }\textbf {\bibinfo {volume} {3}},\ \bibinfo {pages} {982} (\bibinfo {year} {2012})}\BibitemShut {NoStop}%
\bibitem [{\citenamefont {Liu}\ and\ \citenamefont {Fu}(2015)}]{liu2015electrically}%
  \BibitemOpen
  \bibfield  {author} {\bibinfo {author} {\bibfnamefont {J.}~\bibnamefont {Liu}}\ and\ \bibinfo {author} {\bibfnamefont {L.}~\bibnamefont {Fu}},\ }\bibfield  {title} {\bibinfo {title} {Electrically tunable quantum spin hall state in topological crystalline insulator thin films},\ }\href@noop {} {\bibfield  {journal} {\bibinfo  {journal} {Physical Review B}\ }\textbf {\bibinfo {volume} {91}},\ \bibinfo {pages} {081407} (\bibinfo {year} {2015})}\BibitemShut {NoStop}%
\bibitem [{\citenamefont {Murakami}(2006)}]{murakami2006quantum}%
  \BibitemOpen
  \bibfield  {author} {\bibinfo {author} {\bibfnamefont {S.}~\bibnamefont {Murakami}},\ }\bibfield  {title} {\bibinfo {title} {Quantum spin hall effect and enhanced magnetic response by spin-orbit coupling},\ }\href@noop {} {\bibfield  {journal} {\bibinfo  {journal} {Physical Review Letters}\ }\textbf {\bibinfo {volume} {97}},\ \bibinfo {pages} {236805} (\bibinfo {year} {2006})}\BibitemShut {NoStop}%
\bibitem [{\citenamefont {Chen}\ \emph {et~al.}(2023)\citenamefont {Chen}, \citenamefont {Wu}, \citenamefont {Tulu}, \citenamefont {Wang}, \citenamefont {Juanson}, \citenamefont {Watanabe}, \citenamefont {Taniguchi}, \citenamefont {Pettes}, \citenamefont {Campbell}, \citenamefont {Gadre}, \citenamefont {Zhou}, \citenamefont {Chen}, \citenamefont {Cao}, \citenamefont {Jauregui}, \citenamefont {Wu}, \citenamefont {Pan},\ and\ \citenamefont {Sanchez-Yamagishi}}]{chen2023exceptional}%
  \BibitemOpen
  \bibfield  {author} {\bibinfo {author} {\bibfnamefont {L.}~\bibnamefont {Chen}}, \bibinfo {author} {\bibfnamefont {A.~X.}\ \bibnamefont {Wu}}, \bibinfo {author} {\bibfnamefont {N.}~\bibnamefont {Tulu}}, \bibinfo {author} {\bibfnamefont {J.}~\bibnamefont {Wang}}, \bibinfo {author} {\bibfnamefont {A.}~\bibnamefont {Juanson}}, \bibinfo {author} {\bibfnamefont {K.}~\bibnamefont {Watanabe}}, \bibinfo {author} {\bibfnamefont {T.}~\bibnamefont {Taniguchi}}, \bibinfo {author} {\bibfnamefont {M.~T.}\ \bibnamefont {Pettes}}, \bibinfo {author} {\bibfnamefont {M.}~\bibnamefont {Campbell}}, \bibinfo {author} {\bibfnamefont {C.~A.}\ \bibnamefont {Gadre}}, \bibinfo {author} {\bibfnamefont {Y.}~\bibnamefont {Zhou}}, \bibinfo {author} {\bibfnamefont {H.}~\bibnamefont {Chen}}, \bibinfo {author} {\bibfnamefont {P.}~\bibnamefont {Cao}}, \bibinfo {author} {\bibfnamefont {L.~A.}\ \bibnamefont {Jauregui}}, \bibinfo {author} {\bibfnamefont {R.}~\bibnamefont {Wu}}, \bibinfo {author} {\bibfnamefont {X.}~\bibnamefont {Pan}},\ and\
  \bibinfo {author} {\bibfnamefont {J.~D.}\ \bibnamefont {Sanchez-Yamagishi}},\ }\href@noop {} {\bibinfo {title} {Exceptional electronic transport and quantum oscillations in thin bismuth crystals grown inside van der waals materials}} (\bibinfo {year} {2023}),\ \Eprint {https://arxiv.org/abs/2211.07681} {arXiv:2211.07681 [cond-mat.mes-hall]} \BibitemShut {NoStop}%
\bibitem [{\citenamefont {Wu}\ \emph {et~al.}(2018)\citenamefont {Wu}, \citenamefont {Lovorn},\ and\ \citenamefont {Tutuc}}]{wu2018hubbard}%
  \BibitemOpen
  \bibfield  {author} {\bibinfo {author} {\bibfnamefont {F.}~\bibnamefont {Wu}}, \bibinfo {author} {\bibfnamefont {T.}~\bibnamefont {Lovorn}},\ and\ \bibinfo {author} {\bibfnamefont {E.}~\bibnamefont {Tutuc}},\ }\bibfield  {title} {\bibinfo {title} {Hubbard model physics in transition metal dichalcogenide moir{\'e} bands},\ }\href@noop {} {\bibfield  {journal} {\bibinfo  {journal} {Physical review letters}\ }\textbf {\bibinfo {volume} {121}},\ \bibinfo {pages} {026402} (\bibinfo {year} {2018})}\BibitemShut {NoStop}%
\bibitem [{\citenamefont {Lu}\ \emph {et~al.}(2023{\natexlab{b}})\citenamefont {Lu}, \citenamefont {Zhang}, \citenamefont {Wang}, \citenamefont {Gao}, \citenamefont {Yang}, \citenamefont {Guo}, \citenamefont {Gao}, \citenamefont {Ye}, \citenamefont {Han},\ and\ \citenamefont {Liu}}]{lu2023synergistic}%
  \BibitemOpen
  \bibfield  {author} {\bibinfo {author} {\bibfnamefont {X.}~\bibnamefont {Lu}}, \bibinfo {author} {\bibfnamefont {S.}~\bibnamefont {Zhang}}, \bibinfo {author} {\bibfnamefont {Y.}~\bibnamefont {Wang}}, \bibinfo {author} {\bibfnamefont {X.}~\bibnamefont {Gao}}, \bibinfo {author} {\bibfnamefont {K.}~\bibnamefont {Yang}}, \bibinfo {author} {\bibfnamefont {Z.}~\bibnamefont {Guo}}, \bibinfo {author} {\bibfnamefont {Y.}~\bibnamefont {Gao}}, \bibinfo {author} {\bibfnamefont {Y.}~\bibnamefont {Ye}}, \bibinfo {author} {\bibfnamefont {Z.}~\bibnamefont {Han}},\ and\ \bibinfo {author} {\bibfnamefont {J.}~\bibnamefont {Liu}},\ }\bibfield  {title} {\bibinfo {title} {Synergistic correlated states and nontrivial topology in coupled graphene-insulator heterostructures},\ }\href@noop {} {\bibfield  {journal} {\bibinfo  {journal} {Nature Communications}\ }\textbf {\bibinfo {volume} {14}},\ \bibinfo {pages} {5550} (\bibinfo {year} {2023}{\natexlab{b}})}\BibitemShut {NoStop}%
\bibitem [{\citenamefont {Tseng}\ \emph {et~al.}(2022)\citenamefont {Tseng}, \citenamefont {Song}, \citenamefont {Jiang}, \citenamefont {Lin}, \citenamefont {Wang}, \citenamefont {Suh}, \citenamefont {Watanabe}, \citenamefont {Taniguchi}, \citenamefont {McGuire}, \citenamefont {Xiao} \emph {et~al.}}]{tseng2022gate}%
  \BibitemOpen
  \bibfield  {author} {\bibinfo {author} {\bibfnamefont {C.-C.}\ \bibnamefont {Tseng}}, \bibinfo {author} {\bibfnamefont {T.}~\bibnamefont {Song}}, \bibinfo {author} {\bibfnamefont {Q.}~\bibnamefont {Jiang}}, \bibinfo {author} {\bibfnamefont {Z.}~\bibnamefont {Lin}}, \bibinfo {author} {\bibfnamefont {C.}~\bibnamefont {Wang}}, \bibinfo {author} {\bibfnamefont {J.}~\bibnamefont {Suh}}, \bibinfo {author} {\bibfnamefont {K.}~\bibnamefont {Watanabe}}, \bibinfo {author} {\bibfnamefont {T.}~\bibnamefont {Taniguchi}}, \bibinfo {author} {\bibfnamefont {M.~A.}\ \bibnamefont {McGuire}}, \bibinfo {author} {\bibfnamefont {D.}~\bibnamefont {Xiao}}, \emph {et~al.},\ }\bibfield  {title} {\bibinfo {title} {Gate-tunable proximity effects in graphene on layered magnetic insulators},\ }\href@noop {} {\bibfield  {journal} {\bibinfo  {journal} {Nano Letters}\ }\textbf {\bibinfo {volume} {22}},\ \bibinfo {pages} {8495} (\bibinfo {year} {2022})}\BibitemShut {NoStop}%
\bibitem [{\citenamefont {Yang}\ \emph {et~al.}(2023{\natexlab{a}})\citenamefont {Yang}, \citenamefont {Gao}, \citenamefont {Wang}, \citenamefont {Zhang}, \citenamefont {Gao}, \citenamefont {Lu}, \citenamefont {Zhang}, \citenamefont {Liu}, \citenamefont {Gu}, \citenamefont {Luo} \emph {et~al.}}]{yang2023unconventional}%
  \BibitemOpen
  \bibfield  {author} {\bibinfo {author} {\bibfnamefont {K.}~\bibnamefont {Yang}}, \bibinfo {author} {\bibfnamefont {X.}~\bibnamefont {Gao}}, \bibinfo {author} {\bibfnamefont {Y.}~\bibnamefont {Wang}}, \bibinfo {author} {\bibfnamefont {T.}~\bibnamefont {Zhang}}, \bibinfo {author} {\bibfnamefont {Y.}~\bibnamefont {Gao}}, \bibinfo {author} {\bibfnamefont {X.}~\bibnamefont {Lu}}, \bibinfo {author} {\bibfnamefont {S.}~\bibnamefont {Zhang}}, \bibinfo {author} {\bibfnamefont {J.}~\bibnamefont {Liu}}, \bibinfo {author} {\bibfnamefont {P.}~\bibnamefont {Gu}}, \bibinfo {author} {\bibfnamefont {Z.}~\bibnamefont {Luo}}, \emph {et~al.},\ }\bibfield  {title} {\bibinfo {title} {Unconventional correlated insulator in crocl-interfaced bernal bilayer graphene},\ }\href@noop {} {\bibfield  {journal} {\bibinfo  {journal} {Nature Communications}\ }\textbf {\bibinfo {volume} {14}},\ \bibinfo {pages} {2136} (\bibinfo {year} {2023}{\natexlab{a}})}\BibitemShut {NoStop}%
\bibitem [{\citenamefont {Wang}\ \emph {et~al.}(2022{\natexlab{b}})\citenamefont {Wang}, \citenamefont {Gao}, \citenamefont {Yang}, \citenamefont {Gu}, \citenamefont {Lu}, \citenamefont {Zhang}, \citenamefont {Gao}, \citenamefont {Ren}, \citenamefont {Dong}, \citenamefont {Jiang} \emph {et~al.}}]{wang2022quantum}%
  \BibitemOpen
  \bibfield  {author} {\bibinfo {author} {\bibfnamefont {Y.}~\bibnamefont {Wang}}, \bibinfo {author} {\bibfnamefont {X.}~\bibnamefont {Gao}}, \bibinfo {author} {\bibfnamefont {K.}~\bibnamefont {Yang}}, \bibinfo {author} {\bibfnamefont {P.}~\bibnamefont {Gu}}, \bibinfo {author} {\bibfnamefont {X.}~\bibnamefont {Lu}}, \bibinfo {author} {\bibfnamefont {S.}~\bibnamefont {Zhang}}, \bibinfo {author} {\bibfnamefont {Y.}~\bibnamefont {Gao}}, \bibinfo {author} {\bibfnamefont {N.}~\bibnamefont {Ren}}, \bibinfo {author} {\bibfnamefont {B.}~\bibnamefont {Dong}}, \bibinfo {author} {\bibfnamefont {Y.}~\bibnamefont {Jiang}}, \emph {et~al.},\ }\bibfield  {title} {\bibinfo {title} {Quantum hall phase in graphene engineered by interfacial charge coupling},\ }\href@noop {} {\bibfield  {journal} {\bibinfo  {journal} {Nature Nanotechnology}\ }\textbf {\bibinfo {volume} {17}},\ \bibinfo {pages} {1272} (\bibinfo {year} {2022}{\natexlab{b}})}\BibitemShut {NoStop}%
\bibitem [{\citenamefont {Yang}\ \emph {et~al.}(2023{\natexlab{b}})\citenamefont {Yang}, \citenamefont {Xu}, \citenamefont {Feng}, \citenamefont {Schindler}, \citenamefont {Xu}, \citenamefont {Bi}, \citenamefont {Bernevig}, \citenamefont {Tang},\ and\ \citenamefont {Liu}}]{yang2023mathbb}%
  \BibitemOpen
  \bibfield  {author} {\bibinfo {author} {\bibfnamefont {K.}~\bibnamefont {Yang}}, \bibinfo {author} {\bibfnamefont {Z.}~\bibnamefont {Xu}}, \bibinfo {author} {\bibfnamefont {Y.}~\bibnamefont {Feng}}, \bibinfo {author} {\bibfnamefont {F.}~\bibnamefont {Schindler}}, \bibinfo {author} {\bibfnamefont {Y.}~\bibnamefont {Xu}}, \bibinfo {author} {\bibfnamefont {Z.}~\bibnamefont {Bi}}, \bibinfo {author} {\bibfnamefont {B.~A.}\ \bibnamefont {Bernevig}}, \bibinfo {author} {\bibfnamefont {P.}~\bibnamefont {Tang}},\ and\ \bibinfo {author} {\bibfnamefont {C.-X.}\ \bibnamefont {Liu}},\ }\bibfield  {title} {\bibinfo {title} {$\mathbb{Z}_2$-nontrivial moire minibands and interaction-driven quantum anomalous hall insulators in topological insulator based moire heterostructures},\ }\href@noop {} {\bibfield  {journal} {\bibinfo  {journal} {arXiv preprint arXiv:2304.09907}\ } (\bibinfo {year} {2023}{\natexlab{b}})}\BibitemShut {NoStop}%
\bibitem [{\citenamefont {Pai}\ \emph {et~al.}(2018)\citenamefont {Pai}, \citenamefont {Tylan-Tyler}, \citenamefont {Irvin},\ and\ \citenamefont {Levy}}]{pai2018physics}%
  \BibitemOpen
  \bibfield  {author} {\bibinfo {author} {\bibfnamefont {Y.-Y.}\ \bibnamefont {Pai}}, \bibinfo {author} {\bibfnamefont {A.}~\bibnamefont {Tylan-Tyler}}, \bibinfo {author} {\bibfnamefont {P.}~\bibnamefont {Irvin}},\ and\ \bibinfo {author} {\bibfnamefont {J.}~\bibnamefont {Levy}},\ }\bibfield  {title} {\bibinfo {title} {Physics of srtio3-based heterostructures and nanostructures: a review},\ }\href@noop {} {\bibfield  {journal} {\bibinfo  {journal} {Reports on Progress in Physics}\ }\textbf {\bibinfo {volume} {81}},\ \bibinfo {pages} {036503} (\bibinfo {year} {2018})}\BibitemShut {NoStop}%
\bibitem [{\citenamefont {Rezayi}\ and\ \citenamefont {Read}(1994)}]{rezayi1994fermi}%
  \BibitemOpen
  \bibfield  {author} {\bibinfo {author} {\bibfnamefont {E.}~\bibnamefont {Rezayi}}\ and\ \bibinfo {author} {\bibfnamefont {N.}~\bibnamefont {Read}},\ }\bibfield  {title} {\bibinfo {title} {Fermi-liquid-like state in a half-filled landau level},\ }\href@noop {} {\bibfield  {journal} {\bibinfo  {journal} {Physical review letters}\ }\textbf {\bibinfo {volume} {72}},\ \bibinfo {pages} {900} (\bibinfo {year} {1994})}\BibitemShut {NoStop}%
\bibitem [{\citenamefont {Geraedts}\ \emph {et~al.}(2018)\citenamefont {Geraedts}, \citenamefont {Wang}, \citenamefont {Rezayi},\ and\ \citenamefont {Haldane}}]{geraedts2018berry}%
  \BibitemOpen
  \bibfield  {author} {\bibinfo {author} {\bibfnamefont {S.~D.}\ \bibnamefont {Geraedts}}, \bibinfo {author} {\bibfnamefont {J.}~\bibnamefont {Wang}}, \bibinfo {author} {\bibfnamefont {E.}~\bibnamefont {Rezayi}},\ and\ \bibinfo {author} {\bibfnamefont {F.}~\bibnamefont {Haldane}},\ }\bibfield  {title} {\bibinfo {title} {Berry phase and model wave function in the half-filled landau level},\ }\href@noop {} {\bibfield  {journal} {\bibinfo  {journal} {Physical review letters}\ }\textbf {\bibinfo {volume} {121}},\ \bibinfo {pages} {147202} (\bibinfo {year} {2018})}\BibitemShut {NoStop}%
\bibitem [{\citenamefont {Fremling}\ \emph {et~al.}(2018)\citenamefont {Fremling}, \citenamefont {Moran}, \citenamefont {Slingerland},\ and\ \citenamefont {Simon}}]{fremling2018trial}%
  \BibitemOpen
  \bibfield  {author} {\bibinfo {author} {\bibfnamefont {M.}~\bibnamefont {Fremling}}, \bibinfo {author} {\bibfnamefont {N.}~\bibnamefont {Moran}}, \bibinfo {author} {\bibfnamefont {J.}~\bibnamefont {Slingerland}},\ and\ \bibinfo {author} {\bibfnamefont {S.~H.}\ \bibnamefont {Simon}},\ }\bibfield  {title} {\bibinfo {title} {Trial wave functions for a composite fermi liquid on a torus},\ }\href@noop {} {\bibfield  {journal} {\bibinfo  {journal} {Physical Review B}\ }\textbf {\bibinfo {volume} {97}},\ \bibinfo {pages} {035149} (\bibinfo {year} {2018})}\BibitemShut {NoStop}%
\bibitem [{\citenamefont {Wang}\ \emph {et~al.}(2019)\citenamefont {Wang}, \citenamefont {Geraedts}, \citenamefont {Rezayi},\ and\ \citenamefont {Haldane}}]{wang2019lattice}%
  \BibitemOpen
  \bibfield  {author} {\bibinfo {author} {\bibfnamefont {J.}~\bibnamefont {Wang}}, \bibinfo {author} {\bibfnamefont {S.~D.}\ \bibnamefont {Geraedts}}, \bibinfo {author} {\bibfnamefont {E.}~\bibnamefont {Rezayi}},\ and\ \bibinfo {author} {\bibfnamefont {F.}~\bibnamefont {Haldane}},\ }\bibfield  {title} {\bibinfo {title} {Lattice monte carlo for quantum hall states on a torus},\ }\href@noop {} {\bibfield  {journal} {\bibinfo  {journal} {Physical Review B}\ }\textbf {\bibinfo {volume} {99}},\ \bibinfo {pages} {125123} (\bibinfo {year} {2019})}\BibitemShut {NoStop}%
\bibitem [{\citenamefont {Wang}(2019)}]{wang2019dirac}%
  \BibitemOpen
  \bibfield  {author} {\bibinfo {author} {\bibfnamefont {J.}~\bibnamefont {Wang}},\ }\bibfield  {title} {\bibinfo {title} {Dirac fermion hierarchy of composite fermi liquids},\ }\href@noop {} {\bibfield  {journal} {\bibinfo  {journal} {Physical review letters}\ }\textbf {\bibinfo {volume} {122}},\ \bibinfo {pages} {257203} (\bibinfo {year} {2019})}\BibitemShut {NoStop}%
\bibitem [{\citenamefont {Stern}\ and\ \citenamefont {Fu}(2023)}]{stern2023transport}%
  \BibitemOpen
  \bibfield  {author} {\bibinfo {author} {\bibfnamefont {A.}~\bibnamefont {Stern}}\ and\ \bibinfo {author} {\bibfnamefont {L.}~\bibnamefont {Fu}},\ }\bibfield  {title} {\bibinfo {title} {Transport properties of a half-filled chern band at the electron and composite fermion phases},\ }\href@noop {} {\bibfield  {journal} {\bibinfo  {journal} {arXiv preprint arXiv:2311.16761}\ } (\bibinfo {year} {2023})}\BibitemShut {NoStop}%
\bibitem [{\citenamefont {Fang}\ \emph {et~al.}(2012)\citenamefont {Fang}, \citenamefont {Gilbert},\ and\ \citenamefont {Bernevig}}]{fang2012bulk}%
  \BibitemOpen
  \bibfield  {author} {\bibinfo {author} {\bibfnamefont {C.}~\bibnamefont {Fang}}, \bibinfo {author} {\bibfnamefont {M.~J.}\ \bibnamefont {Gilbert}},\ and\ \bibinfo {author} {\bibfnamefont {B.~A.}\ \bibnamefont {Bernevig}},\ }\bibfield  {title} {\bibinfo {title} {Bulk topological invariants in noninteracting point group symmetric insulators},\ }\href@noop {} {\bibfield  {journal} {\bibinfo  {journal} {Physical Review B}\ }\textbf {\bibinfo {volume} {86}},\ \bibinfo {pages} {115112} (\bibinfo {year} {2012})}\BibitemShut {NoStop}%
\bibitem [{\citenamefont {Wang}\ \emph {et~al.}(2021{\natexlab{b}})\citenamefont {Wang}, \citenamefont {Yuan},\ and\ \citenamefont {Fu}}]{wang2021moire}%
  \BibitemOpen
  \bibfield  {author} {\bibinfo {author} {\bibfnamefont {T.}~\bibnamefont {Wang}}, \bibinfo {author} {\bibfnamefont {N.~F.}\ \bibnamefont {Yuan}},\ and\ \bibinfo {author} {\bibfnamefont {L.}~\bibnamefont {Fu}},\ }\bibfield  {title} {\bibinfo {title} {Moir{\'e} surface states and enhanced superconductivity in topological insulators},\ }\href@noop {} {\bibfield  {journal} {\bibinfo  {journal} {Physical Review X}\ }\textbf {\bibinfo {volume} {11}},\ \bibinfo {pages} {021024} (\bibinfo {year} {2021}{\natexlab{b}})}\BibitemShut {NoStop}%
\bibitem [{\citenamefont {Cano}\ \emph {et~al.}(2021)\citenamefont {Cano}, \citenamefont {Fang}, \citenamefont {Pixley},\ and\ \citenamefont {Wilson}}]{cano2021moire}%
  \BibitemOpen
  \bibfield  {author} {\bibinfo {author} {\bibfnamefont {J.}~\bibnamefont {Cano}}, \bibinfo {author} {\bibfnamefont {S.}~\bibnamefont {Fang}}, \bibinfo {author} {\bibfnamefont {J.}~\bibnamefont {Pixley}},\ and\ \bibinfo {author} {\bibfnamefont {J.~H.}\ \bibnamefont {Wilson}},\ }\bibfield  {title} {\bibinfo {title} {Moir{\'e} superlattice on the surface of a topological insulator},\ }\href@noop {} {\bibfield  {journal} {\bibinfo  {journal} {Physical Review B}\ }\textbf {\bibinfo {volume} {103}},\ \bibinfo {pages} {155157} (\bibinfo {year} {2021})}\BibitemShut {NoStop}%
\bibitem [{\citenamefont {Lu}\ \emph {et~al.}(2010)\citenamefont {Lu}, \citenamefont {Shan}, \citenamefont {Yao}, \citenamefont {Niu},\ and\ \citenamefont {Shen}}]{lu2010massive}%
  \BibitemOpen
  \bibfield  {author} {\bibinfo {author} {\bibfnamefont {H.-Z.}\ \bibnamefont {Lu}}, \bibinfo {author} {\bibfnamefont {W.-Y.}\ \bibnamefont {Shan}}, \bibinfo {author} {\bibfnamefont {W.}~\bibnamefont {Yao}}, \bibinfo {author} {\bibfnamefont {Q.}~\bibnamefont {Niu}},\ and\ \bibinfo {author} {\bibfnamefont {S.-Q.}\ \bibnamefont {Shen}},\ }\bibfield  {title} {\bibinfo {title} {Massive dirac fermions and spin physics in an ultrathin film of topological insulator},\ }\href@noop {} {\bibfield  {journal} {\bibinfo  {journal} {Physical review B}\ }\textbf {\bibinfo {volume} {81}},\ \bibinfo {pages} {115407} (\bibinfo {year} {2010})}\BibitemShut {NoStop}%
\bibitem [{\citenamefont {Liu}\ \emph {et~al.}(2010)\citenamefont {Liu}, \citenamefont {Qi}, \citenamefont {Zhang}, \citenamefont {Dai}, \citenamefont {Fang},\ and\ \citenamefont {Zhang}}]{liu2010model}%
  \BibitemOpen
  \bibfield  {author} {\bibinfo {author} {\bibfnamefont {C.-X.}\ \bibnamefont {Liu}}, \bibinfo {author} {\bibfnamefont {X.-L.}\ \bibnamefont {Qi}}, \bibinfo {author} {\bibfnamefont {H.}~\bibnamefont {Zhang}}, \bibinfo {author} {\bibfnamefont {X.}~\bibnamefont {Dai}}, \bibinfo {author} {\bibfnamefont {Z.}~\bibnamefont {Fang}},\ and\ \bibinfo {author} {\bibfnamefont {S.-C.}\ \bibnamefont {Zhang}},\ }\bibfield  {title} {\bibinfo {title} {Model hamiltonian for topological insulators},\ }\href@noop {} {\bibfield  {journal} {\bibinfo  {journal} {Physical Review B}\ }\textbf {\bibinfo {volume} {82}},\ \bibinfo {pages} {045122} (\bibinfo {year} {2010})}\BibitemShut {NoStop}%
\bibitem [{\citenamefont {Cano}\ \emph {et~al.}(2017{\natexlab{b}})\citenamefont {Cano}, \citenamefont {Bradlyn}, \citenamefont {Wang}, \citenamefont {Hirschberger}, \citenamefont {Ong},\ and\ \citenamefont {Bernevig}}]{Cano_2017}%
  \BibitemOpen
  \bibfield  {author} {\bibinfo {author} {\bibfnamefont {J.}~\bibnamefont {Cano}}, \bibinfo {author} {\bibfnamefont {B.}~\bibnamefont {Bradlyn}}, \bibinfo {author} {\bibfnamefont {Z.}~\bibnamefont {Wang}}, \bibinfo {author} {\bibfnamefont {M.}~\bibnamefont {Hirschberger}}, \bibinfo {author} {\bibfnamefont {N.~P.}\ \bibnamefont {Ong}},\ and\ \bibinfo {author} {\bibfnamefont {B.~A.}\ \bibnamefont {Bernevig}},\ }\bibfield  {title} {\bibinfo {title} {Chiral anomaly factory: Creating weyl fermions with a magnetic field},\ }\bibfield  {journal} {\bibinfo  {journal} {Physical Review B}\ }\textbf {\bibinfo {volume} {95}},\ \href {https://doi.org/10.1103/physrevb.95.161306} {10.1103/physrevb.95.161306} (\bibinfo {year} {2017}{\natexlab{b}})\BibitemShut {NoStop}%
\bibitem [{Note1()}]{Note1}%
  \BibitemOpen
  \bibinfo {note} {Assume it is $(c_1k_x+c_2k_y)\sigma _zs_x+(c_3k_x+c_4k_y)\sigma _zs_y$, then rotational symmetry requires $c_1=c_4$, and $c_2=c_3=0$}\BibitemShut {NoStop}%
\end{thebibliography}%



\newpage
\clearpage
\setcounter{section}{0}
\setcounter{figure}{0}
\let\oldthefigure\thefigure
\renewcommand{\thefigure}{S\oldthefigure}

\setcounter{table}{0}
\renewcommand{\thetable}{S\arabic{table}}

\renewcommand{\thesection}{S\arabic{section}}
\renewcommand{\thesubsection}{\thesection.\arabic{subsection}}
\renewcommand{\thesubsubsection}{\thesubsection.\arabic{subsubsection}}

\onecolumngrid
\begin{appendix}

\section{More on magic near parent band inversion}
We measure all the quantities in units of $E_D$ and $k_D$ and supress these two symbols in this section.
\subsection{Real space charge density migration}
\begin{figure}
\begin{center}
\includegraphics[width=0.8\textwidth]{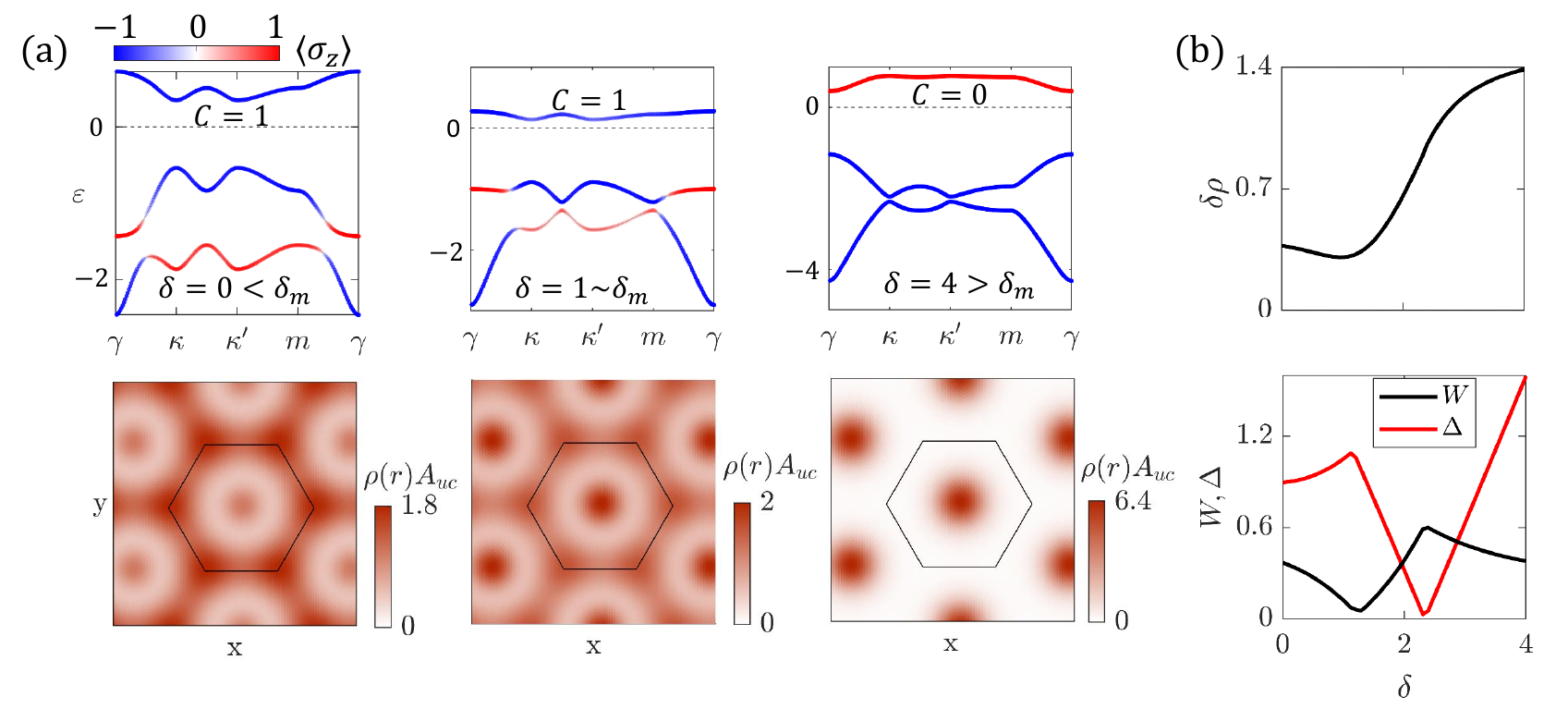}
\end{center}
\caption{
 Evolution of the real-space charge density. \textbf{(a)} Band structure (top) and real space charge density of the first conduction miniband (bottom) as $\delta$ is tuned. 
    The color of the bands indicates the $\langle\sigma_z\rangle$ expectation value.
    The Chern numbers of the first conduction miniband is labeled. We showed three scenarios: $\delta=0<\delta_m$ (deep-inverted), $\delta=1.0\sim\delta_m$ (the magic value),$\delta=4.0>\delta_m$ ( before inversion). 
    \textbf{(b)}  The top panel is the $\delta \rho$ as $\delta$ is tuned through the magic value. The bottom panel is the corresponding direct gap $\Delta$ and bandwidth $W$. The parameter used is $V_0=1$ and $g=2$, $\phi=\pi$(The same as the main text figures. So the bottom panel is a section of the main text figures.).
      }
\label{SPfig:realdensity}
\end{figure}

As we argued in the main text, at large $\delta$ with $\phi=60^{\circ}$, the wave function of first conduction miniband will be localized on the triangular minimum of the potential. At deep inversion $\delta<\delta_m$, the first conduction miniband will be localized on the honeycomb maximum of the potential. The interpolation between these two extreme scenario shows a magic point where the real space charge density is most uniform. We define a metric for this non-uniformality
\begin{equation}
(\delta \rho)^2=A_{uc}\int_{unit cell} d\bm{r}(\rho(\bm{r})-\bar{\rho})^2
\end{equation}
As plotted in Fig.\ref{SPfig:realdensity}, indeed we observe such interpolation and the corresponding minimum of $\delta \rho$ occur at approximately at $\delta_m$.

\subsection{Quantum geometry at $g=2$ and $g=3$}
\begin{figure}
    \centering
    \includegraphics[width=0.6\linewidth]{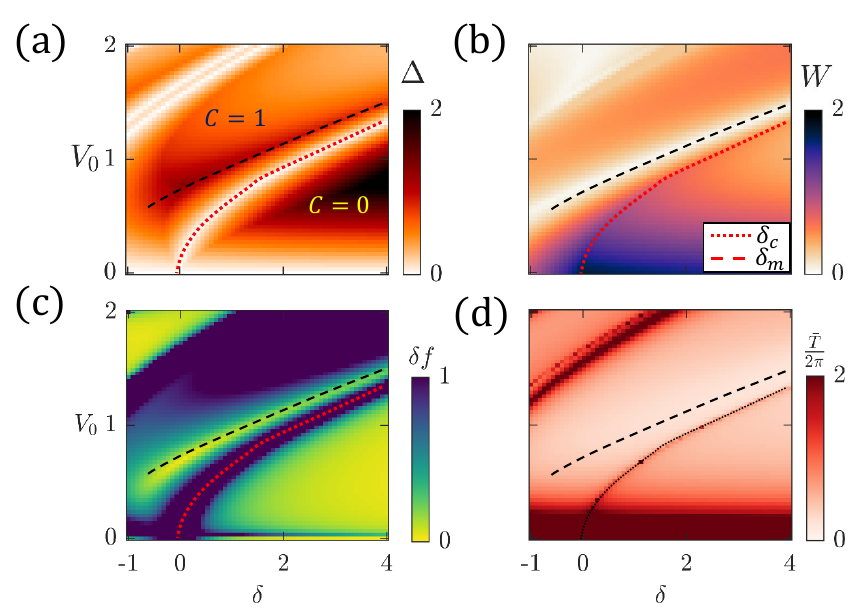}
    \caption{ 
    Energetic and geometric properties of the flat band at $g=2$.
    The \textbf{(a)} direct gap $\Delta$, \textbf{(b)} bandwidth $W$, \textbf{(c)} Berry curvature standard deviation $\delta f$, and \textbf{(d)} trace-condition violation $\bar{T}$ of the first conduction miniband as a function of potential strength $V_0$ and inversion depth $\delta$ of the band inversion model  introduced in the main text in a triangular lattice potential with $g=2$, $\phi=\pi$. 
    The miniband inversion $\delta_c$ and the magic $\delta_m$ are shown as dotted and dashed lines. 
    }
    \label{SPfig:g2phase}
\end{figure}

\begin{figure}[h]
    \centering
    \includegraphics[width=0.6\linewidth]{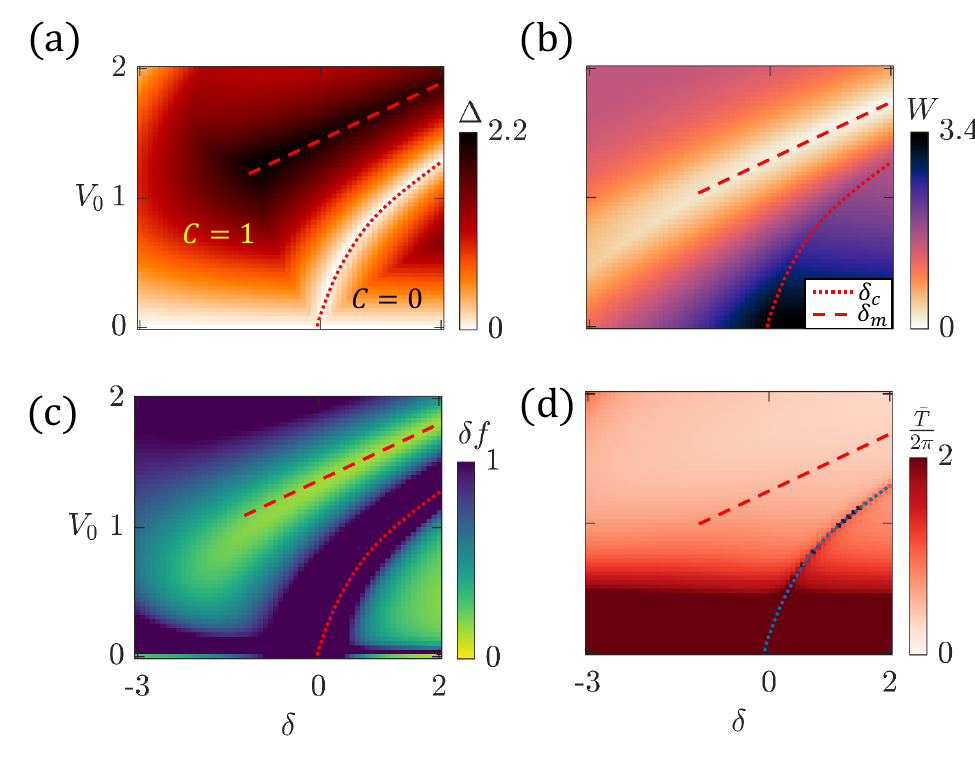}
    \caption{
    Same as Fig~\ref{SPfig:g2phase} but with $g=3$.
    }
    \label{SPfig:g3phase}
\end{figure}

We mentioned that the mechanism for generating magic bands is naturally associated with the spreading out of the berry curvature. We showed only the bandwidth and the gap in the main text.
Here, we further show two quantum geometric indicators, the Berry curvature (BC) non-uniformity $\delta f$ and trace condition violation $\overline{T}$, defined
\begin{equation}
(\delta f)^2 = \frac{A_{\mathrm{mBZ}}}{4\pi^2}\int (f(k)-\overline{f})^2 d\bm{k}
\end{equation}
\begin{equation}
\overline{T}=\int \left(\Tr[g^{FS}_{\mu\nu}(k)]-|f(k)|\right)d\bm{k}
\end{equation}
They are shown as a function of $\delta$ and $V_0$ for $g=2$ in Fig~\ref{SPfig:g2phase}, and for $g=3$ in Fig~\ref{SPfig:g3phase}.
Both $\delta f$ and $\overline{T}$ are very small near the magic line at which bandwidth is minimized, indicating favorable conditions for fractional states.
Fig~\ref{SPfig:g3phase} also serves to demonstrate that the precise choice of $g=2$ to show in the main text is not a fine-tuned choice.

\subsection{Magic at $\phi=\pi/6$}

In the main text, we have discussed the case with a perfect triangular potential $\phi=\pi $ (and also honeycomb by particle-hole transformation.). 
Here, we consider a $\phi=\pi/6$ potential, which is easily achievable with a control layer made of twisted hBN \cite{zhao2021universal,kim2023electrostatic}. 
This potential has one potential maximum and one minimum per unit cell, which forms two offset triangular lattices.
The origin of this potential arises from 
the spatially periodic out-of-plane electric polarization in twisted hBN which points downward at the AB regions, and upward at the BA regions.
This electric polarization leads to an electrostatic potential felt by a nearby layer with both attractive and repulsive potential sites.

With this kind of potential, at large $\delta$, i.e. uninverted, the first conduction/valence will have their minimum/maximum point at $\gamma$. Both bands will have $C=0$ as a result of their $C_3$ eigenvalue at the high symmetry points of mBZ. Notice also that these two bands only have the same $C_3$ eigenvalue at one of the valleys (say $\kappa$).\par
\begin{figure}[h]
\begin{center}
\includegraphics[width=0.8\textwidth]{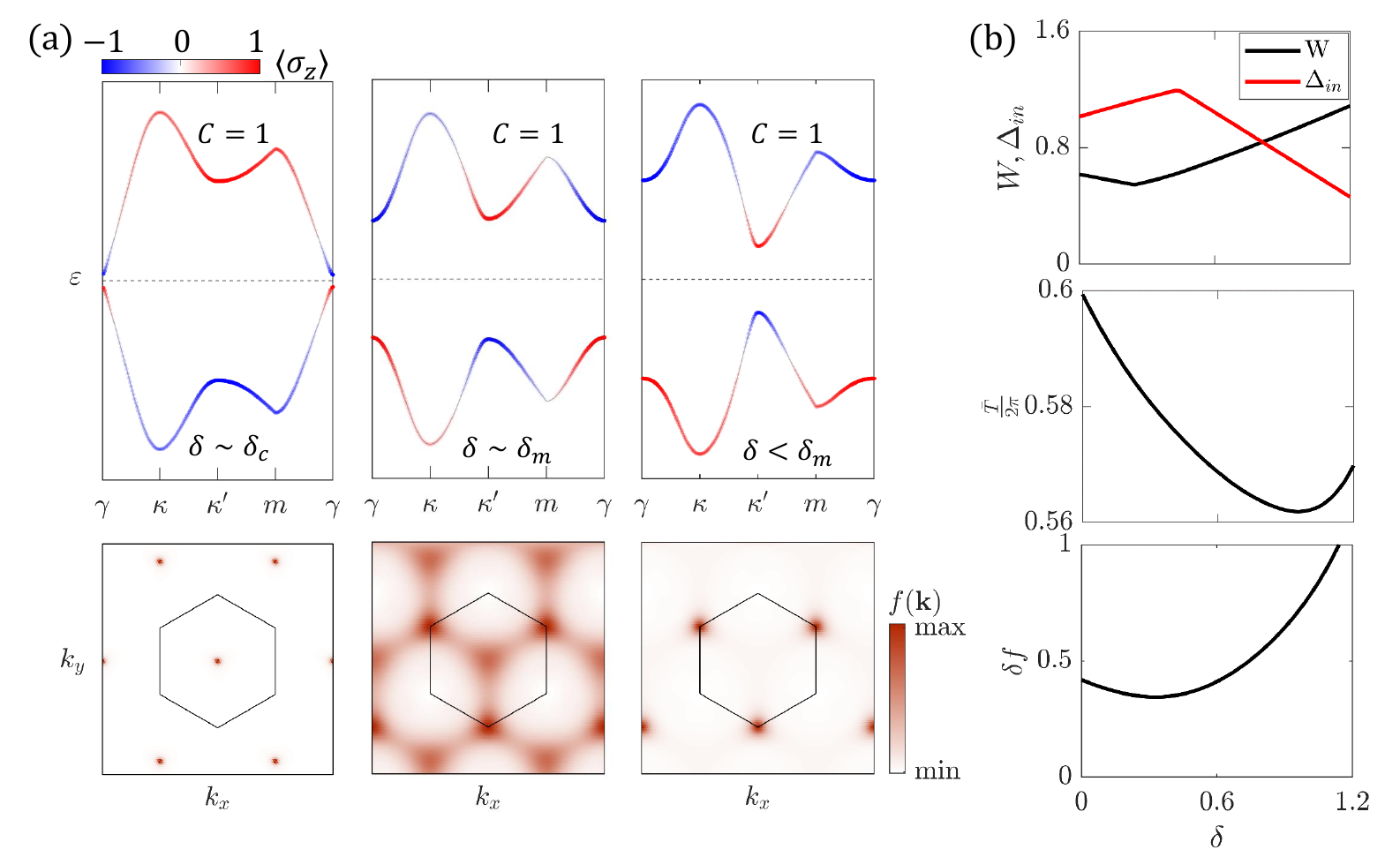}
\end{center}
\caption{
 Tuning to the flat band with a $\phi=\pi/6$ potential.
 \textbf{(a)} Schematic figure illustrating the miniband inversion as $\delta$ is tuned.
    The color indicates the $\langle\sigma_z\rangle$ expectation value.
    The Chern numbers of the first conduction miniband is labeled. We showed three scenarios: $\delta\sim\delta_c$ (right after inversion), $\delta=\delta_m$ (the magic value),$\delta<\delta_m$ (below the magic value). The bottom panel is the corresponding BC distribution.
    \textbf{(b)} Bandwidth $W$, indirect gap $\Delta_{in}$, trace condition violation $\bar{T}$, BC non-uniformality $\delta f$ as a function of $\delta$. The parameter used is $V_0=1$, $g=2.5$, $\phi=\frac{\pi}{6}$.
      }
\label{SPfig:semimagic}
\end{figure}

As we slowly decrease $\delta$, these two bands will first invert at $\gamma$ (first column of Fig. \ref{SPfig:semimagic} (a)), resulting in the miniband having a concentrated BC at $\gamma$. As we further decrease $\delta$, the two bands also get inverted at $\kappa$ and $\kappa'$. At $\kappa'$, due to the difference in $C_3$ eigenvalue, the first conduction band will then have a concentrated BC around $\kappa'$ (third column of Fig. \ref{SPfig:semimagic} (a)). The interpolation between these two gives us the optimal point where the various band quantum geometry quantities are optimized (second column of Fig. \ref{SPfig:semimagic} (a)). 

These quantities are summarized in Fig~\ref{SPfig:semimagic}(b).
While there is still an optimal point, we find that the bandwidth-to-bandgap ratio, $\delta f$, and $\overline{T}$ are not as ideal as for the perfect triangular lattice potential.

\section{Symmetry Analysis}
We claim in the main text that the starting point of our Hamiltonian is the most general one consistent with all the symmetry constraint. We prove the claim in this section. We denote the Pauli matrices in the spin subspace by $s$ and the Pauli matrices in the orbital subspace by $\sigma$.

The symmetry constraint of the system is parity symmetry $P$, time reversal symmetry $T^2=-1$, and rotational symmetry $R$. 
Although we consider full continuous rotation symmetry, discrete rotation symmetry $C_n$ with $n>2$ is sufficient in restricting the form of the effective $k\cdot p$ theory up to $O(k^3)$.
We take their action in the spin/orbital subspace to be the following
\begin{equation}
P=\sigma_x\otimes I\quad T= I\otimes is_y K \quad R_\theta=I \otimes \exp(i\frac{\theta}{2}s_z)
\end{equation}
In addition, $P$ sends $k\to -k$, $T$ sends $k$ to $-k$, and $R_\theta$ sends $\vec{k}\rightarrow \mathbf{R}_\theta \vec{k}$ when acting on Bloch states, where $\mathbf{R}_\theta$ is the rotation matrix by angle $\theta$.
By requiring $PT$ to be a symmetry of the Hamiltonian, among the 16 possible choices $\sigma_{\alpha}\otimes s_{\beta}$ $\alpha,\beta=0,1,2,3$, only the following are permissible (commutes with $PT$)
\begin{equation}
\begin{split}
\sigma_z \otimes s_{\beta=1,2,3} &\quad p=-1\quad t=-1\\
\sigma_x \otimes I &\quad p=1\quad t=1\\
\sigma_y \otimes I &\quad p=-1\quad t=-1\\
I\otimes I &\quad p=1\quad t=1
\end{split}
\end{equation}
The $p,t=\pm 1$ after each lines indicates whether that operator commutes with $P$ $(p=1)$ or anticommutes with $P$ $(p=-1)$. $I\otimes I$ is trivial, so we will ignore it from now on. Then when doing $k\cdot p$ expansion of the Hamiltonian, the most general form is the following (ignoring the constant term, and up to $k^2$ term)
\begin{equation}
H(k)=H_x k_x+H_y k_y+H_{xx}k_x^2+H_{yy}k_y^2+H_{xy}k_xk_y
\end{equation}

\begin{equation}
H=\sum_k (a_{k\uparrow}^{\dagger},b_{k \uparrow}^{\dagger},a_{k\downarrow}^{\dagger},b_{k \downarrow}^{\dagger})H(k)(a_{k\uparrow},b_{k \uparrow},a_{k\downarrow},b_{k \downarrow})^{T}
\end{equation}

Given that $P$ sends the momentum of the operator from $k$ to $-k$, then $H_x$ and $H_y$ can only be expressed in terms of those matrices with $p=-1$. Similarly, $H_{xx}$,$H_{yy}$, $H_{xy}$ can only be expressed in terms of those matrices with $p=1$. Without further invoking rotation symmetry, this is as far as we can go for constructing the Hamiltonian.

Let us first work with the $k^2$ order terms. We have argued that the Hamiltonian must of the following form
\begin{equation}
H_2 (k)=(C+Dk_x^2+E k_y^2+ Fk_xk_y)\sigma_x\otimes I
\end{equation}
Since $R$ acts trivially in the orbital subspace, and $\sigma_x \otimes I$ is trivial in the spin subspace, rotational symmetry requires 
\begin{equation}
Dk_x^2+E k_y^2+ Fk_xk_y=Dq_x^2+E q_y^2+ Fq_xq_y
\end{equation}
where $\vec{q}=\mathbf{R}_{\theta}\vec{k}$.
This then tells us that $D=E$, $F=0$. (From the proof here, it is clear that continuous rotational symmetry is not necessary, any discrete rotational symmetry $C_n$ for $n\geq 3$ is sufficient)

We then proceed to the linear terms. Here $H_x, H_y$ can only be written in terms of linear combinations of $\sigma_z \otimes s_\beta=1,2,3$ and $\sigma_y \otimes I$. Notice that $\sigma_y \otimes I$ commutes with rotation operator $R$. Then rotational symmetry forbids the presence of $\sigma_y \otimes I$. By forbidding, we mean the following. Assume that we write the following Hamiltonian
\begin{equation}
H_{assume}=\sum_{k}  (a_{k\uparrow}^{\dagger},b_{k \uparrow}^{\dagger},a_{k\downarrow}^{\dagger},b_{k \downarrow}^{\dagger})(c_1 k_x+c_2  k_y) \sigma_y\otimes I (a_{k\uparrow}^{\dagger},b_{k \uparrow}^{\dagger},a_{k\downarrow}^{\dagger},b_{k \downarrow}^{\dagger})^{\dagger}
\end{equation}
By requiring $RH_{assume}R^{-1}=H_{assume}$, we arrive at the conclusion $c_1=c_2=0$. By the same reasoning $\sigma_z \otimes s_z$ is also forbidden.

Then we are only left with the possibility of $\sigma_z\otimes s_{x,y}$. By essentially the same reasoning we had when dealing with $H_{assume}$, we find that the linear term must be of the following form \footnote{Assume it is $(c_1k_x+c_2k_y)\sigma_zs_x+(c_3k_x+c_4k_y)\sigma_zs_y$, then rotational symmetry requires $c_1=c_4$, and $c_2=c_3=0$}
\begin{equation}
H_1(k)=(k_x\sigma_zs_x+k_y\sigma_z s_y)
\end{equation}
Then putting together our Hamiltonian $H_1(k)$ and $H_2(k)$, we find that the most general Hamiltonian consistent with all the symmetry is the following
\begin{equation}
H(k)=A(k_x\sigma_z\otimes s_x+k_y\sigma_z \otimes s_y)+(C+Dk^2)\sigma_x\otimes I
\end{equation}

Let us define a new set of Pauli matrices $\tau$ and $\mu$ in the following manner
\begin{equation}
\tau_z=\sigma_x s_z\quad \mu_z=\sigma_x \quad \mu_x=\sigma_ys_y \quad \mu_y=\sigma_z s_y
\end{equation}
Notice that $\mu$ anticommutes within themselves and all three $\mu$ commutes with $\tau$. In terms of the newly defined Pauli matrices, the Hamiltonian can be written as
\begin{equation}
H=A(\tau_z\mu_xk_x+k_y\mu_y)+(C+Dk^2)\mu_z
\end{equation}
This is exactly the form of the Hamiltonian we presented in the main text. As a matter of fact, there are lots of degrees of freedom when defining the $\mu$ and $\tau$. For example, we could have also defined it in the following manner
\begin{equation}
\mu_x=\sigma_z \otimes s_x \quad \mu_y=-\sigma_y\otimes s_x \quad \mu_z=\sigma_x \otimes I\quad \tau_z=\sigma_x\otimes s_z
\end{equation}
Then $H(k)$ will be slightly different in terms of $\tau$ and $\mu$, but the physics is invariant.
Finally, a term proportional to $(k_x^2+k_y^2)I\otimes I$ is always allowed, which will lead to a particle-hole mass asymmetry ($\alpha_1\neq\alpha_2$ in the main text).

\section{Many Body Calculation Extended data}

Here, we study many-body physics in the first valence band of Cd$_3$As$_2$ and expand on the results presented in Fig. 4 of the main text.

\begin{figure*}
\begin{center}
\includegraphics[width=0.8\textwidth]{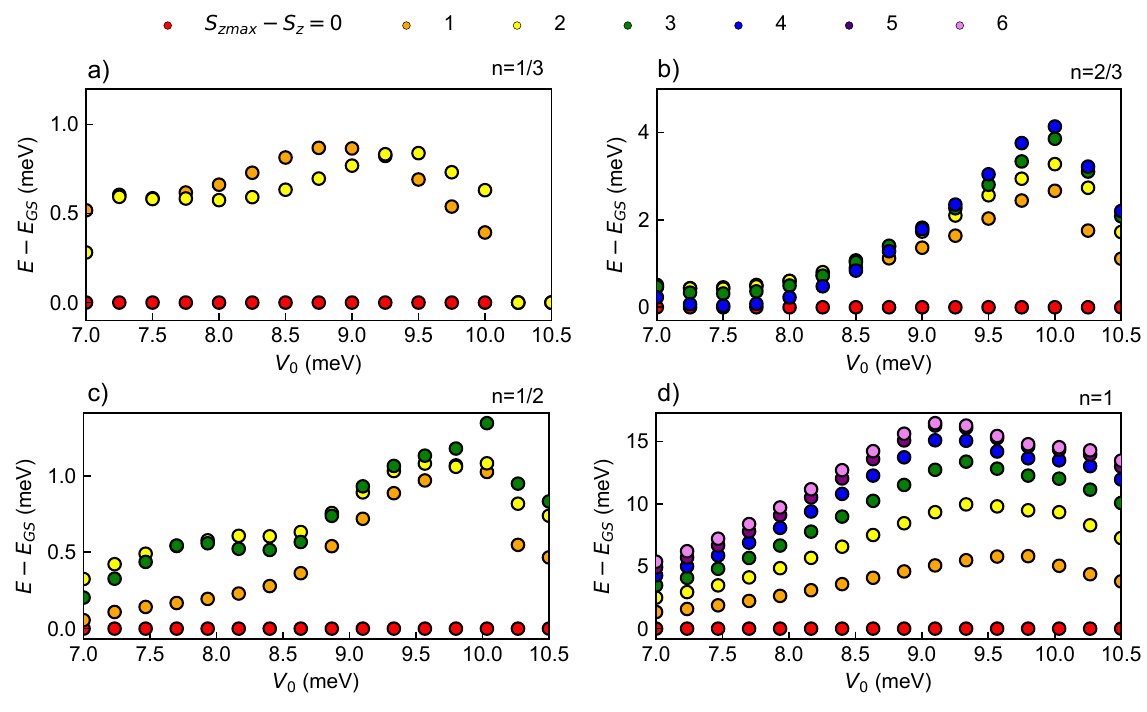}
\end{center}
\caption{
Magnetism of Cd$_3$As$_2$ from numerical diagonalization.
Lowest energy level within each $S_z$ eigensector as a function of superlattice potential strength $V_0$ at $n=\frac{1}{3}$ \textbf{(a)}, $\frac{2}{3}$ \textbf{(b)}, and $\frac{1}{2}$ \textbf{(c)}, and $1$ \textbf{(d)} hole filling of the first Cd$_3$As$_2$ valence band, demonstrating full $S_z$ polarization within a range of potential strengths . The finite system used contains 12 superlattice unit cells (same as defined in the Supplementary Material of Ref. \cite{reddy2023fractional}). Physical parameters are same as in Fig. 4 of the main text.
      }
\label{SPfig:magnetismED}
\end{figure*}

\begin{figure*}
\begin{center}
\includegraphics[width=0.5\textwidth]{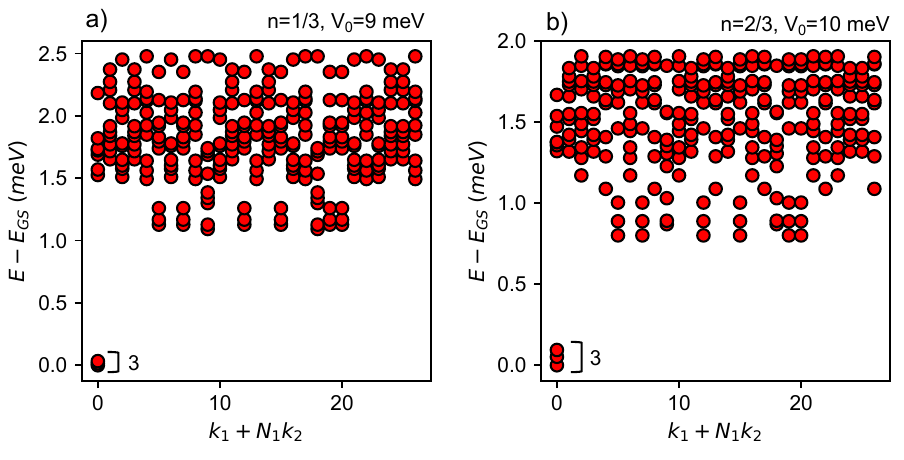}
\end{center}
\caption{ Evidence of FCIs in Cd$_3$As$_2$.  Many-body spectra within maximum $S_z$ sector at and $n=\frac{1}{3}$ \textbf{(a)} and $n=\frac{2}{3}$ \textbf{(b)} hole filling of the first Cd$_3$As$_2$ valence miniband, demonstrating gapped, three-fold quasidegenerate FCI ground states. All model parameters are the same as stated in Fig. (4) of the main text. The finite system used contains 27 superlattice unit cells (same as defined in the Supplementary Material of Ref. \cite{reddy2023fractional})
      }
\label{SPfig:oneTwoThirdsED}
\end{figure*}

In Fig. \ref{SPfig:magnetismED}, we show that full $S_z$ polarization occurs at all filling factors studied in Fig. (4) of the main text (Fig. \ref{SPfig:magnetismED}). Also, we show many-body spectra at $n=\frac{1}{3}$, $\frac{2}{3}$ within the fully spin-polarized sector to confirm the ground states' FCI nature (Fig. \ref{SPfig:oneTwoThirdsED}).

\begin{figure*}
\begin{center}
\includegraphics[width=0.5\textwidth]{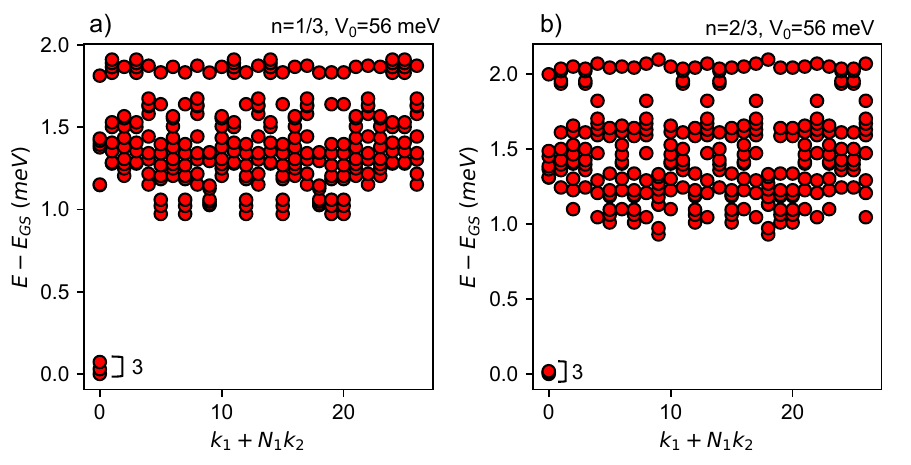}
\end{center}
\caption{Evidence of FCIs in Sb$_2$Te$_3$. Analogous to Fig. \ref{SPfig:oneTwoThirdsED} except for the first conduction miniband of Sb$_2$Te$_3$ with layer thickness $L=2.5$nm. Parameters entering Eq. 1 of the main text are $\delta=200.76$meV, $v=337.86$meV nm, $\alpha_1=374.6$meV nm$^2$, and $\alpha_2=567.0$meV nm$^2$. The superlattice potential parameters are $V_0=56$meV, $\phi=\pi$, $a=21$nm. $\epsilon=10$.}
\label{SPfig:SbTeOneTwoThirdsED}
\end{figure*}

In Fig. ~\ref{SPfig:oneTwoThirdsED}, we show example many-body spectra at $n=\frac{1}{3}$, $\frac{2}{3}$ to demonstrate the appearance of fractional Chern insulator ground states. We note that fact that the center-of-mass momentum quantum numbers ($k_1+N_1k_2$) of the ground states are all the same is inconsistent spontaneous translation symmetry breaking, ruling out the possibility of spontaneous translation symmetry breaking. In Fig. \ref{SPfig:SbTeOneTwoThirdsED}, we show similar evidence for FCIs in the first conduction miniband of Sb$_2$Te$_3$ (parameters derived in Sec~\ref{sec:RealMaterials}).

\begin{figure*}
\begin{center}
\includegraphics[width=0.5\textwidth]{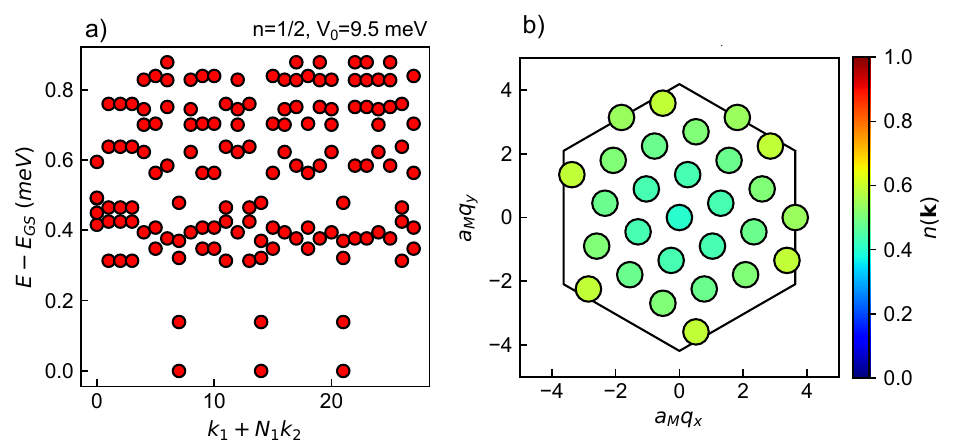}
\end{center}
\caption{
Evidence of anomalous composite Fermi liquid in Cd$_3$As$_2$.
\textbf{(a)} Many-body spectrum within maximum $S_z$ sector at and $n=\frac{1}{2}$. \textbf{(b)} Corresponding momentum distribution function $n(\bm{k})$. Physical parameters are the same as in Fig. 4 of the main text.}
\label{SPfig:CFL}
\end{figure*}

In Fig. ~\ref{SPfig:CFL}, we provide evidence for an anomalous composite Fermi liquid at $n=\frac{1}{2}$. Fig. ~\ref{SPfig:CFL} (a) shows that in each of the three momentum sectors $k_1 + N_1 k_2 = 7,14,21$ (which are the $m$ points of the mini Brillouin zone), there are two low-lying quasi-degenerate ``ground states". These ground states' center-of-mass momenta obey a ``compact composite Fermi sea" rule established in previous numerical studies of composite Fermi liquids~\cite{rezayi1994fermi, geraedts2018berry,fremling2018trial,wang2019lattice,wang2019dirac,wang2021exact} and are identical to those obtained through numerical diagonalization of the lowest Landau level on an identical system geometry~\cite{reddy2023toward}.
Fig. ~\ref{SPfig:CFL}(b) shows the corresponding momentum distribution function ($n(\bm{k})=\frac{1}{N_{GS}}\sum_{i}\bra{\Psi_i}c^{\dag}_{\bm{k}}c_{\bm{k}}\ket{\Psi_i}$ where $i$ runs over the $N_{GS}$ ground states), which lacks the Fermi surface discontinuity characteristic of  Landau Fermi liquid. This numerical evidence indicates an anomalous (zero-field) composite Fermi liquid state at $n=\frac{1}{2}$~\cite{goldman2023zero,dong2023composite,stern2023transport}.

For a detailed discussion of our methods, we refer the reader to the Supplementary Material of Ref. ~\cite{reddy2023fractional}.

 \section{Massive Dirac model}

\begin{figure}
    \centering
    \includegraphics[width=\textwidth]{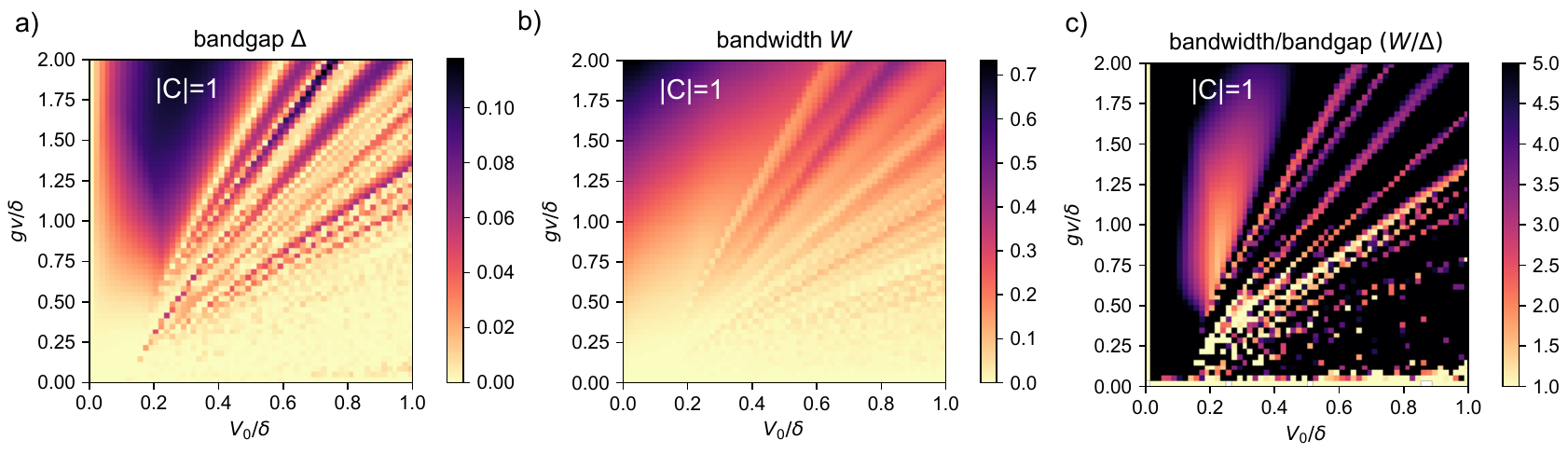}
    \caption{
    Bandgap and bandwidth of the massive Dirac fermion model in a periodic potential.
    (a) Minimum bandgap isolating the lowest conduction band, (b) lowest conduction bandwidth, and (c) their ratio as a function of the two dimensionless parameters characterizing the massive Dirac model given fixed $\phi=0$.}
    \label{SPFig:massiveDiracPhaseDiagram}
\end{figure}

Here, we discuss on topological minibands in the massive Dirac model with a triangular Bravais lattice periodic scalar potential,
\begin{align}
    \begin{split}
        H^{\tau}_0(\bm{k}) = v(\tau k_x\sigma_x +k_y\sigma_y) + \frac{\delta}{2}\sigma_z  + V(\bm{r})
    \end{split}
\end{align}
where
\begin{align}
    \begin{split}
        V(\bm{r})=2V_0\sum_{n=1}^{3} \cos(\bm{g}_n\cdot\bm{r}+\phi)
    \end{split}
\end{align}
as in Eq. (2) in the main text. We find that the model fails to exhibit topological minibands with small bandwidth-to-bandgap ratios. We also find that it is unlikely to exhibit integer or fractional quantized anomalous Hall effects at the filling $n\leq 1$, where $n$ is the number of carriers per superlattice unit cell. 

This model has been studied previously in Ref. \cite{Su_2022,suri2023superlattice}. It was shown that topological bands occur in the conduction band when the phase parameter $\phi \mod \frac{2\pi}{3} \sim 0$, corresponding to a potential with honeycomb minima. This result can be understood by calculating the $C_3$ eigenvalues at $C_3$-symmetric points of the mini-Brillouin zone, which fix the Chern number mod $3$ \cite{fang2012bulk}, via degenerate perturbation theory in the weak superlattice potential limit. 
This model is characterized, up to an overall energy scale, by the dimensionless parameters $gv/\delta$ and $V_0/\delta$, and $\phi$ where $g=|
\bm{g}_1|$.

While a superlattice potential with $\phi=0^{\circ}$ produces a topological conduction miniband, we find numerically that the ratio its bandwidth to the gap separating it from the next conduction miniband cannot be made small (Fig. \ref{SPFig:massiveDiracPhaseDiagram}). We interpret this result as follows. In the limit $\delta\rightarrow\infty$, the Hamiltonian reduces to a standard $\bm{p}^2$ kinetic energy Hamiltonian with a periodic potential. The lowest two minibands have honeycomb lattice character with gapless mini-Dirac points at $\kappa,\,\kappa'$. As $\delta$ decreases from $\infty$, these mini Dirac points gap out, resulting in a $|C|=1$ lowest conduction miniband.

On the other hand, the miniband spectrum of the massless ($\delta=0$) Dirac model is gapless. In particular, when $\delta=0$, there are mini-Dirac points between the lowest two conduction minibands each of the three inequivalent mini-Brillouin zone edge midpoints, $m$, $m'$, $m''$. \cite{wang2021moire, cano2021moire}. One can show that the presence of a psuedo-time reversal symmetry ($\tilde{\mathcal{T}} = -i\sigma_y\mathcal{K}$, $[\tilde{\mathcal{T}},H]=0$) when $\delta=0$ guarantees double degeneracies for a fixed $\tau$ at all crystal momenta $\bm{k}$ invariant under $\bm{k}\rightarrow-\bm{k}$ (that is, $\gamma$ and $m$ points). Here $\mathcal{K}$ is the complex conjugation operator and we emphasize the prefix ``pseudo" since we have not specified the physical meaning of $\sigma$. The size of the direct gaps at $\kappa,\,\kappa'$ and $m,\,m',\,m''$ behave oppositely with $\delta$, the bandgap resists becoming large as $\delta$ is tuned (see Fig. \ref{SPFig:massiveDiracPhaseDiagram} (a-d)).

In Fig. \ref{SPFig:massiveDiracPhaseDiagram}, we show the small direct bandgap $\Delta$, bandwidth $W$, and their ratio for the lowest conduction minibands as a function of the two dimensionless parameters $gv/\delta$ and $V_0/\delta$ with fixed $\phi=0$. While we observe a large region of parameter space in which $|C|=1$, the bandwidth-to-bandgap ratio in this regime does not become small, achieving a minimum value near $\sim 1.4$.

\begin{figure}
    \centering
    \includegraphics[width=\textwidth]{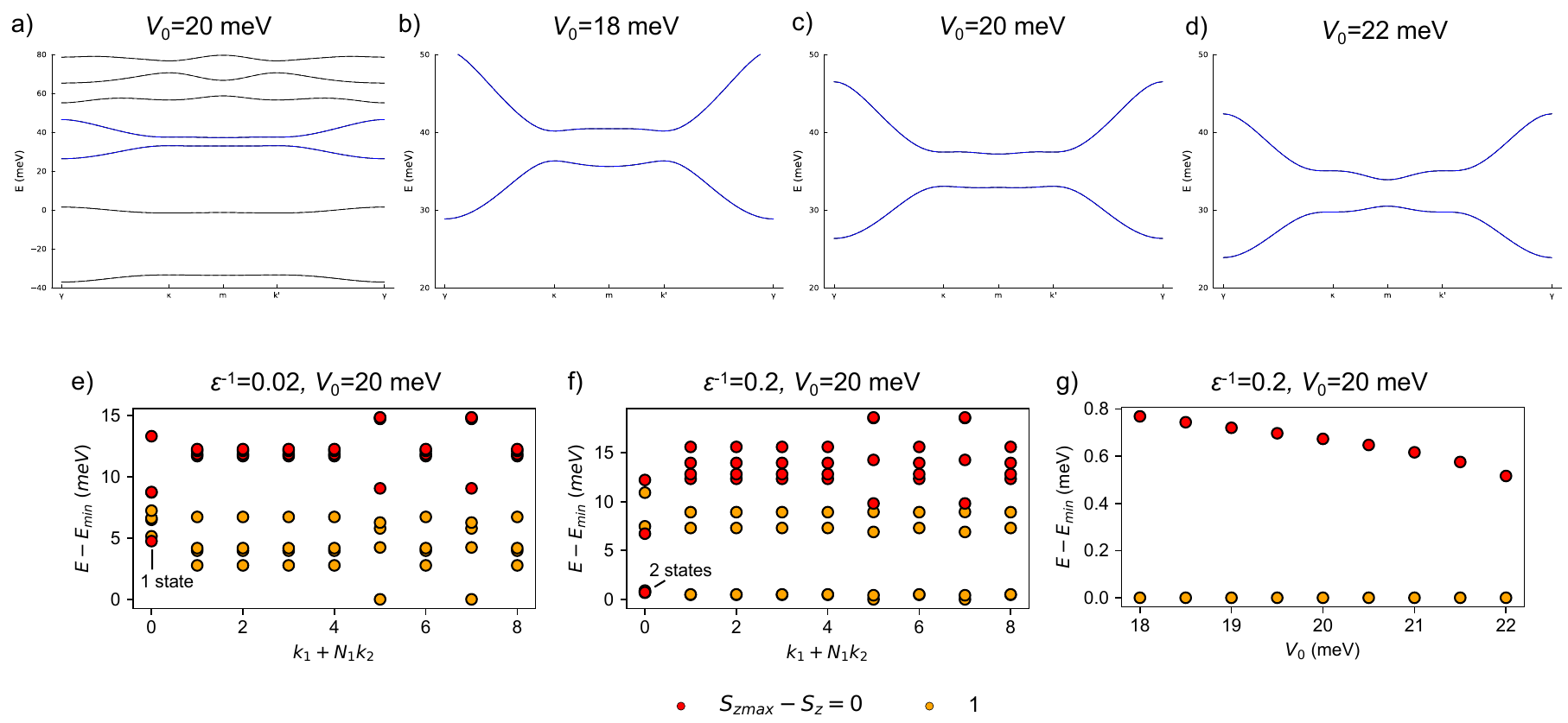}
    \caption{
    Absence of QAH in the massive Dirac fermion model from two-band numerical diagonalization.
    (a-d) Example band structures of lowest two conduction minibands corresponding to the $|C|=1$ region of Fig. \ref{SPFig:massiveDiracPhaseDiagram}(d) where the bandwidth/bandgap ratio is maximized. (e-g) Many-body calculations demonstrating absence of IQAH state (see text). Parameters: $\delta=90$ meV, $v=200$ meV nm, $a_M=20$ nm, $\phi=0^{\circ}$, corresponding to $V_0/\delta =0.22$, $gv/\delta = 0.81$.}
    \label{SPFig:massiveDiracED}
\end{figure}

We now examine the possibility of an integer quantized anomalous Hall state by performing exact diagonalization calculations in Fig~\ref{SPFig:massiveDiracED}. One might expect that, given sufficiently strong interactions, a calculation projected only to the lowest conduction miniband in each valley would indeed give a $\tau$-polarized IQAH groundstate, and this is indeed what we observe (not shown). However, the small value of the bandgap brings the reliability of such a calculation into question. We thus expand our Hilbert space to also include the second lowest conduction miniband within each $\tau$ sector and find that IQAH state never becomes the ground state. Due to Hilbert space dimension limitations, we compare only the $S_{zmax}$ and $S_{zmax}-1$ sectors (see main text). In the absence of interactions, the unique lowest energy state within the maximally $\tau$ polarized sector is the full-band Chern insulator state. However, a finite bandwidth means that a non-valley polarized Fermi liquid state is lower in energy. As we increase interactions, we observe a level crossing within the maximally $\tau$-polarized sector \emph{before} we observe full $\tau$ polarization. After the level crossing, there are two nearly degenerate low-energy states within the $S_{zmax}$ sector. These two energy states likely correspond in the thermodynamic limits to topologically trivial insulating states in which charge polarized to one sublattice of the honeycomb lattice of superlattice potential minima. Exact diagonalization calculations at fractional filling, assuming $S_{z}$ polarization and projecting only to the lowest band, also fail to exhibit evidence of fractional quantum anomalous Hall states (not shown). 

\section{Real Materials}\label{sec:RealMaterials}
In this section, we discuss the procedure for modeling real material candidates for the mechanism presented in the main text.
\subsection{From 3D bulk Hamiltonian to 2D thin film}
The starting point is the low-energy 3D Hamiltonian~\cite{lu2010massive,liu2010model}
\begin{equation}
\begin{split}
H(\bm{k})&=(C+D_1 k_z^2+D_2(k^2_x+k^2_y))+(M-B_1k_z^2-B_2(k^2_x+k^2_y))\sigma_z+A_1k_z\tau_z\sigma_x+A_2(k_x\tau_x\sigma_x+k_y\tau_y\sigma_x)
\end{split}
\end{equation}

We consider Cd$_3$As$_2$ (shown in the main text) and Sb$_2$Te$_3$ (not shown in the main text).
We use parameters for $\text{Sb}_2\text{Te}_3$ from Ref.\cite{liu2010model}, and parameters for $\text{Cd}_3\text{As}_2$ from Ref.\cite{Cano_2017}. 
For completeness, we reproduce them here.
For $\text{Cd}_3\text{As}_2$,
\begin{equation}
\begin{split}
M_1=0.0205 \text{eV} \quad A_1=0.0 \text{eV}\AA \quad A_2=0.889 \text{eV}\AA \quad B_1=18.77 \text{eV}\AA^2\\
\quad B_2=13.5 \text{eV}\AA^2 \quad C=-0.0145 \text{eV} \quad D_1=10.59 \text{eV}\AA^2 \quad D_2=11.5 \text{eV}\AA^2
\end{split}
\end{equation}
and for $\text{Sb}_2\text{Te}_3$,
\begin{equation}
\begin{split}
M_1=-0.22 \text{eV}\quad A_1=0.84 \text{eV}\AA \quad A_2=3.4 \text{eV}\AA \quad B_1=-19.64 \text{eV}\AA^2  \\
B_2=-48.51 \text{eV}\AA^2  \quad C=0.001 \text{eV} \quad D_1=-12.39 \text{eV}\AA^2 \quad D_2=-10.78 \text{eV}\AA^2
\end{split}
\end{equation}

To construct low energy Hamiltonian for the thin film, we substitute $k_z$ by $-i\partial_z$ in  $H(\bm{k})$. For film of thickness $L$, we choose basis states along the z direction as
\begin{equation}
\ket{n_z,\tau_z,\sigma_z}=\sqrt{\frac{2}{L}} \sin(\frac{n_z \pi}{L}z)
\label{eq:sinbasis}
\end{equation}
where $n_z=1,2,\dots $ are positive integers, and $L$ is the thickness of the film.
Then, the Hamiltonian can be constructed by computing the matrix elements of $H(k_x,k_y,k_z\to-i\partial_z )$ in the basis of $\ket{n_z,\tau_z,\sigma_z}$, and the eigenstates for any $(k_x,k_y)$ can be obtained. 
We notice that there is an isospin symmetry $I_1\equiv \tau_z\sigma_z (-1)^n$ such that $H(k_x,k_y,k_z\to -i\partial_z)$ does not mix between different isospin sectors. 
Note that for Cd$_3$As$_2$, there is an higher symmetry and $I_2=\tau_z\sigma_z$ is also a conserved isospin.
For the remainder of this section, we focus on the $I_1=-1$ sector for Sb$_2$Te$_3$ and $I_2=-1$ sector for Cd$_3$As$_2$ (note that this choice of $I_2$ or $I_1$ for Cd$_3$As$_2$ makes no difference here, but will be useful in Sec~\ref{sec:zdept}).

\begin{figure}
    \centering
    \includegraphics[width=0.8\textwidth]{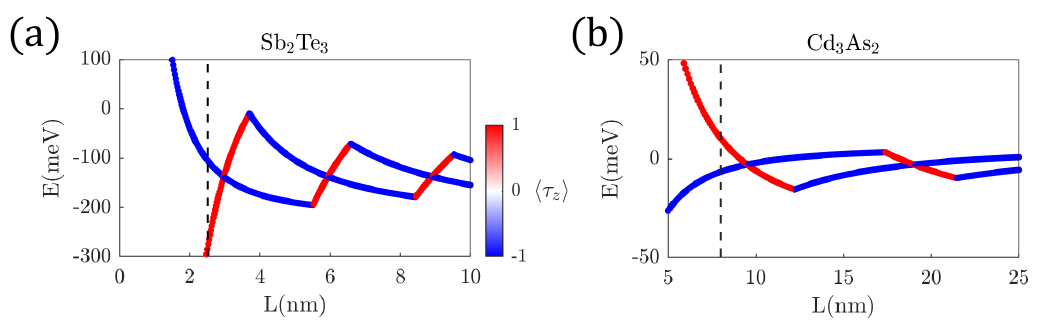}
    \caption{
    Energy and $\tau_z$ eigenvalues of the two eigenstates of $H(k_x=0,k_y=0,k_z\to-i\partial_z)$ that are closest to charge neutrality. (a) for $\text{Sb}_2\text{Te}_3$ thin film in $I_1=-1$ sector, (b) for $\text{Cd}_3\text{As}_2$ thin film in $I_2=-1$ sector. $E$ is the energy of the state, $L$ is the thickness of the thin film. The dashed line is the thickness we use for all our single particle and ED calculation. The difference in $E$ of the two states is the parent gap. }
    \label{SPfig_tauz}
\end{figure}

Fig~\ref{SPfig_tauz} shows the energy of the valence and conduction bands as a function of film thickness $L$ for these two materials.  
Since $H(0,0,-i\partial_z)$ commutes with $\tau_z$, these states can be labeled by their $\tau_z$ eigenvalue.
Based on the criterion discussed in the main text, we choose thicknesses $L=\SI{2.5}{nm}$ for Sb$_2$Te$_3$ and $L=\SI{8}{nm}$ for Cd$_3$As$_2$.
For this choice of $L$, the conduction and valence bands have opposite $\tau_z$ eigenvalue.  

To obtain the two-band model of the thin film, used in the main text, we do a further $k\cdot p$ expansion of this Hamiltonian.  
Specifically, let $\ket{1}$ and $\ket{2}$ be the two states closest to charge neutrality in the $I_1=-1$ sectors of $H(0,0,-i\partial_z)$ with energies $E_1$ and $E_2$.
In terms of these two states, we may define the following 2-by-2 matrices,
\begin{equation}
\begin{split}
(H_{0})_{i,j} & = \braket{i}{H(0,0,-i\partial_z) | j}\\
(H_{x})_{i,j} & = \braket{i}{\partial_{k_x}H(0,0,-i\partial_z) | j}\\
(H_{y})_{i,j} & =\braket{i}{\partial_{k_y}H(0,0,-i\partial_z) | j}\\
(H_{xx})_{i,j} &= \braket{i}{\partial^2_{k_x}H(0,0,-i\partial_z) | j}\\
(H_{yy})_{i,j} &= \braket{i}{\partial^2_{k_y}H(0,0,-i\partial_z) | j}
\end{split}
\end{equation}
where 
\begin{equation}
\begin{split}
H(0,0,-i\partial_z)&=C-D_1\partial_z^2+(M+B_1 \partial_z^2)\sigma_z -iA_1\partial_z \tau_z \sigma_x\\
\partial_{k_x}H(0,0,-i\partial_z)&=A_2 \tau_x \sigma_x\\
\partial_{k_y}H(0,0,-i\partial_z)&=A_2 \tau_y \sigma_x\\
\partial^2_{k_x}H(0,0,-i\partial_z)&=D_2-2B_2\sigma_z\\
\partial^2_{k_y}H(0,0,-i\partial_z)&=D_2-2B_2\sigma_z 
\end{split}
\end{equation}
Then, the two-band thin film Hamiltonian, given by a Taylor expansion near $\bm{k}=0$, is
\begin{equation}
h_{\text{thin-film}}(k_x,k_y)=H_0+H_x k_x+H_y k_y+H_{xx}\frac{k^2_x}{2}+H_{yy}\frac{k^2_y}{2}
\end{equation}
The fact that $\ket{1}$ and $\ket{2}$ have opposite $\tau_z$ eigenvalues implies that $H_{xx}$ and $H_{yy}$ will be diagonal, $H_{x}$ and $H_y$ will be off-diagonal. Thus, $h_{\text{thin-film}}$ will be exactly the form of the Hamiltonian describing topological band inversion, as used in the main text.

For the Hamiltonian in the main text Eq.1, the parameters we obtain are:
for Cd$_3$As$_2$,
\begin{equation}
\begin{split}
\delta=\SI{16.9}{meV},\quad v=\SI{88.9}{meV nm},\quad \alpha_1=\SI{250}{meV nm^2},\quad \alpha_2=\SI{20}{meV nm^2}
\end{split}
\end{equation}
and for Sb$_2$Te$_3$,
\begin{equation}
\begin{split}
\delta=\SI{200.76}{meV},\quad v=\SI{337.86}{meV nm},\quad \alpha_1=\SI{374.6}{meV nm^2},\quad \alpha_2=\SI{567.0}{meV nm^2}
\end{split}
\end{equation}


\subsection{$\text{Sb}_2\text{Te}_3$}

The properties of Cd$_3$As$_2$ are discussed at length in the main text.
Here, to show that Cd$_3$As$_2$ is not a special case, we also show results for another material candidate Sb$_2$Te$_3$.

\begin{figure}
    \centering
    \includegraphics[width=0.98\textwidth]{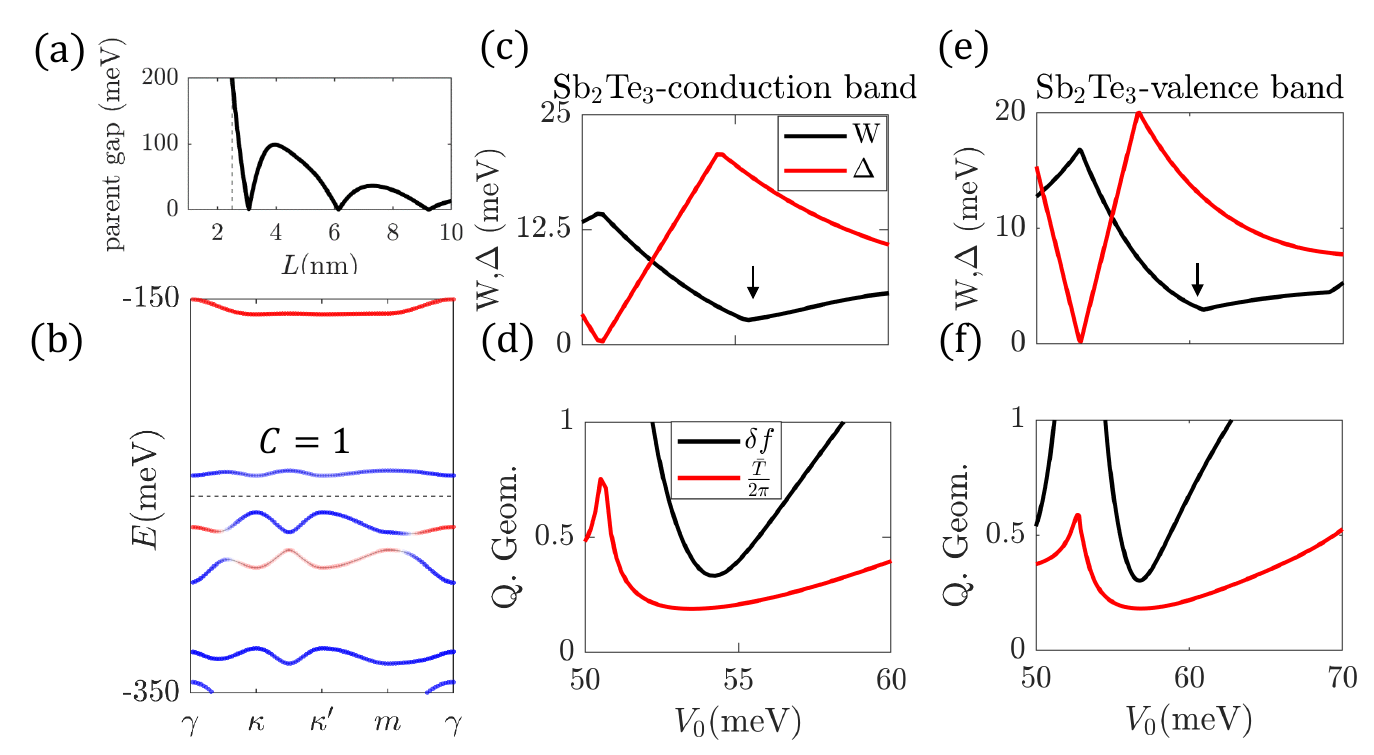}
    \caption{
    Electronic and geometric properties of Sb$_2$Te$_3$ thin film in a superlattice potential.  \textbf{(a)} Gap of the pristine $\text{Sb}_2\text{Te}_3$ thin film as a function of film thickness $L$.  The gray dashed line is at $L=\SI{2.5}{nm}$, which is the thickness we used for (b-f). \textbf{(b)} Minibands with  $a=\SI{21}{nm}$, $V_0$=55 meV,  $\phi=\pi$. 
    Color of the bands indicates orbital content of the band (cf. main text Fig. 1 and Fig. 3).  \textbf{(c,d)} Properties of the first conduction band as a function of $V_0$ with $\phi=\pi$ and $a$=21nm. \textbf{(e,f)} Properties of the first valence band as a function of  $V_0$ with $\phi=0$ and $a$=21nm. The arrow indicates the point where bandwidth/gap, and quantum geometry, is optimized to host fractionalized phases.}
    \label{SPfig_SbTe}
\end{figure}

The properties of the resulting first conduction (valence) minibands of Sb$_2$Te$_3$ at $L=\SI{2.5}{nm}$, in the presence of a triangular (honeycomb) potential, are shown in Fig.\ref{SPfig_SbTe}(c-f) for a superlattice period $a=\SI{21}{nm}$.
As expected from our results in the main text, a dip in the bandwidth and corresponding quantum geometric indicators is observed at $V_0\sim 55\text{meV}$ as we scan through $V_0$.
This indicates that the band has become most ideal for hosting fractionalized state near this point. 
The minibands at this dip are shown in Fig.\ref{SPfig_SbTe} (b) for the triangular lattice potential.

\subsection{$\text{Cd}_3\text{As}_2$ and influence of the thickness of the thin film}
Here, we show further extended data on Cd$_3$As$_2$ thin film, beyond what is shown in the main text.
\subsubsection{Other superlattice potential periods}
\begin{figure}
    \centering
    \includegraphics[width=0.6\textwidth]{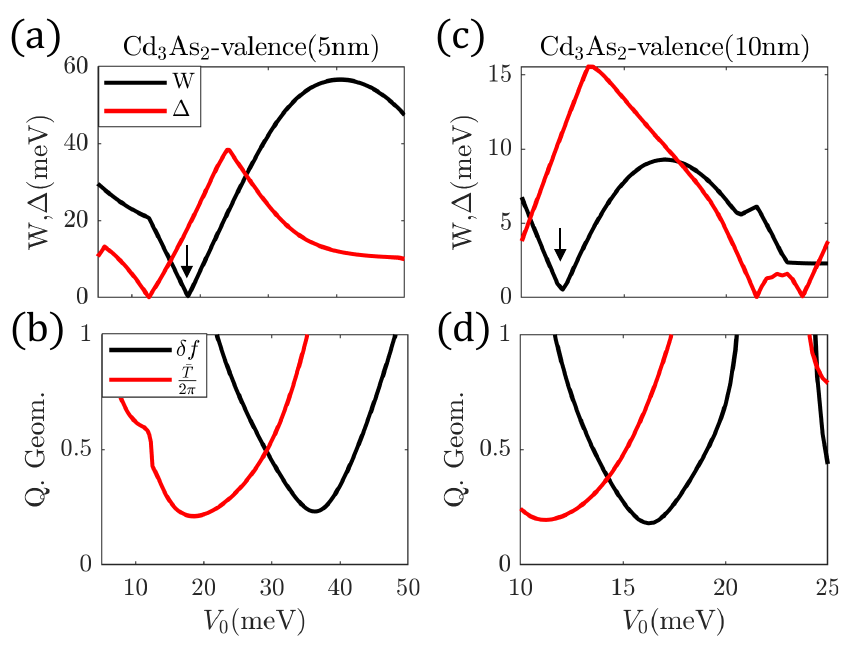}
    \caption{ Properties of the first valence miniband of Cd$_3$As$_2$ for various superlattice periods. The bandwidth $W$, direct gap $\Delta$, Berry curvature standard deviation $\delta f$, trace-condition violation $\bar{T}$, as a function of potential strength $V_0$ for $\phi=0$. (a,b) for the potential period $a=5\text{nm}$, (c,d) for the for $a=10\text{nm}$. Black arrow indicates the optimal point where bandwidth is minimized.}
    \label{SPfig_CdAs_10nm}
\end{figure}

In the main text, we showed the  data for $\text{Cd}_3\text{As}_2$ when the potential period is $a=20\text{nm}$. 
Here, we demonstrate that this is not a fine-tuned result, and that a similarly flat band also emerges for a wide range of superlattice periods.
Fig.~\ref{SPfig_CdAs_10nm} shows the properties of the first valence band for a superlattice potential $a=\SI{5}{nm}$ and $a=\SI{10}{nm}$.  
Both of these  show a pronounced dip in the bandwidth and quantum geometric properties, similar to that observed for $a=\SI{20}{nm}$ used in the main text.


\subsubsection{Finite thickness of the film}\label{sec:zdept}

In the main text, we model the effect of the electrostatic potential as a scalar potential that couples only to the total in-plane density. 
However, when the potential arises from a 2D substrate on one side\cite{kim2023electrostatic}, it will decay exponentially in the $z$ direction as $\exp(-Gz)$, where $G$ is the length of the primitive reciprocal lattice vector of the potential, and $z$ is the distance from the  substrate. 
If the film thickness is small so $GL\ll 1$, then it is a reasonable approximation to ignore this z-dependence. 
However, we chose $L=\SI{8}{nm}$ as the thickness for $\text{Cd}_3\text{As}_2$, while the potential period is $a=\SI{20}{nm}$.  
It may therefore be important to take this effect into account.

Without considering the $z$-dependence, the $C_3$ symmetric scalar potential is
\begin{equation}
V(\bm{r})=2V_0\sigma_0\sum_{n=1}^{3} \cos(\bm{g}_n\cdot\bm{r}+\phi)
\end{equation}
where $\sigma_0$ acts on the orbital space of the Hamiltonian describing the topological band inversion, as defined in the main text. 
When the z-dependence of the potential is taken into account, the $I_1=\tau_z\sigma_z(-1)^n$ isospin is no longer conserved. 
However, the $I_2$ isospin, defined earlier for Cd$_3$As$_2$, is still a conserved quantum number even in the presence of such a potential.
We therefore focus on the $I_2=-1$ sector for the following.

To model the $z$ dependence of the potential, we  define the following 2-by-2 matrix
\begin{equation}
   (H_V)_{i,j}\equiv \braket{i}{\exp(-G z)|j} 
\end{equation}
where $z$ should be understood as an operator acting on the original $\ket{n_z,\tau_z,\sigma_z}$ basis (defined in Eq~\ref{eq:sinbasis}), and
 $\ket{1},\ket{2}$ are the two eigenstates of $H(k_x=0,k_y=0,k_z\to -i\partial_z)$ closest to charge neutrality in the $I_2=-1$ sector.
We substitute $\sigma_0$ by $H_V$ in  $V(\bm{r})$ to account for the $z$-dependence. 

\begin{figure}
    \centering
    \includegraphics[width=0.6\textwidth]{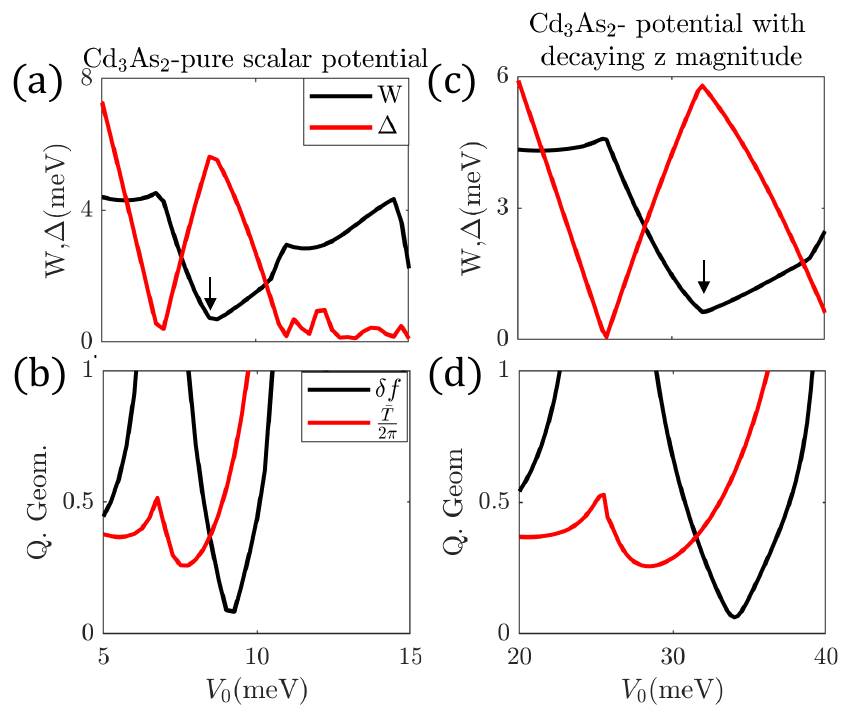}
    \caption{Taking into account the finite thickness of the Cd$_3$As$_2$ film.
    Parameters used are potential period is $a$=20nm, potential shape $\phi=0$. (a,b) ignores the decaying magnitude of the potential along the $z$ direction, while (c,d) takes this decay into consideration by the method specified in the text. Black arrow indicates the point where bandwidth is minimized.}
    \label{SPfig_CdAs}
\end{figure}

As we show in Fig. \ref{SPfig_CdAs}, the primary effect of this $z$-dependence is to shift the position of the dip in bandwidth and quantum geometrical quantities.
However, the fact that such a minimum exists is robust.   
We scan $V_0$ for both the case with and without considering the effect of the decaying $z$ potential. 
For pure scalar potential case, the dip occurs at $V_0\sim 10\text{meV}$. Taking into consideration of the z-decay shifts $V_0$ to around 30meV.

\begin{figure}
    \centering
    \includegraphics[width=0.6\textwidth]{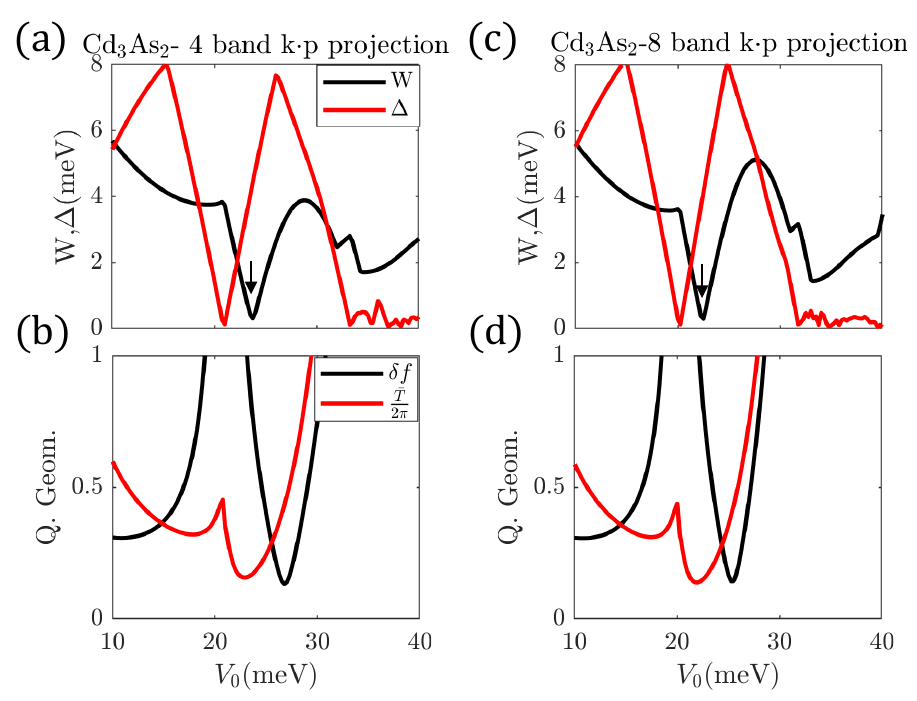}
    \caption{Same as Fig.\ref{SPfig_CdAs}(c,d), except that we keep more bands in when doing the $k\cdot p$ expansion. (a,b) shows the results for keeping 4 bands, and (c,d) shows the results for 8 bands.}
    \label{SPfig_morebands}
\end{figure}

\subsubsection{Beyond two-band approximation}
In the above, we have presented results when we only project to the two bands closest to charge neutrality. 
In principle, the number of bands we use when doing the $k\cdot p$ projection should be determined by convergence of the resulting minibands. 
When the z-dependence is ignored, the results presented above for $\text{Cd}_3\text{As}_2$ and $\text{Sb}_2\text{Te}_3$ are already well converged in the two-band approximation. However, $\exp(-Gz)$ will have non-negligible elements between the two lowest bands and other remote bands, making it necessary to keep more bands when doing the $k\cdot p$ projection.

Fig.\ref{SPfig_morebands} shows the results of keeping more bands when doing the $k\cdot p$ expansion for $\text{Cd}_3\text{As}_2$, including the effect of z-dependence. 
We find that the dip in bandwidth and quantum geometrical quantities only shifts in position, demonstrating its robustness as expected based on the physical picture of its origin presented in the main text.

\end{appendix}

\end{document}